\documentclass{article}

\usepackage{amsfonts}
\usepackage{amsmath}
\usepackage{amssymb}
\usepackage{amsbsy}
\usepackage{epsfig}
\usepackage{epstopdf}
\usepackage{subfigure}
\usepackage{graphicx}
\usepackage{mathptmx}
\usepackage[round,sort,authoryear]{natbib}
\usepackage{color}
\usepackage[colorlinks,bookmarksopen,bookmarksnumbered,citecolor=red,urlcolor=red]{hyperref}
\usepackage{lscape}
\usepackage{makecell}
\usepackage{array}

\begin{document}

\title{Hemodynamic assessment of pulmonary hypertension in mice: A model based analysis of the disease mechanism}
\author{M. Umar Qureshi \and Mitchel Colebank \and Mihaela Paun\and Laura Ellwein \and Naomi Chesler\and Mansoor A. Haider\and Nicholas A. Hill\and Dirk Husmeier\and Mette S. Olufsen}
\date{\vspace{-5ex}}
\newcommand{\Addresses}{{
  \bigskip
  \footnotesize

 M.~Umar~Qureshi (Corresponding author), \textsc{Department of Mathematics, North Carolina State University,
    Raleigh, North Carolina 27695, USA}\par\nopagebreak
  \textit{E-mail address}: \texttt{muquresh@ncsu.edu}

  \medskip

  Mitchel Colebank,  Mansoor A. Haider, Mette S. Olufsen \textsc{Department of Mathematics, North Carolina State University,
    Raleigh, North Carolina 27695, USA}
  \medskip 

Laura Ellwein  \textsc{ Department of Mathematics and Applied Mathematics, Virginia Commonwealth University, Richmond, VA 23284, USA}
  \medskip

  Naomi Chesler \textsc{Department of Biomedical Engineering and Madison, University of Wisconsin at
Madison, WI, 53706-1609, USA}
  \medskip

  Mihaela Paun, Dirk Husmeier, Nicholas A. Hilll \textsc{School of Mathematics and Statistics, University of Glasgow, Glasgow G12 8QQ, UK}\par\nopagebreak

}}

\maketitle

\begin{abstract}
This study uses a one dimensional fluid dynamics arterial network model to infer changes in hemodynamic quantities associated with pulmonary hypertension in mice. Data for this study include blood flow and pressure measurements from the main pulmonary artery for 7 control mice with normal pulmonary function and 5 hypertensive mice with hypoxia induced pulmonary hypertension. Arterial dimensions for a 21 vessel network are extracted from micro-CT images of lungs from a representative control and hypertensive mouse. Each vessel is represented by its length and radius. Fluid dynamic computations are done assuming that the flow is Newtonian, viscous, laminar, and has no swirl. The system of equations is closed by a constitutive equation relating pressure and area, using a linear model derived from stress-strain deformation in the circumferential direction assuming that the arterial walls are thin, and also an empirical nonlinear model. For each dataset, an inflow waveform is extracted from the data, and nominal parameters  specifying the outflow boundary conditions are computed from mean values and characteristic time scales extracted from the data. The model is calibrated for each mouse by estimating parameters that minimize the least squares error between measured and computed waveforms. Optimized parameters are compared across the control and the hypertensive groups to characterize vascular remodeling with disease. Results show that pulmonary hypertension is associated with stiffer and less compliant proximal and distal vasculature with augmented wave reflections, and that elastic nonlinearities are insignificant in the hypertensive animal.
\end{abstract}


\section{Introduction}\label{intro}

Pulmonary hypertension (PH) is defined as an invasively measured mean pulmonary arterial blood pressure (mPAP) greater than 25\,mmHg \citep{Simonneau2013}. It is associated with vascular remodeling, which adversely affects the properties of the cardiopulmonary system, including pulmonary vascular resistance (PVR), proximal and distal arterial stiffness, compliance, and amplitude of wave reflections \citep{Nichols2011}. The mPAP and PVR are conventionally used diagnostic markers for PH but are not good indicators of disease severity \citep{Wang2011}. Here we use  proximal arterial stiffness and the wave reflection amplitude as  physiomarkers for detecting disease progression \citep{Castelain2001,Hunter2011}. In particular, the proximal arterial stiffness is an excellent predictor of mortality in patients with pulmonary arterial hypertension \citep{Gan2007}. Quantifying relative distributions of proximal and distal arterial stiffness (or compliance) and wave reflections in elevating the mPAP and PVR is vital for understanding  disease mechanisms.

In this study, we setup and calibrate a mathematical model predicting wave propagation in the pulmonary vasculature in C57BL6/J male mice with normal pulmonary function (control group (CTL), n = 7) and in mice with hypoxia-induced pulmonary hypertension (hypertensive group (HPH), n = 5) \citep{Tabima2012,Vanderpool2011}. The novelty of this study is the integration of high fidelity morphometric and hemodynamic data from multiple mice with a one dimensional (1D) model of large pulmonary arteries coupled with a zero dimensional (0D) model of the vascular beds. This is achieved by incorporating available data at each stage of the modeling including network extraction, parameter estimation and model validation. The outcome is used to infer disease progression by quantifying relative changes in PVR, proximal and distal arterial stiffness, compliance, and amplitudes of wave reflections, across the two groups (CTL and HPH). Moreover, we investigate the influence of presumed elastic nonlinearities in the wall model on the parameter inference. This approach allows us to study disease impact on the distal vasculature by predicting waveforms from multiple locations in the simulated network, which are difficult to obtain experimentally.

\paragraph{Experimental studies.} To understand the relation between hemodynamics and vascular remodeling, it is essential to analyze morphometric and hemodynamic time-series data during disease progression. Morphometric data can be obtained by non invasive procedures, like magnetic resonance imaging (MRI) or computed tomography (CT) scans \citep{Meaney2012}, but dynamic pulmonary blood pressure can only be obtained from right heart catheterization \citep{Rich2011}. Moreover, in humans, the disease takes years to develop making it difficult to study its progression. An alternative means to gain understanding is to study disease progression in mouse models. An advantage is that mice have a relatively short lifespan and it is feasible to generate specific disease groups (e.g. mice with HPH) within a short time-span ($<$1 month). Experimental studies in mice are typically done within a specific genetic strain, in order to limit variation among individuals. In most studies (e.g. \cite{Pursell2016,Tabima2012,Vanderpool2011}), hypoxia is used to induce pulmonary hypertension. This type of PH (Group III) is clinically manifested in human patients with hypoxia and lung disease \citep{Simonneau2013}, believed to be initiated by remodeling of the vascular beds (e.g microvascular vasoconstriction and rarefaction \citep{Wang2011}) followed by progressive remodeling of the large arteries \citep{Tuder2007a}. Therefore, investigation of pulmonary hypertension in mice with hypoxia may provide vital understanding of disease mechanisms in humans with similar pathology.

\paragraph{Modeling studies.}  Examples of previous modeling efforts include lumped 0D \citep{Lankhaar2006,Lumens2009}, distributed 1D \citep{Acosta2017,Lungu2014,Qureshi2014} and locally complex three dimensional (3D) \citep{Tang2012,Yang2016} models. \cite{Lankhaar2006} combined a 0D 3-element Windkessel model with hemodynamic data from PH patients to quantify right ventricular afterload.  \cite{Lumens2009} developed a geometric heart model coupled with a closed loop 0D model of the entire circulation to predict ventricular hypertrophy in patients with PH. \cite{Tang2012} and \cite{Yang2016} constructed patient specific 3D models of pulmonary arteries to study shear stress \citep{Tang2012}, and PVR in pre and post operative situations \citep{Yang2016}. To our knowledge, only \cite{Lungu2014} used a coupled 1D-0D model of the main pulmonary artery (MPA) to study diagnostic parameters including PVR, mPAP, arterial stiffness and compliance in patients with and without PH. Studies by \cite{Acosta2017} and \cite{Qureshi2014} used a 1D framework to develop distributed models of the pulmonary arteries and veins to study cardiopulmonary coupling \citep{Acosta2017} and microvascular remodeling \citep{Qureshi2014} during PH. Yet, none of these studies used actual pressure data for parameter estimation and model validation. Other notable studies using 1D models, but not investigating pulmonary hypertension, include \cite{Blanco2014,Mynard2015,Olufsen2000,Reymond2009,Willemet2015}. See \cite{Boileau2015,Safaei2016,Vosse2011} for an overview of modeling approaches and computational tools, and \cite{Hunter2011,Kheyfets2013,Tawhai2011} for focused reviews on modeling pulmonary hypertension.

Most of the above studies consider application to humans. Only a few of 1D modeling studies have investigated wave propagation in systemic \citep{Aslanidou2015} and pulmonary \citep{Pilhwa2016,QureshiCMBE17} arteries in mouse networks. To our knowledge, no studies combine disease specific imaging and hemodynamics data to predict changes in disease. In this study, we expand upon prior results from \cite{QureshiCMBE17} by developing a 1D fluid dynamics network model of pulse wave propagation in the large pulmonary arteries of control and hypertensive mice. We do so by extracting detailed network information and by optimizing hemodynamic predictions. To set up the 1D domain, we extract the arterial networks from micro-CT images of a representative CTL mouse and a mouse with HPH. We combine these networks with hemodynamic data, measured in the MPA of each mouse, to predict pressure and flow waveforms. For each vessel in the network, we solve (numerically) fluid dynamics equations derived from the Navier--Stokes equations coupled with a constitutive equation relating pressure and vessel area (i.e. describing the vessel wall mechanics). We use the measured flow waveform from each mouse as the inflow boundary condition and use 0D Windkessel models \citep{Westerhof2009} to characterize the impedance of the vascular beds. 
 
As the disease progresses, vascular remodeling changes the composition of wall constituents \citep{Humphrey2008}. This affects the local stress-strain response, which is important for inferring arterial wall stiffness and compliance \citep{Hunter2011}. We focus on understanding how wall deformation changes under hypertensive conditions. Specifically, we aim to study if  presumed elastic nonlinearities have a significant impact on disease progression. To this end we use a well established linear wall model \citep{Willemet2015,Safaei2016} as well as an empirical nonlinear  wall model.

The advantage of the linear wall model is that it is easy to derive from first principles and it has been shown to be successful within physiological pressure/area values for systemic arteries \citep{Olufsen2000}. Yet, it does not account for the fact that arteries stiffen with pressure. On the other hand, detailed, structural hyperelastic wall models can be derived from first principles, e.g. \cite{HO2010}, but they are difficult to analyze due to the large number of parameters. Inspired by \cite{Langewouters1985}, we introduce a simple empirical nonlinear model with three key properties: (a) it predicts vessel stiffening with pressure so that the lumen area approaches a finite limit as pressure increases, (b) it incorporates an undeformed area (or radius) corresponding to zero transmural pressure, and (c) it reduces to the linear wall model under a small strain assumption providing a basis for model comparison and nominal parameter estimation. 

The process of modeling requires {\it a priori} specification of  parameters, including proximal arterial stiffness, total vascular resistance, and peripheral compliance, which are known to vary across individuals. This creates the need for methods to estimate parameters that predict observed hemodynamics across individual mice.  A few recent studies have performed parameter estimation in blood flow models, including the study by \cite{Eck2017}, which used polynomial chaos expansion to analyze a stochastic model of pressure waves in the large systemic arteries, the study by \cite{Andrea2017}, which used an ensemble Kalman filter (EnKF) to estimate the unknown inflow to a single vessel, representing the ovine aorta, and the study by \cite{Tran2017}, which used a Bayesian Markov chain Monte Carlo (MCMC) approach to estimate parameters in a multi-scale three dimensional model of coronary arteries. Similarly, in the recent study by \cite{Paun2018}, we used MCMC  to estimate parameters for a 1D model of mouse pulmonary arteries. It should be noted that MCMC algorithms come at a substantial computational cost, making them infeasible for multi-subject studies. Other studies estimating pressure dynamics using optimization algorithms were done using either 0D \citep{Valdez2011,Williams2014} or 1D \citep{Lungu2014} models.  Yet, none of these studies investigated hemodynamic variation in the pulmonary arterial network.
 
In this study, we estimate global network parameters that allow prediction of observed dynamics in both CTL and HPH groups. We first determine a {\it priori} parameter values for the wall models and boundary conditions. This is done by combining available data and existing results in the literature \citep{Alastruey2016,Reymond2009}. Second, we solve the overall model using linear and nonlinear wall models and conduct constrained nonlinear optimization to estimate parameters predicting the observed dynamics. A similar approach combining  {\it a-priori} nominal parameters with iterative tuning was used by \cite{Alastruey2016} in a model of the human aorta. In this study blood flow data were available for all terminal vessels, making it easier to compute downstream resistance. To our knowledge, our approach estimating parameters for a 1D network model of the pulmonary circuation using morphometric and dynamic pressure and flow data is novel relative to these studies.

Similar to previous studies \citep{Pilhwa2016,Qureshi2014,QureshiCMBE17}, we compare the hemodynamic signatures in the time and frequency domains for the control and hypertensive animals. We follow \cite{Acosta2017,Lankhaar2006,Lungu2014} to analyze the estimated parameters, inferring HPH progression, and to investigate, using a model selection criterion \citep{Burnham2001,Schwarz1978}, the extent to which the nonlinear wall model enables accurate prediction of the observed dynamics.\\

The manuscript is organized as follows: Sec.~\ref{sec:Methods} presents experimental and mathematical methods including data extraction procedures, governing equations, parameter estimation and numerical simulations. In Sec.~\ref{sec:Results} we present results comparing CTL and HPH hemodynamics, analyzing: waveforms predictions along the arterial network, estimated parameters and their sensitivities to hypertensive conditions, and impedance and wave reflection analysis. Key findings are discussed in Sec.~\ref{sec:discussion}, followed by limitations in Sec.~\ref{sec:Limitations}. Finally, we state key conclusions in Sec.~\ref{sec:conclusion}.

\section{Methods}\label{sec:Methods}

\subsection{Experimental Methods}\label{sec:data}
This study uses existing hemodynamic data and micro computed tomography (micro-CT) images from control and hypertensive mice. Detailed experimental protocols for extracting the hemodynamic and image data can be found in \cite{Tabima2012} and \cite{Vanderpool2011}, respectively. Both procedures were approved by the University of Wisconsin Institutional Animal Care and Use Committee. Below we summarize these protocols and highlight the data analyzed herein.

\begin{figure}[h]
\centering
\includegraphics[scale = 0.23]{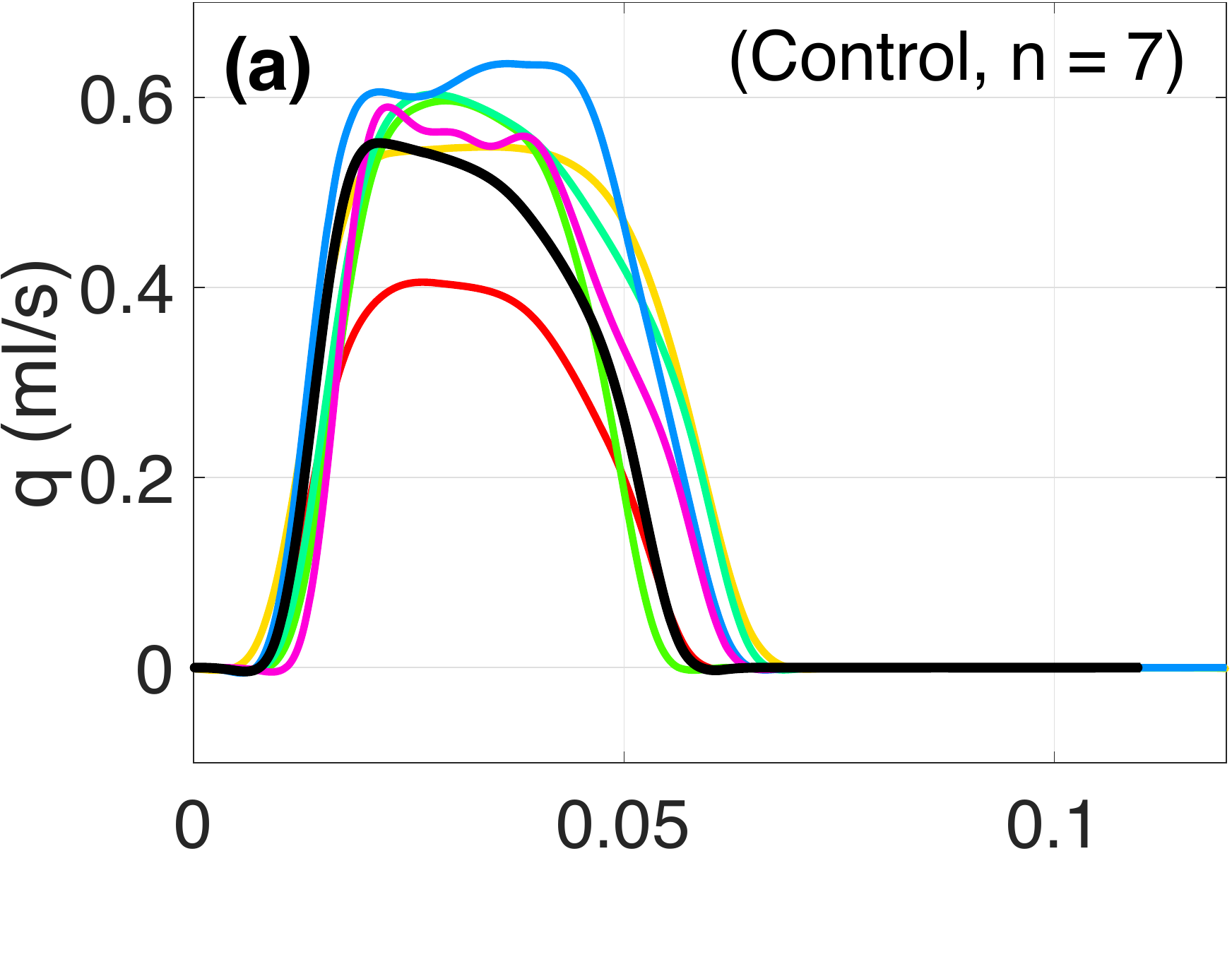}
\includegraphics[scale = 0.23]{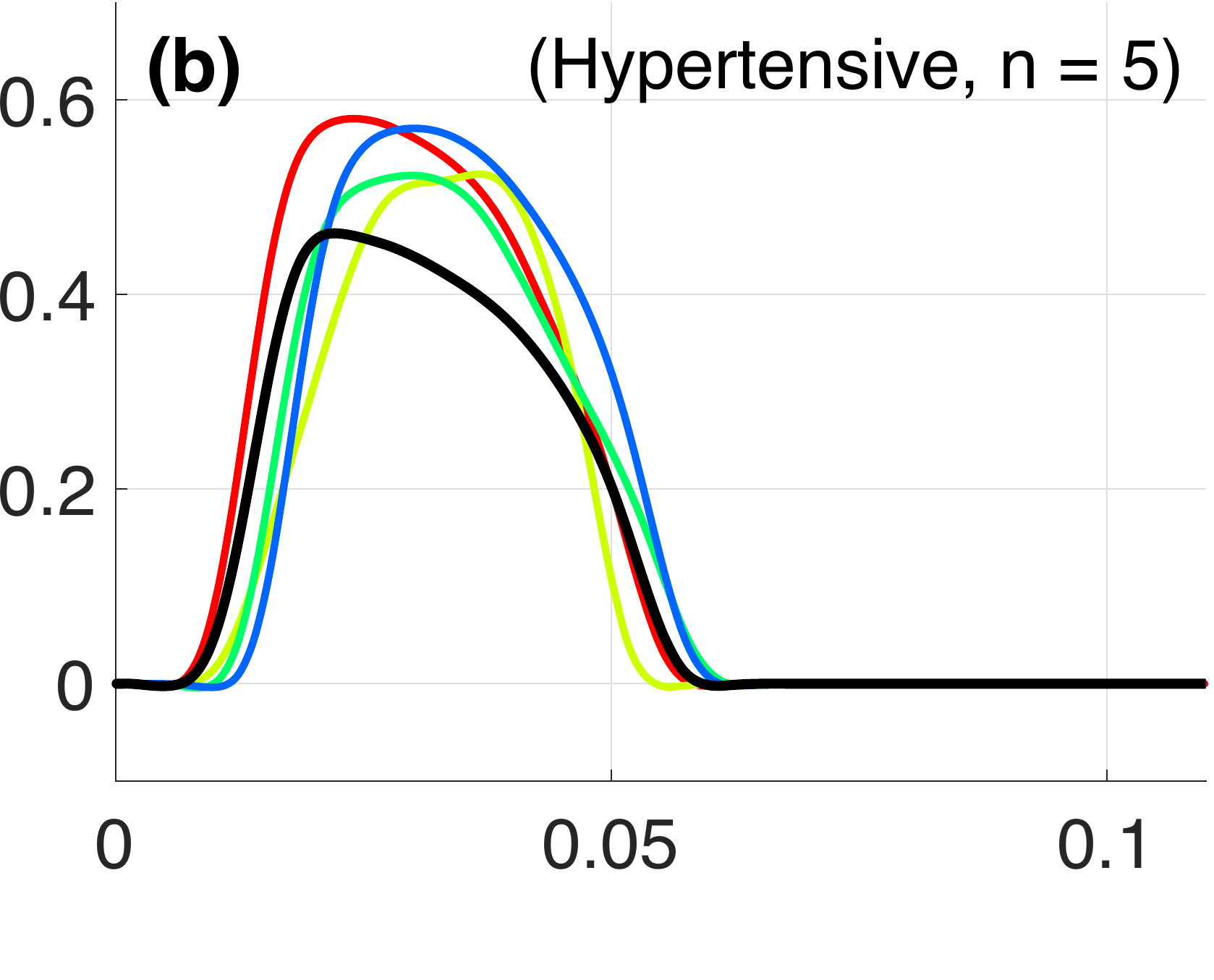}\vspace{-0.2cm}
\includegraphics[scale = 0.23]{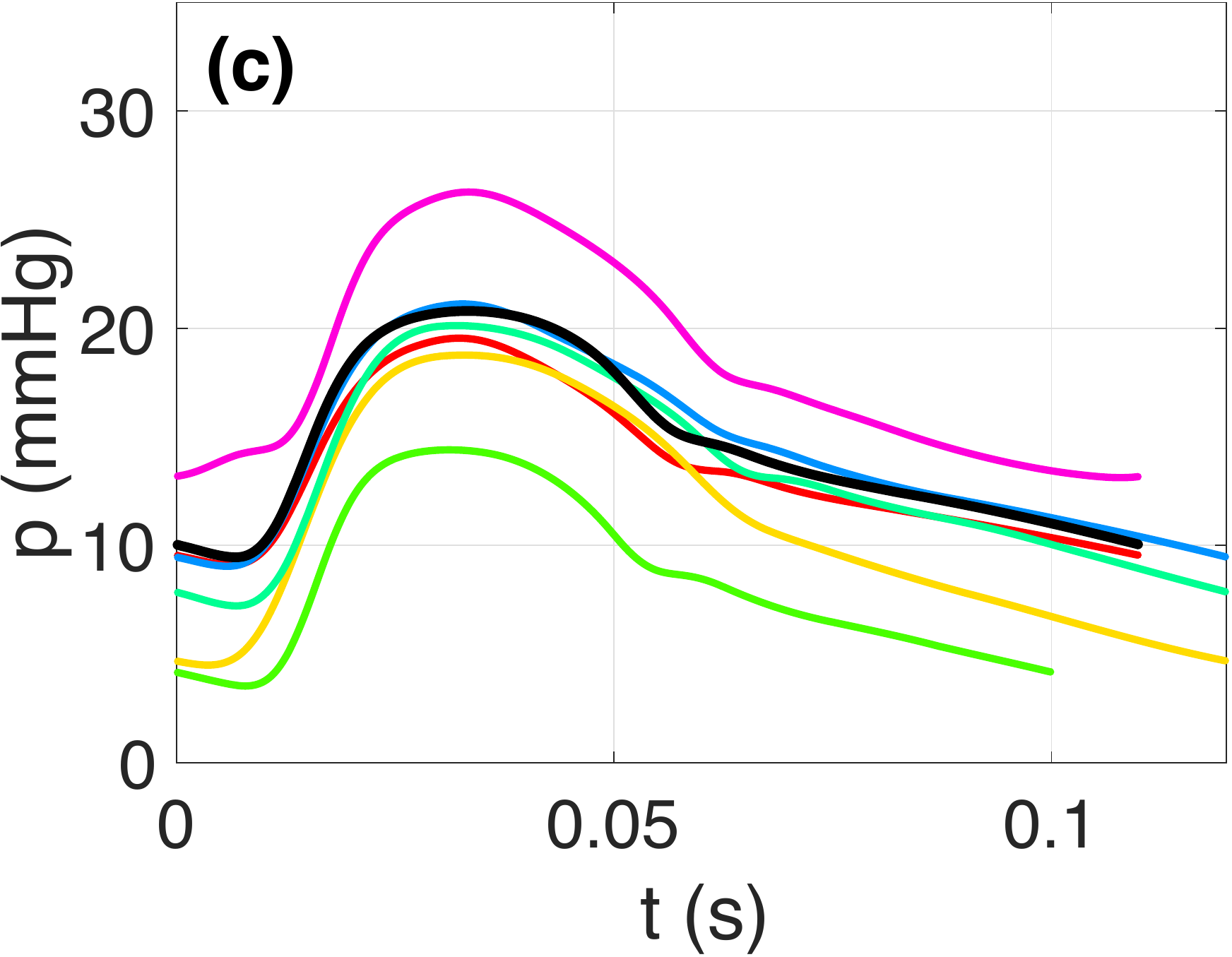}
\includegraphics[scale = 0.23]{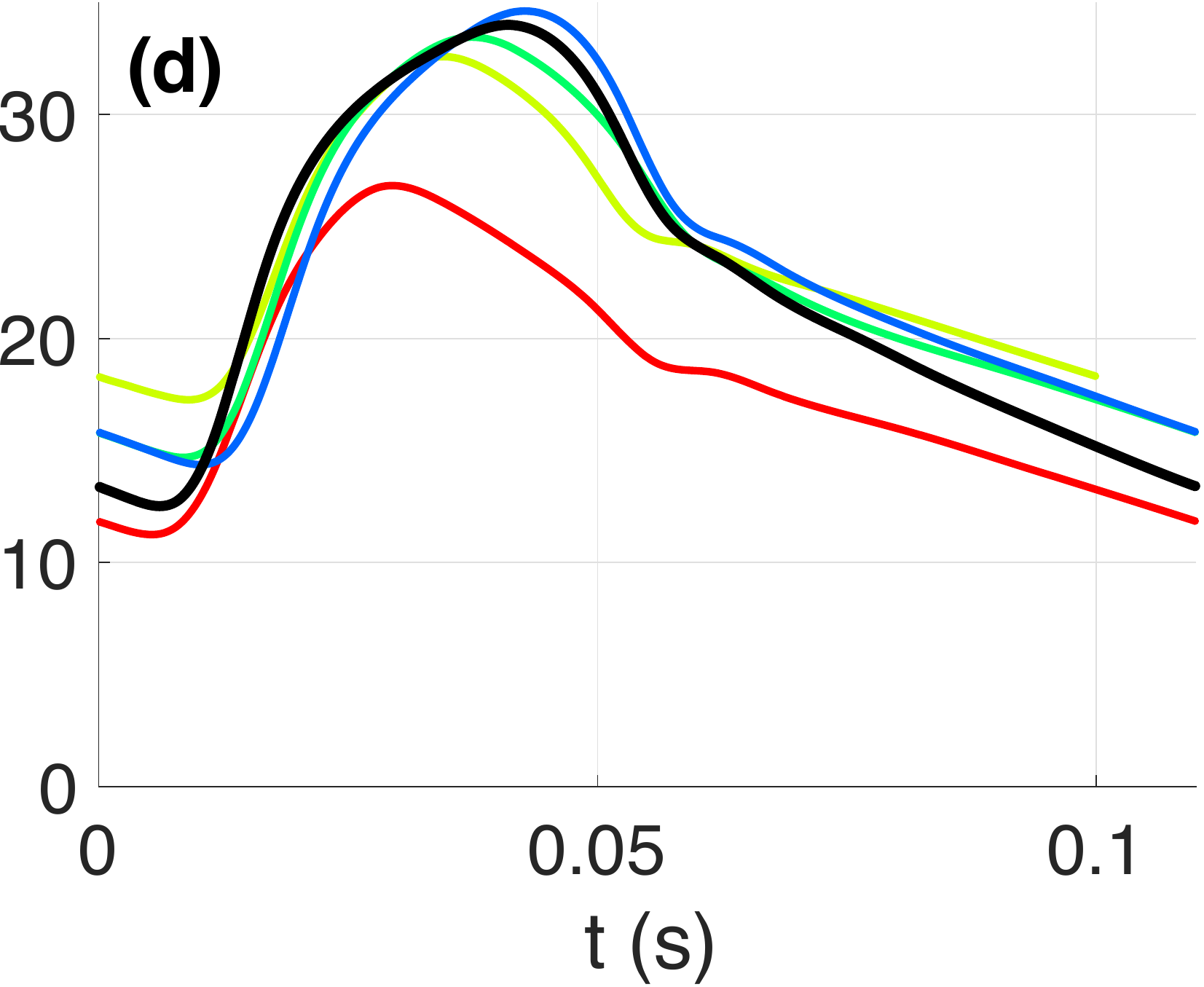}\vspace{0.2cm}
\includegraphics[scale = 0.23]{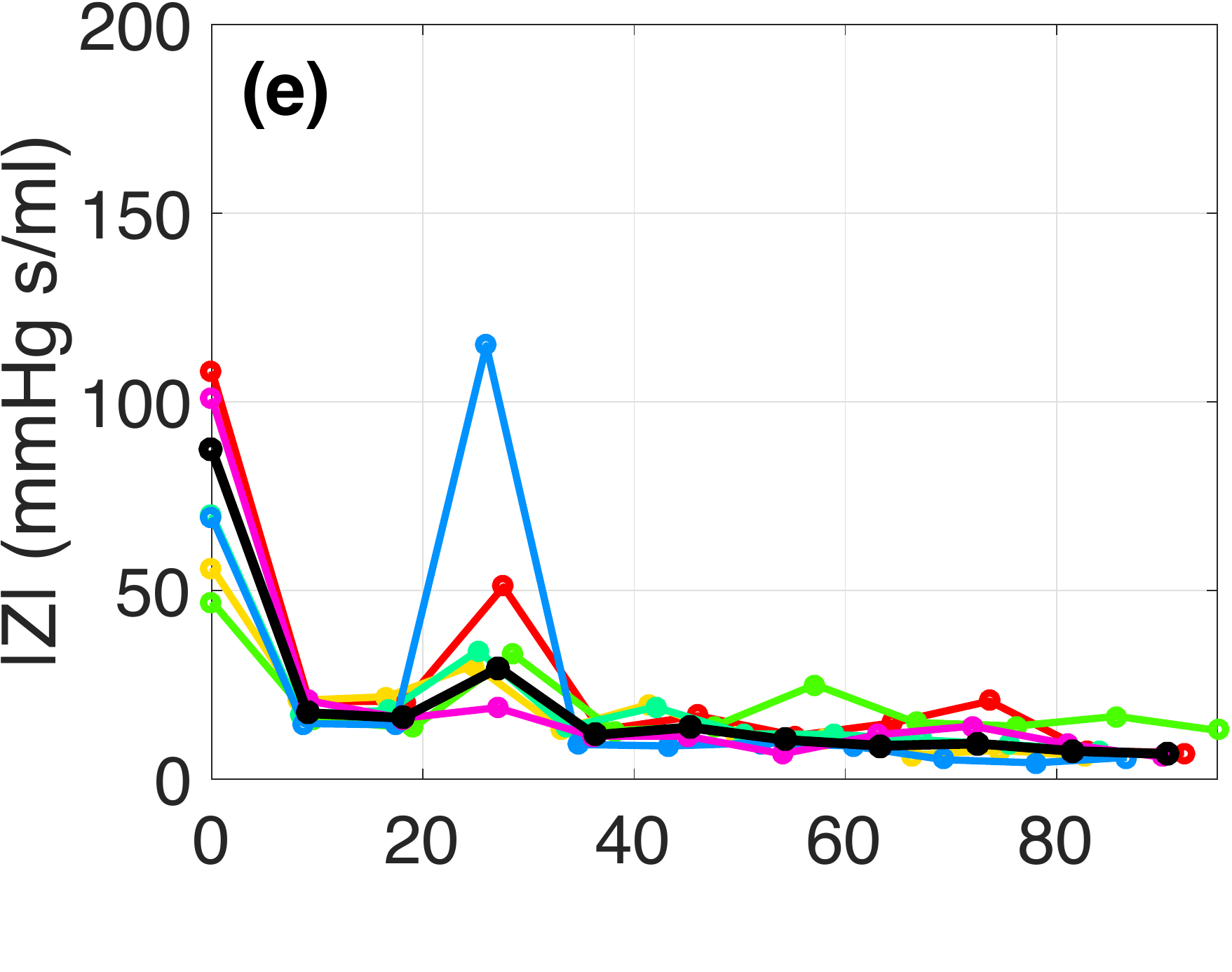}
\includegraphics[scale = 0.23]{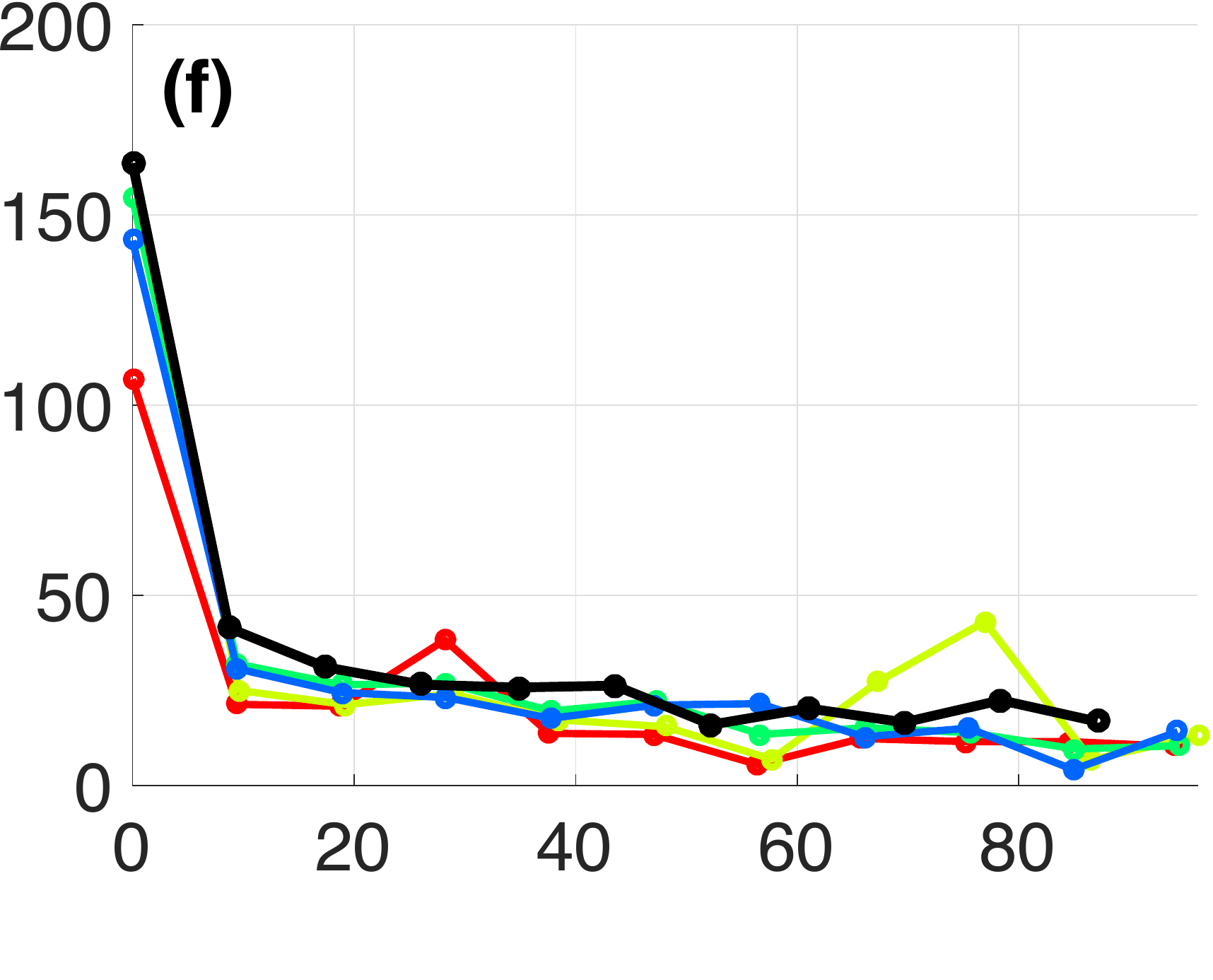}
\includegraphics[scale = 0.23]{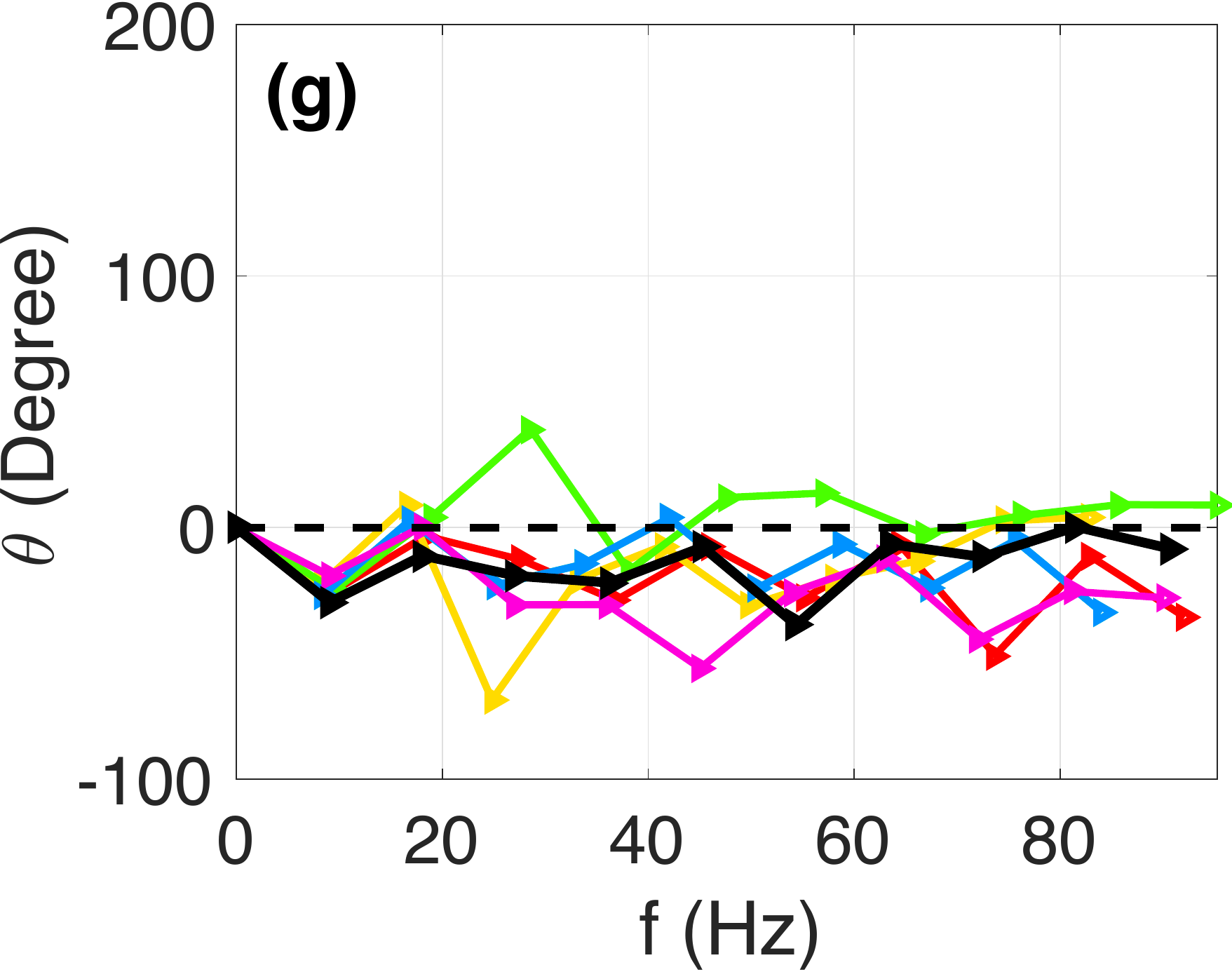}
\includegraphics[scale = 0.23]{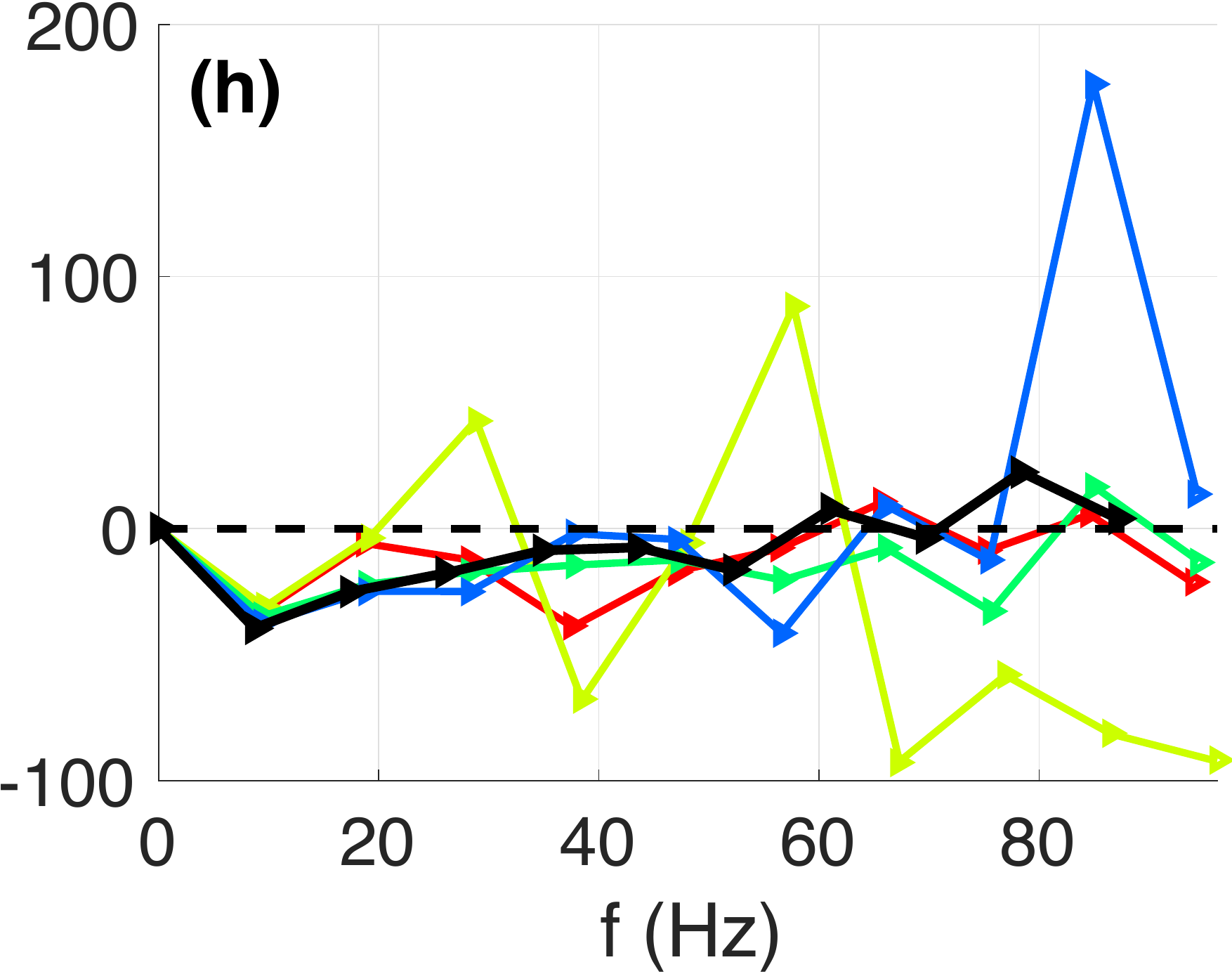}\\
\caption{\footnotesize{(a--d) Simultaneously measured flow and pressure waveforms in MPA of control and hypertensive mice. Each waveform is averaged over 20 consecutive cardiac cycles. (e--h) shows associated frequency domain signatures, where $f$ is the frequency, $|Z|$ is impedance modulus and $\theta$ is the associated phase angle (see eq.~(\ref{eq:Zin})). Thick black curves represent the representative control and hypertensive animal for which the simulations are presented in this study.}}\label{fig:hemoData}
\end{figure}

\paragraph{\bf Hemodynamic data.}
The hemodynamic data include dynamic pressure and flow waveforms from male C57BL6/J mice, average age 12-13 weeks and average body weight of 24\,g. The mice were divided into CTL (n = 7) and HPH groups (n = 5). The mice in the HPH group were exposed to 21 days of chronic hypoxia (10\% O$_2$ partial pressure) and both groups were exposed to a 12 hour light-dark cycle. Mice were instrumented to obtain dynamic pressure and flow waveforms in the main pulmonary artery as described previously \citep{Tabima2012}. In brief, mice were anesthetized with an intraperitoneal injection of urethane solution (2mg/g body weight) and placed on a heating pad to maintain physiological heart rate. After intubation for ventilation, the chest wall was removed to expose the right ventricle. A 1.0F pressure-tip catheter (Millar Instruments, Houston, TX) was inserted into the apex of right ventricle and advanced to the MPA. Flow was measured with ultrasound (Visualsonics, Toronto, Ontario, Canada) using a 40 MHz probe during catheterization and recorded synchronously with pressure and ECG (Cardiovascular Engineering, Norwood, MA). Pressure and flow waveforms were signal-averaged using the ECG as a fiducial point. Twenty consecutive cardiac cycles were averaged to produce average pressure and average flow waveforms. Representative hemodynamic data and associated frequency domain signatures are shown in Figure~\ref{fig:hemoData} and essential cardiovascular parameters are summarized in Table~\ref{Tab:HemoStats}. 

\begin{table}[h]
\centering
\caption{{Average hemodynamic characteristics (mean $\pm$ SD) for the control and hypertensive animals.}}\label{Tab:HemoStats}
{\footnotesize
\begin{tabular}{lll}
\hline\noalign{\smallskip}
&\textbf{Healthy (n = 7)}& \textbf{Hypoxia (n = 5)}\\
\noalign{\smallskip}\hline\noalign{\smallskip}
HR (beats/min)&533$ \ \pm \ $27&559$ \ \pm \ $21\\[2pt]
CO (ml/s)&0.18$ \ \pm \ $0.03&0.15$ \ \pm \ $0.01\\[2pt]
mPAP (mmHg)&13.4$ \ \pm \ $3.1&22.1$ \ \pm \ $2.3\\[2pt]
sPAP (mmHg)&20.1$ \ \pm \ $3.5&32.3$ \ \pm \ $3.1\\[2pt]
dPAP (mmHg)&8.0$ \ \pm \ $3.3&14.0$ \ \pm \ $2.3\\[2pt]
pPAP (mmHg)&12.2$ \ \pm \ $1.4&18.3$ \ \pm \ $2.8\\[2pt]
PVR (mmHg.s/ml) &77.0$ \ \pm \ $22.7&146.1$ \ \pm \ $23.6\\ 
\noalign{\smallskip}\hline
\end{tabular}}

\vspace{0.1cm}
{\footnotesize Abbreviations: Heart rate (HR), cardiac output (CO),  mean pressure (mPAP), systolic pressure (sPAP), diastolic pressure (dPAP),  pulse pressure (pPAP) in the main pulmonary artery, total pulmonary vascular resistance (PVR).} 
 \end{table}

\paragraph{\bf Imaging data.} Stacked planar X-ray micro-CT images of pulmonary arterial trees were obtained from male C57BL6/J mice, average age 10-12 weeks, under control (healthy mice) and 10 days of hypobaric hypoxia at 10\% O$_2$ partial pressure protocols. Detailed descriptions of animal handling and lung preparation can be found in \cite{Vanderpool2011}, whereas details of the micro-CT image acquisition are described in \cite{Karau2001}. In brief, mice were anesthetized with intraperitoneal injection of pentobarbital sodium  (52 mg/kg body weight) and then euthanized by exsanguination. The trachea and the main pulmonary artery were cannulated and the heart was dissected away. Pulmonary artery cannula (PE-90 tubing, 1.27mm external and 0.86mm internal diameter) was positioned well above the first bifurcation. Excised lungs were treated with Rho kinase inhibitor and ventilated and perfused with perfluorooctyl bromide (PFOB), a vascular contrast agent, and placed in the imaging chamber. The arterial trees were then imaged under a static filling pressure of 6.3 mmHg, while rotating the lungs in the X-ray beam at 1$^{\circ}$ increments to obtain 360 planar images. Each planar image was averaged over seven frames to minimize noise and maximize vascular contrast. Fo each lung, isometric 3D volumetric dataset (497$\times$497$\times$497 pixels) was obtained, by reconstructing the 360 planer images using the Feldkamp cone-beam algorithm \citep{Feldkamp1984}, and converted into Dicom 3.0.

\begin{figure*}
\centering
\includegraphics[scale = 0.4]{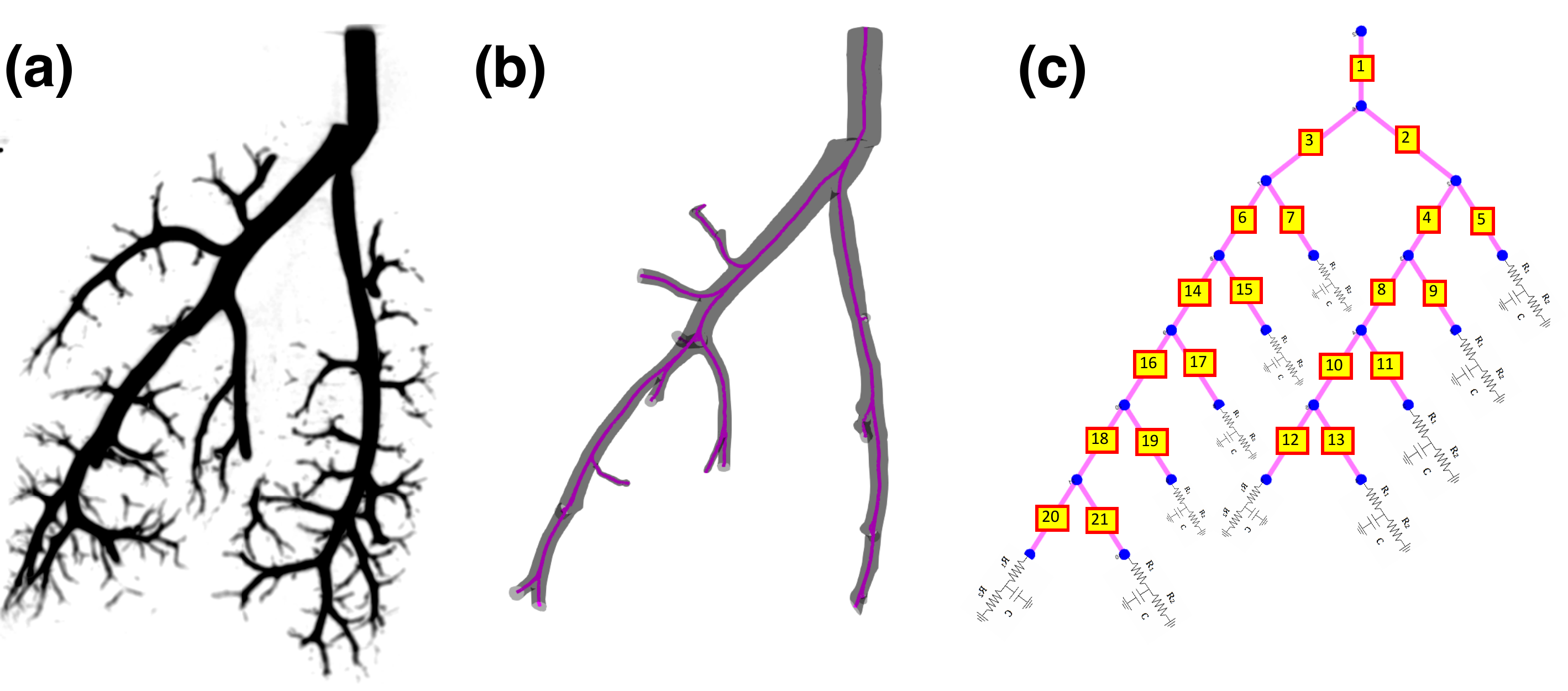}\\
\caption{\footnotesize 1D network segmentation: (a) the micro-CT image, (b) 3D smoothed network for centerline extraction, and (c) the directed graph reflecting connectivity of vessels in the network.The size of segments in the graph do not reflect their dimensions.}\label{fig:Network}
\end{figure*}

For this study, two representative networks with 21 vessels were extracted from images of the control and hypertensive mice. The 21 vessel network was chosen since it was the most expansive network that could be identified with a one-to-one vessel map in both control and hypertensive animals. Network dimensions and connectivities were obtained using the segmentation protocol described by \cite{Ellewin2015}. This protocol uses ITK-SNAP \citep{ITKSNAP} to create a 3D geometry from Dicom 3.0 files, using semi-automated ``snake evolution" in the regions of interest (the 21 vessels). The image pixel threshold was set at 45 to reduce artifact detection. Paraview (Kitware; Clifton Park, NY) was used to convert file types to vtk polygonal data (.vtp) allowing us to compute centerlines and connectivity using the Vascular Modeling ToolKit (VMTK \citep{VMTK}). The output from VMTK is a $n\times4$ matrix representing each vessel by a unique set of coordinates ${\bf{x}}_i\in\mathbb{R}^3\,(i =0,\dots, n-1)$ and the associated radius value, $r_i$, computed from the maximally inscribed sphere within the 3D vessel. Known internal diameter of the contrast filled cannula (PE-90 tubing) was used to compute a scaling factor to convert voxels into cm. For the MPA, the radius $\overline{r}_0$ was computed as a mean of all slices $r_i$ between the cannulated region and the first bifurcation, whereas the radii of other vessels were computed as the mean of all slices along the vessels but away from the junctions. For each vessel, the length, $L$, was calculated as the sum of the shortest distances ($l_i$) between successive points, i.e.
\begin{equation}\label{eq:dist}
\overline{r}_0 = \frac{1}{n}\sum_{i} r_i, \quad L = \sum_i l_i , \quad{\textrm{where}}\quad l_i = \|{\mathbf{ x}}_{i+1} - {\bf{ x}}_{i}\|,\\
\end{equation}
$i = 0,\dots,n-1$ and $n$ is the total number of samples for the vessel. The network, in its load free state (i.e. at zero transmural pressure) is then constructed by assigning $L$ and $r_0=\overline{r}_0$ to each vessel. The shared coordinates information is used to generate a connectivity map of the 3D structure (see Figure~\ref{fig:Network}(c)) using the ``digraph" function in MATLAB (version 16a). Both control and hypertensive networks have the same connectivity illustrated in Figure~\ref{fig:Network}(c), but the individual vessel radii and length vary as shown in Table~\ref{Tab:Network}. 

\begin{table}
\centering
\caption{Dimensions of vessels in the 21-vessel network.}\label{Tab:Network} 
 {\footnotesize 
 \setlength\tabcolsep{3.5pt}
 \begin{tabular}{cccccc}
   \hline\noalign{\smallskip}
  \multicolumn{4}{c}{\bf \hspace{3cm}{Control}} & \multicolumn{2}{c}{\bf Hypertensive}\\ [2pt]
  \hline\noalign{\smallskip}
   vessel & connectivity & $r_0\times 10^{-1} $ &$ L\times 10^{-1}$ & $r_0\times 10^{-1} $ & $ L\times 10^{-1}$ \\
   index   & (daughters) & (cm)               & (cm)           &  (cm)              & (cm)\\
   \noalign{\smallskip}\hline\noalign{\smallskip}
      1$^*$ & (2,3)	&  0.47  &  4.10 & 0.51& 3.58\\[2pt]
      2 & (4,5)	&  0.26  &  4.45 &0.26& 4.03\\[2pt]
      3 & (6,7)	& 0.37  &  3.72 & 0.37& 3.08\\[2pt]
      4 & (8,9)	&  0.24 &  2.41 & 0.25&2.92\\[2pt]
      5 &  --       &  0.13 &  0.52& 0.17 &0.65\\[2pt]
      6 & (14,15) &  0.32  &  2.02 & 0.28 &1.60\\[2pt]
      7 & --	       &  0.17  &  2.12 & 0.19 &0.93\\[2pt]
      8 & (10,11)&  0.23  &  3.11 & 0.24 &2.06\\[2pt]
      9 & --	       &  0.17  &  1.77 & 0.17 &0.51\\[2pt]
     10 &(12,13)&  0.20  &  2.62 & 0.22 &2.37\\[2pt]
     11 & --	      &  0.16  &  0.69 & 0.17 &0.88\\[2pt]
     12 & --	     &  0.15  &  1.40 & 0.19 &1.27\\[2pt]
     13 & --	     &  0.14  &  0.62 & 0.15 &0.51\\[2pt]
     14 &(16,17)  &  0.26  &  0.81 & 0.27 &1.20\\[2pt]
     15 & --	  &  0.19  &  1.84 & 0.19 &1.55\\[2pt]
     16 & (18,19)&  0.25  &  0.83 & 0.26 &0.71\\[2pt]
     17 &-- 	 &  0.15  &  3.02 & 0.18 &1.68\\[2pt]
     18 & (20,21)&  0.24  &  4.69 & 0.24 &3.55\\[2pt] 
     19 &-- 	&  0.15  &  1.77 & 0.18 &1.86\\[2pt]
     20 &-- 	&  0.22  &  1.78 & 0.23 &2.24\\[2pt]
     21 &-- 	&  0.18  &  0.55 & 0.19 &1.07\\[2pt] 
    \noalign{\smallskip}\hline\\[-6pt]
  \end{tabular}}
  
   {\footnotesize * root vessel. Connectivity $(i,j)$, $i$ denotes the left daughter and $j$ the right daughter.
   Vessels indicated by -- are terminal.}
\end{table}

\subsection{Fluid Dynamics Model}\label{sec:fluids}

Assuming that blood is incompressible, flow is Newtonian, laminar and axisymmetric, and has no swirl, conservation of mass and momentum \citep{Olufsen2000} are given by
\begin{equation}\label{eq:mom}
\frac{\partial A}{\partial t} + \frac{\partial q}{\partial x} = 0,\quad \frac{\partial q}{\partial t} + \frac{\partial}{\partial x}\left(\frac{q^2}{A}\right)+\frac{A}{\rho}\frac{\partial p}{\partial x} = -\frac{2\pi\nu r}{\delta}\frac{q}{A},
\end{equation}
where $x$ and $t$ are the axial and temporal coordinates, $p(x,t) = p_{in}(x,t)-p_{ex}$ (mmHg) is the transmural blood pressure, $p_{in}$ and $p_{ex}$ are the pressures acting in and outside the arterial wall, respectively, $q(x,t)$ (ml/s) is the volumetric flow rate, $A(x,t)=\pi r^2$\,(cm$^2$) is the cross-sectional area and $r(x,t)$ (cm) is the vessel radius. The blood density $\rho$\,(g/ml) and the kinematic viscosity $\nu$\,(cm$^2$/s) are assumed constant. The momentum equation is derived under the no-slip condition assuming that the wall is impermeable, and that the velocity of the fluid at the wall equals the velocity of the wall. To satisfy this condition, we assume that the  velocity profile over the lumen area is flat, decreasing linearly within the boundary layer with thickness $\delta = \sqrt{\nu T/2\pi}$ \citep{Vosse2011}.

\subsection{Wall Model}\label{sec:wallmodel}
To close the system of equations, a constitutive equation relating pressure and cross-sectional area is needed. In this study, we compare two models. A linear elastic model \citep{Safaei2016} derived from balancing circumferential stress and strain, and an empirical nonlinear wall model inspired by \cite{Langewouters1985}.

\paragraph{The linear wall model} is derived under the assumptions that the vessels are cylindrical and purely elastic, that the walls are thin ($h/r_0\ll1$), incompressible and homogeneous, that the loading and deformation are axisymmetric, and that the vessels are tethered in the longitudinal direction. Under these conditions, the external forces can be reduced to stresses in the circumferential direction, and Laplace's law \citep{Nichols2011} yields the linear stress-strain relation 
\begin{equation}\label{eq:linWall}
 p = \beta \left(\sqrt{\frac{A}{A_0}}-1\right),\quad\text{where}\quad \beta = \frac{Eh}{(1-\kappa^2)r_0}
\end{equation}
is the stiffness parameter, defined in terms of Young's modulus $E$ in the circumferential direction. The associated Poisson ratio $\kappa$, the wall thickness $h$, and the undeformed radius $r_0$ at zero transmural pressure ($p=0$). $A_0=\pi r_0^2$ (cm$^2$) denotes the undeformed cross sectional area. Similar to previous studies (e.g.~ \cite{Olufsen2000,Safaei2016}), we use $\kappa = 0.5$.

\paragraph{The nonlinear wall model} relates pressure and area as
\begin{equation}\label{eq:nlinWall}
p =  p_1\tan\left[\frac{\pi}{\gamma}\left(\frac{A}{A_0} -1\right)\right],
\end{equation}
where $p_1>0$ (mmHg) is a material parameter that describes the half-width pressure, and $\gamma>0$ is a scaling parameter that determines the maximal lumen area $A_\infty$ as $p\rightarrow\infty$, giving 
\begin{equation}\label{eq:Am}
A_\infty= (1+\gamma/2)A_0.
\end{equation}
This model is formulated to ensure that $A= A_0$ at $p = 0$.

\subsection{Boundary Conditions}\label{sec:BCs}
Since the system of equations is hyperbolic, boundary conditions must be specified at the inlet and outlet of each vessel, i.e. the network needs an inflow condition, junction conditions, and outflow conditions.

At the network inlet we specify a flow waveform extracted from hemodynamic data~(see Figure~\ref{fig:hemoData}). At junctions (all bifurcations in the network studied), we impose pressure continuity and conservation of flow, i.e.
\begin{equation}\label{eq:bifcond}
p_p(L,t) = p_{d_i}(0,t) \quad\textrm{and}\quad q_p(L,t) = \sum_iq_i(0,t),
\end{equation} 
where, the subscripts $p$ and $d_i$ ($ i = 1,2$) denote the parent and daughter vessels, respectively. 

At the terminal vessels, a Windkessel model (represented by an $R_1C_pR_2$ circuit) is used to prescribe the outflow boundary condition (see Fig.~\ref{fig:Network}c) by computing input impedance $Z_{\textrm{wk}}(L,\omega)$ as
\begin{equation}\label{eq:Zwk}
  Z_{\textrm{wk}}(L,\omega)\equiv\frac{P(L,\omega)}{Q(L,\omega)} = R_1+\frac{R_2}{1+\dot{\iota}\omega R_2C_p},
\end{equation}
where $P(L,\omega)$ and $Q(L,\omega)$ are the pressure and flow in the frequency domain, $\omega=2\pi/T$ is the angular frequency and $T$ is the length of the cardiac cycle. $R_1,R_2$ (mmHg\,s/ml) denote the two resistances, and $C_p$ (ml/mmHg) the capacitance. Moreover, the total peripheral resistance is $R_T = R_1 +R_2$, where $R_1$ represents the resistance of the proximal vasculature and $R_2$ is the resistance of the distal vasculature, whereas $C_p$ denotes the total peripheral compliance of the vascular region in question. Similar to \cite{QureshiCMBE17}, $Z_{\textrm{wk}}(L,\omega)$ relates the pressure and flow at the outlet of each terminal vessel via a convolution integral over $T$.

\subsection{Parameter Values}\label{sec:ParInit}

The model parameters are divided into three groups: hemodynamics $\Phi_h =\{T, \nu, \rho, \delta\}$, vessel wall stiffness $\Phi_{w.\text{lin}} = \{\beta\}$ for the linear wall model and $\Phi_{w.\text{nlin}}=\{p_1, \gamma\}$ for the nonlinear wall model, Windkessel $\Phi_\text{wk}=\{R_1,R_2,C_p\}$ for the vascular beds. 

\paragraph{Hemodynamic parameters} are assumed constant. For each mouse (control and hypertensive), the length of the cardiac cycle $T=$\,1/HR (s) is extracted from the data (mean HR values for each group are given in Table~\ref{Tab:HemoStats}). The blood density $\rho=1.057$\,g/ml \citep{Riches1973} and the kinematic viscosity $\nu= 0.0462$\,cm$^2$/s, measured at  a shear rate of 94\,s$^{-1}$ \citep{Windberger2003}. These values represent average values for mice. As discussed earlier the boundary layer thickness is $\delta = \sqrt{2\pi\nu/T}$.

\begin{figure}[h]
\centering
\includegraphics[scale = 0.3]{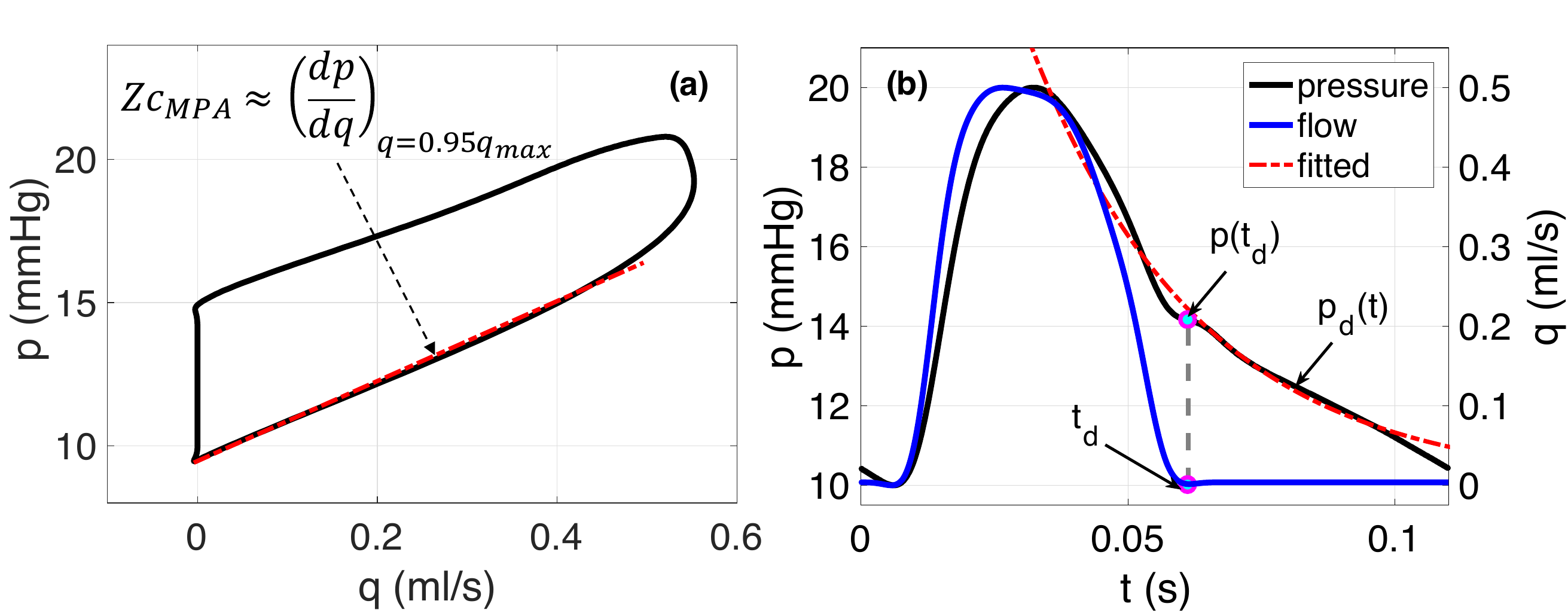}
\includegraphics[scale = 0.22]{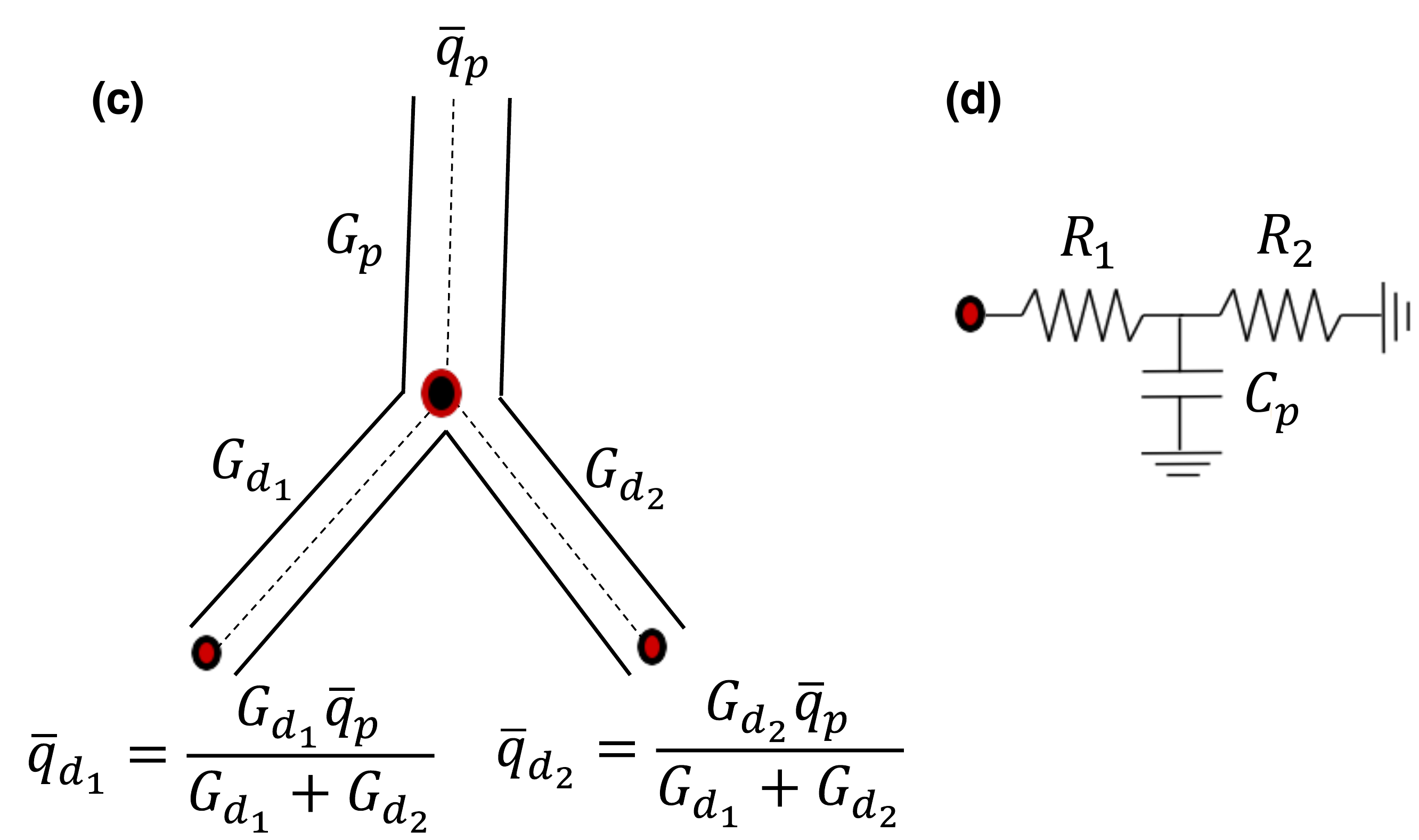}
\caption{\footnotesize Illustration of nominal parameter estimation for the linear elastic wall model and the Windkessel outflow conditions. (a): approximation of $Z_c$ from the slope of the pressure-flow loop during early ejection,  (b): estimation of time constant $\tau$ by curve fitting, (c): flow distribution across a bifurcation to compute the resistance at the terminal points,  (d): Windkessel model attached at the outlet of terminal vessels.}\label{fig:parEst}
\end{figure}

\paragraph{The nominal vessel wall parameters} for the linear and nonlinear wall models are approximated by combining an analytical approach, involving linearization of the fluid dynamics model, with available pressure and flow data. Since the hemodynamic data is available from one location only, we assume a constant $\beta$ throughout the network. 

Linearization of (\ref{eq:mom}) and (\ref{eq:linWall}) about the reference state $(A,p)=(A_0,0)$ provides an expression for the characteristic impedance $Z_c$ \citep{Nichols2011} as
\begin{equation}\label{eq:Zc}
 Z_c = \frac{\rho\, c_0}{A_0}\quad \Rightarrow \quad c_0 = \frac{A_0 Z_c}{\rho},
\end{equation}
where $c_0$ is the wave speed at $p=0$ (see Appendix~\ref{app:PWV} for the definition of $c_0$ for the linear ($c_{0.\textrm{lin}}$) and the nonlinear ($c_{0.\textrm{nlin}}$) wall models). For each mouse, $Z_c$ is esitmated from the slope of  the pressure-flow loop during early ejection, using the `up-slope method' (see Figure~\ref{fig:parEst}a) \citep{Dujardin1981} including 95\% of the flow during ejection phase.

For the {\it linear wall model}, substituting $c_0$ in (\ref{eq:Zc}) with $c_{0.\textrm{lin}}$ gives
\begin{equation}\label{eq:Ehr0Zc}
\sqrt{\frac{\beta}{2\rho}}=\frac{A_0 Z_c}{\rho} \quad \Rightarrow  \quad \beta = \frac{2(A_0 Z_c)^2}{\rho}.
\end{equation}

Similarly, for the {\it nonlinear wall model}, substituting $c_0$ in (\ref{eq:Zc}) with $c_{0.\textrm{nlin}}$ and using (\ref{eq:Ehr0Zc}), gives
\begin{equation}\label{eq:nomnLin}
\frac{p_1}{\gamma} = \frac{(A_0 Z_c)^2}{\pi\rho}\quad \Leftrightarrow \quad \frac{p_1}{\gamma} = \frac{\beta}{2\pi },
\end{equation}
where $\beta$ is given by (\ref{eq:Ehr0Zc}) and $p_1/\gamma$ is the `stiffness' of the nonlinear wall model. To fully specify the nominal (initial) values for the parameter inference (see Sec. \ref{sec:Simulation} and Table \ref{Tab:ParNom}), we set $\gamma=2$ and $p_1=\beta/\pi$. These values give $A_\infty = 2A_0$ cm$^2$. Note that (\ref{eq:nomnLin}) is only valid if both the linear and nonlinear models incorporate the same value of $A_0$ at $p=0$.

\paragraph{Nominal parameters for outflow boundary conditions} must be computed for each terminal vessel. {\it A priori} values for these parameters can be obtained from distributing the total peripheral resistance computed in the MPA ($R_T = \frac{\overline{p} - p_c}{\overline{q}}$, the ratio of the mean pressure gradient $\overline{p} - p_c$ to mean flow) to each terminal vessel $j$ as
\[
  R_{T,j} = \frac{\overline{p}}{\overline{q}_j},
\]
where $\overline{q}_j$ is the mean flow in vessel $j$. This expression is valid, under the assumption  that capillary pressure $p_c$ drops to zero and that mean pressure $\overline{p}$ remains constant across the arterial network. The flow to vessel $j$ is estimated by applying Poiseuille's law recursively at each junction, giving
\begin{equation}\label{eq:WKRnom}
  \overline{q}_{d_i}  =  \frac{ G_{d_i}}{\sum_i G_{d_i}}\overline{q}_p,\ \textrm{where} \ G_{d_i} = \left(\frac{\pi r_0^4}{8\mu L}\right)_{d_i} \  \textrm{for} \quad  i = 1,2,
\end{equation}
is the vessel conductance. Here $q_{d_i}$ denotes the mean flow to vessel $i$. Similar to previous studies, e.g. \cite{Reymond2009}, the total resistance is distributed as  $R_{1,j}  = aR_{T,j}$ and $R_{2,j} = R_{T,j}-R_{1,j}$, where the {\it a priori} value of $a=0.2$ \citep{McDonald:UK}.

Finally, as suggested by \cite{Stergiopulos1995}, the total vascular compliance $C_{T}$ (defined in Appendix~\ref{app:compliance}) can be calculated by estimating the time-constant $\tau = R_TC_T$ by fitting data to an exponential function of the form
\begin{equation}
p_d(t) = p(t_d)\exp(-(t-t_d)/\tau),
\end{equation}
where the diastolic pressure $p_d(t)$ decays in time (see Figure~\ref{fig:parEst}b). Assuming that $\tau$ is constant throughout the network,  the compliance  $C_{p,j}$ for each Windkessel can be computed as
\begin{equation}
C_{p,j}=\frac{\tau}{R_{T,j}}.
\end{equation}

\subsection{Numerical Simulations}\label{sec:Simulation}
Similar to previous studies \citep{Olufsen2000,Olufsen2012}, we use the two-step Lax Wendroff method to solve the model equations in Sections~\ref{sec:fluids} and \ref{sec:wallmodel}, for the arterial networks presented in Table~\ref{Tab:Network}. We combine data from Sec.~\ref{sec:data} with methods described in Sec.~\ref{sec:ParInit} to estimate nominal parameters, specifying arterial wall stiffness and boundary conditions (Sec.~\ref{sec:BCs}). Convergence to steady state is ensured by running simulations until the least square error between consecutive cycles of pressure is less than $10^{-8}$, taking about 6 cycles on average. For this study, we fixed the number of cycles to 7 for all mice.

To match the model to data, parameters inferred using optimization include $\Phi_\text{lin}=\{\beta,R_{T,j},\,a,\tau\}$ for the linear model and $\Phi_\text{nlin}=\{p_1,\gamma,R_{T,j},\,a,\tau\}$ for the nonlinear model. We define the objective function using sum of squared errors as
\begin{equation}
S(\Phi) = \sum_{n = 1}^N \left(p_n - p(t_n;\Phi)\right)^2 + \left(q_n - q(t_n;\Phi)\right)^2,
\label{eq:RSS}
\end{equation}
where $N$ is the length of the time vector spanning one cardiac cycle, $p_n$ and $q_n$ are the measured pressure and flow, $p(t_n;\Phi)$ and $q(t_n;\Phi)$ are the computed pressure and flow from the inlet of the MPA, and $\Phi$ are the parameters to be optimized using the linear or the nonlinear wall models. We use the function {\it fmincon} in MATLAB under the Sequential Quadratic Programming (SQP) gradient-based method \citep{Boggs2000} to solve the associated least-squares estimation problem
$$\widetilde{\Phi} = \textrm{arg\,min} \ S(\Phi).$$
Optimization was run in parallel on an iMac (3.1 GHz Intel Core i5, 16GB RAM, OS 10.12.6) initializing the parameters with four initial values including the nominal and nearby values. Nominal values for individual mice are given in Table~\ref{Tab:ParNom} (Appendix~\ref{app:ParametersNom}), whereas upper and lower bounds as well as averaged optimal parameters are given in Table~\ref{Tab:Bounds} and~\ref{Tab:Parameters} in Appendix~\ref{app:Parameters}. The algorithm was iterated until the convergence criterion was satisfied with a tolerance {$<10^{-8}$}.

\subsection{Model Analysis}\label{sec:WIA&IA} 

Similar to \cite{Qureshi2015} and \cite{QureshiCMBE17}, we analyze wave intensity and impedance spectra for further insight. Wave intensity analysis allows us to quantify the type and nature of the reflected waves in the time domain whereas the impedance analysis provides a frequency domain signature, which as shown in Figure~\ref{fig:hemoData} differs between the two groups.

\paragraph{Wave intensity analysis} allows us to separate the simulated waveforms into their incident ($+$) and reflected ($-$) components assuming negligible frictional losses. By setting $q=Au$, where $u$ is the fluid velocity, the incident and reflected waves can be approximated by
\begin{eqnarray}
\quad \Gamma_{\pm}(t)&=&\Gamma_0 + \int_{0}^{T} d \Gamma_{\pm};\qquad \Gamma = p \ \ \mbox{or} \ \ u.\\ 
 \mbox{Here} \ \ dp_{\pm}&=&\frac{1}{2}\left(dp\pm\rho c\,du\right), \ \ du_{\pm}=\frac{1}{2}\left(du\pm\frac{ dp}{\rho c}\right), \nonumber
\end{eqnarray}
and $c$ is the PWV computed by (\ref{eq:PWV}) (Appendix~\ref{app:PWV}). Time-normalized wave intensity $\textrm{WI}_{\pm}$\,(W/cm$^2$s$^2$) is defined as 
\begin{equation}
\textrm{WI}_{\pm}=(dp_{\pm}/d t) \,\, (du_{\pm}/d t).
\end{equation}
$\textrm{WI}_{+}$ along with $d p_+>0$ or $d p_+<0$ characterize the incident waves as compressive or decompressive, while $\textrm{WI}_{-}$ and $d p_->0$ or $d p_-<0$ characterize the reflected waves as compressive or decompressive, respectively. Finally, we compute the wave reflection coefficient $I_R$ as the ratio of the amplitudes of the reflected $\Delta p_-$ to the incident $\Delta p_+$ pressure waves, as
\begin{equation}\label{eq:IR}
I_R = \frac{\Delta p_-}{\Delta p_+}.
\end{equation}

\paragraph{Impedance analysis (IA).}
Under the assumptions of periodicity and linearity, the pulsatile pressure and flow waveforms can be approximated by a Fourier series of the form
\begin{equation}\label{eq:FourierSeries}
\tilde{s}(t_n) =\bar{S} + \sum\limits_{k=1}^{K}\operatorname{Re}[S_ke^{i(\omega_kt_n+\varphi_k)}];\quad n = 0,\dots, N,
\end{equation}
where $\tilde{s}(t_n)$ is the Fourier series approximation of the original waveform $s(t_n)$, $t_n = n/\rm{Fs}$ is the time vector for a given sampling rate Fs, $N=T\times\rm{Fs}$ is the length of the signal $s(t_n)$, $\omega_k = 2k\pi/T\,(k=1,\dots,K$) are the angular frequencies, $\bar{S}$ is the mean of $s(t_n)$, and $S_k$ and $\varphi_k$ (rad) are the moduli and phase spectra, associated with each harmonic $k$, and $K$ is the smallest resolution of harmonics required for the impedance analysis. Both, $S_k$ and $\varphi_k$, are defined in terms of $a_k$ and $b_k$, the coefficients of basic trigonometric Fourier series, i.e.
\[
  \quad S_k = \sqrt{a_k^2+b_k^2},\quad \varphi_k = \tan^{-1}(b_k/a_k).
\]

Setting $s(t_n)$ as $p(t_n)$ and $q(t_n)$ in~(\ref{eq:FourierSeries}), the impedance spectrum Z$(\omega_k)$, can be computed as ratios of harmonics of pressure to flow by
\begin{eqnarray}\label{eq:Zin}
Z(\omega_k)=\frac{P(\omega_k)}{Q(\omega_k)}&\equiv& \frac{\operatorname{Re}[P_k e^{i(\omega_k t_n + \alpha_k)}]}{\operatorname{Re}[Q_k e^{i(\omega_k t_n + \beta_k)}]}\nonumber \\ 
&=&\frac{P_k}{Q_k}\operatorname{Re}[e^{i(\alpha_k - \beta_k)}] \equiv Z_k\operatorname{Re}[{e^{i\theta_k}}],
\end{eqnarray}
where $P_k$ and $Q_k$ are the moduli and $\alpha_k$ and $\beta_k$ the phase angles of the pressure and flow harmonics, respectively. $Z_k$s are the impedance moduli and $\theta_k=\alpha_k-\beta_k$ the corresponding phases at a given frequency. Note if $\theta_k < 0$ then the $k^{\textrm{th}}$ pressure harmonic lags the $k^{\textrm{th}}$ flow harmonic, and vice versa. The zeroth harmonic is the total pulmonary vascular resistance (PVR).

\subsection{Statistical analysis}\label{sec:MethStat}

We implement statistical analysis methods to study parameter interference and devise a model selection criteria to determine the ability of the linear and nonlinear wall models to predict hemodynamics.

\paragraph{Parameter inference.} Optimized parameters and hemodynamic quantities are compared to assess changes with hypertension. For this analysis we compare predictions of total vascular resistance ($R_T$), total vascular compliance ($C_T$), the compliance ratio ($C_p/C_T$), the resistance ratio ($R_1/R_T$), the wave reflection coefficient ($I_R$), and characteristic time scales ($\tau$). Impact of the disease on a given quantity $\chi$, averaged across the two groups, is inferred by computing an importance index $\eta$ as a relative change in $\chi$ due to HPH as
\begin{equation}\label{eq:sensetivity}
\eta = \frac{\chi_\text{\tiny{HPH}} - \chi_\text{\tiny{CTL}}}{\chi_\text{\tiny{CTL}}}.
\end{equation}

\paragraph{Model selection criterion.} The 1D fluid dynamics model is coupled with a linear and a nonlinear wall model, leading to 4 dimensional (4D) and  5 dimensional (5D) parameter spaces, respectively. To identify the model that is more consistent with the data, we employ a statistical criterion that trades off goodness of fit versus model complexity (i.e. the number of estimated parameters). To this end, we use the corrected Akaike Information Criterion (AICc) \citep{Burnham2001} and the Bayesian Information Criterion (BIC) \citep{Schwarz1978} defined in (\ref{eq:LogLAIC}) and (\ref{eq:LogLBIC}). The model with the lower AICc and BIC score is preferred.
\begin{eqnarray}
  \text{AICc} &=& -2\log(\mathcal{L}) + 2D+\frac{2D(D+1)}{N-D-1}, \label{eq:LogLAIC} \\ 
 \nonumber \\ 
   \text{BIC}  &=& -2\log(\mathcal{L}) + D\log N. \label{eq:LogLBIC}
\end{eqnarray}
Here  $\log(\mathcal{L})$ is the maximum log-likelihood, $D$ is the number of parameters in the model and $N$ is the total number of measurements.
If the noise is white, i.e. independent and identically Gaussian distributed, then $\log(\mathcal{L})$ is just a rescaled version of the sum-of-squares function $S(\Phi)$, defined in (\ref{eq:RSS}).
However, a plot of the residuals (see Figure~\ref{fig:Residuals} in the Results section) indicates that the independence assumption is violated and that the correlation structure of the noise needs to be taken into account. Under the assumption of multivariate normal noise, the log likelihood is given by
\begin{equation}\label{eq:loglik}
\log(\mathcal{L}) = -\frac{1}{2}\log({\det{(2\pi \Sigma)}}) - \frac{1}{2}\mathbf{r}^T \Sigma^{-1} \mathbf{r},
\end{equation}
where $\Sigma$ is the covariance matrix and $\mathbf{r}$ is the vector of residuals. For the estimation of $\Sigma$, we tried different approaches. We fitted an autoregressive moving average model, ARMA (p,q), to the time series of residuals, and identified the optimal parameters p and q by minimising the BIC score. We then used the standard procedure proposed by Box \& Jenkins \citep{Box1970} to estimate the covariance matrix. Alternatively, we fitted a Gaussian process (GP) to the time series, using a variety of standard kernels: squared exponential, Mat\'ern 5/2, Mat\'ern 3/2, neural network kernel and a periodic kernel; see  \cite{Rasmussen2006} for details.

For the BIC and AICc scores, we need the maximum likelihood configuration of the  parameters. This would require an iterative optimization scheme, where for each parameter adaptation, the covariance matrix would have to be recomputed. As this would lead to a substantial increase in the computational costs, we approximated the maximum likelihood parameters by the parameters that minimize the residual sum-of-squares error of equation (\ref{eq:RSS}). 

Obtaining $\log(\mathcal{L})$ requires inverting the covariance matrix, see (\ref{eq:loglik}). To avoid numerical instabilities in the inversion of the covariance matrix, we introduce a regularization scheme:
$\Sigma^*(\lambda) = (1-\lambda) \times \Sigma + \lambda \times \sigma^2 \times \mathbb{I}$.
This adds jitter to the diagonal elements for improving the conditioning number of the matrix while leaving the marginal variance $\sigma^2$ (i.e. the diagonal elements of $\Sigma$) invariant. To minimize the bias, the interpolation parameter $\lambda \in [0,1]$ is chosen to be the minimum value subject to the constraint that the inversion of $\Sigma^*$ is numerically stable, as measured by a conditioning number below $10^4$. We discarded covariance matrices that required values of $\lambda$ above 0.1, due to the fact that in these cases the numerical stabilization incurred too high a bias. 

\section{Results}\label{sec:Results}

In this section, we present results of numerical simulations predicting pressure and flow waveforms along the pulmonary arterial network for a representative control and hypertensive mouse followed by a comparison of estimated parameter values. 

\begin{figure*}[t]
\centering
\includegraphics[scale = 0.4]{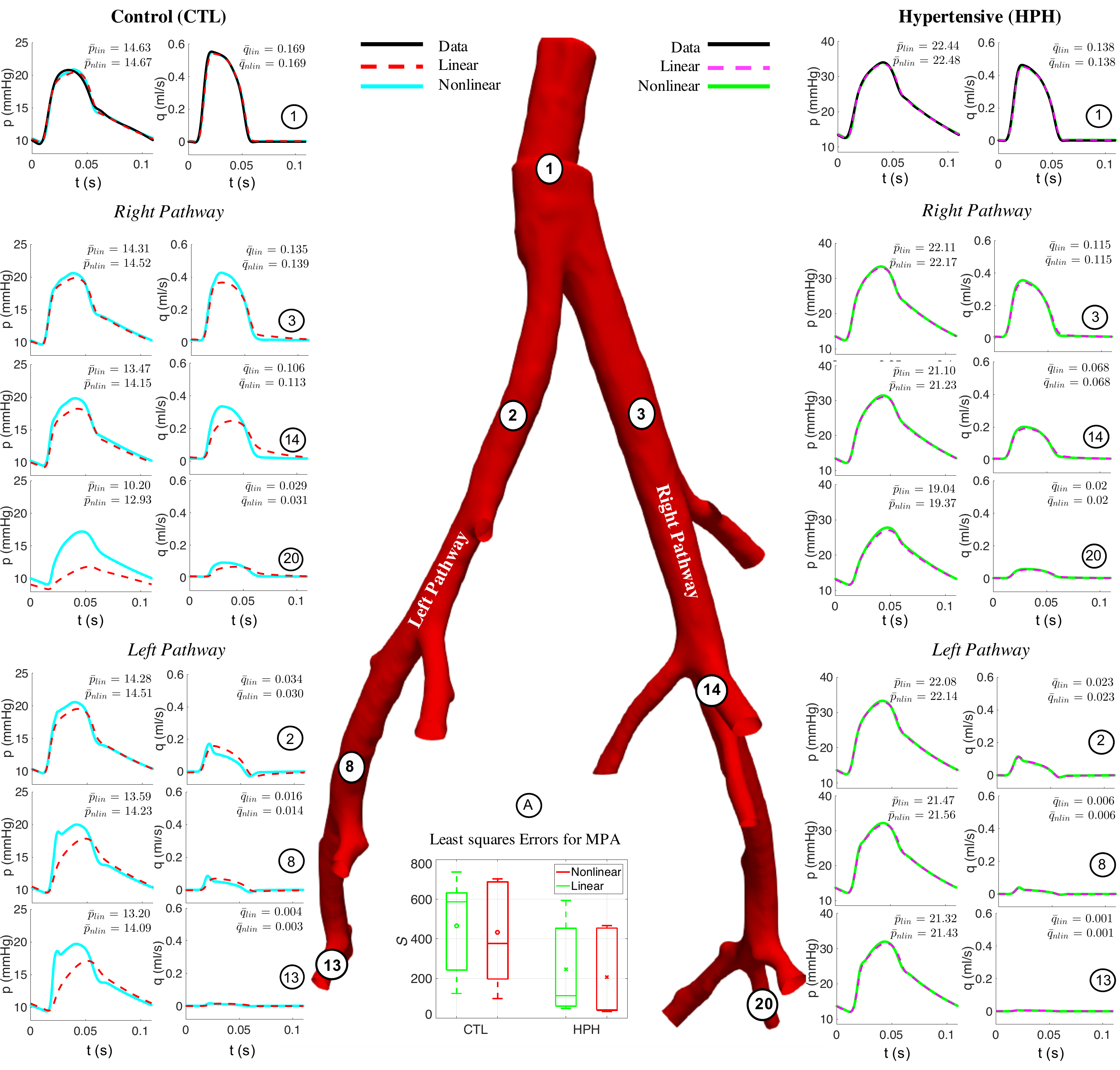}
\caption{\footnotesize Pressure and flow predictions along the pulmonary arterial network using linear (dashed line `- -') and non-linear (solid lines `--') wall models. Results are shown  for a representative control (left) and hypertensive (right) mouse. The center panel (a) show the least squares error avenged across CTL and HPH  groups. }\label{fig:PQsim}
\end{figure*}

\subsection{Hemodynamics}\label{sec:NetHemoDisc}

Figure~\ref{fig:PQsim} shows optimized pressure and flow waveforms for a representative control (left) and hypertensive (right) animal using the linear (dashed) and nonlinear (solid) wall models. Results were obtained estimating the least squares error between measured and computed waveforms in the MPA. The box and whisker plot (Figure~\ref{fig:PQsim}A) summarizes the least squares errors $S$ (reported in Table~\ref{Tab:Parameters} in Appendix~\ref{app:Parameters}) comparing the measured and computed waveforms in the MPA (vessel 1). Results show that both wall models (linear and nonlinear)  are able to fit the data well, and the least squares error is smaller for the hypertensive animals.

Even though both models provide similar fits in the MPA, the two wall models lead to different predictions in the downstream vasculature. For the control mouse, the nonlinear wall model leads to higher pressure in the down-stream vasculature for the control mouse. With the linear wall model, the mean pressure drop ($\Delta \overline{p}_\text{lin}\approx 4$) mmHg from the MPA to the terminal vessels, whereas for the nonlinear model $\Delta \overline{p}_\text{nlin} \approx 2$ mmHg. Moreover,  the nonlinear wall model predicts a notch in the pressure and flow waveforms during the ejection phase,  not observed with the linear wall model (displayed in panels 2, 8 and 13).  Finally it should be noted that these differences cannot be observed in predictions from the hypertensive animal, likely since the wall is almost rigid.

\begin{figure}[h]
\centering
\includegraphics[scale = 0.23]{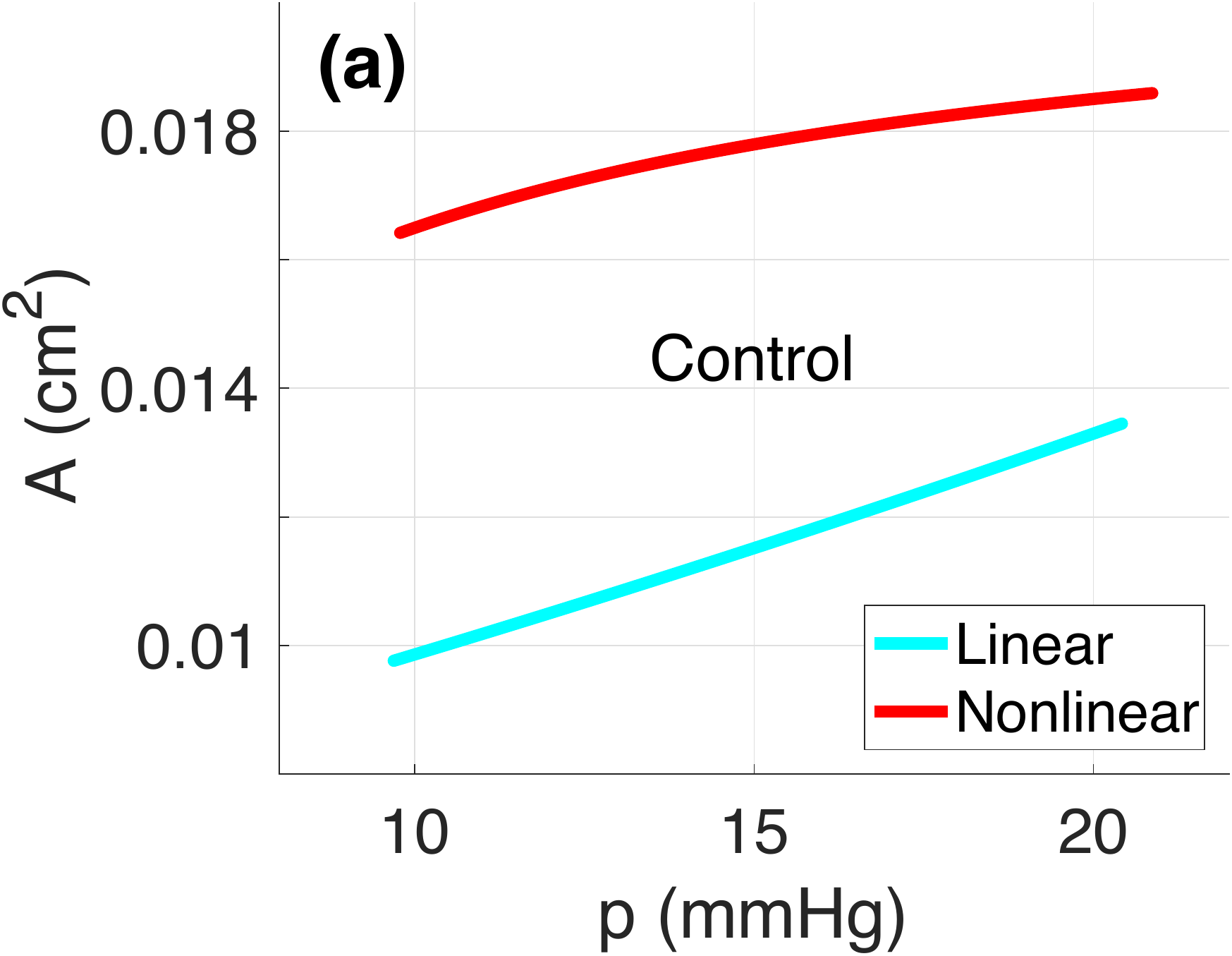}
\includegraphics[scale = 0.23]{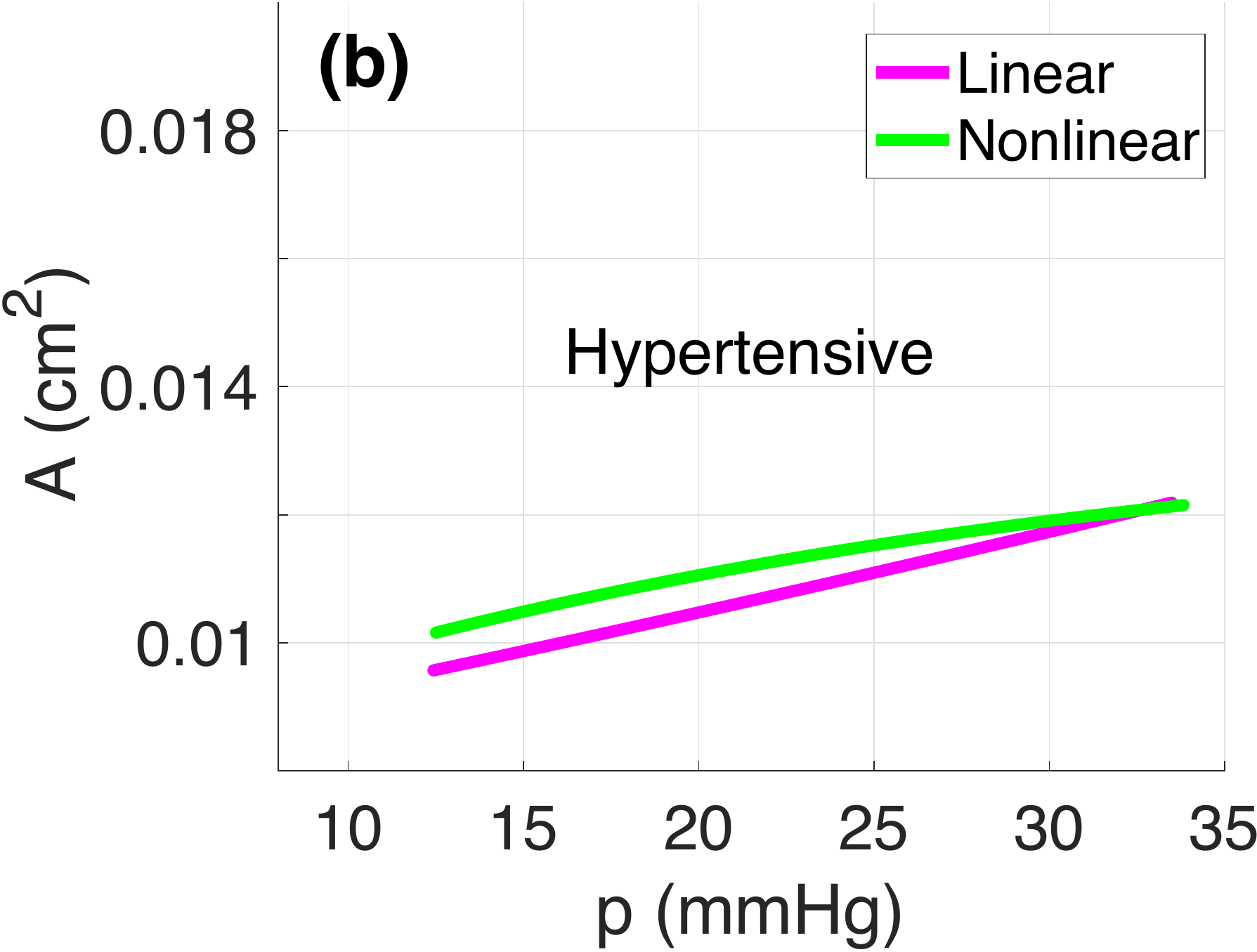}
\caption{Pressure-area plots corresponding to the linear and nonlinear wall models at the midpoint of the MPA for the control mouse (a) and hypertensive mouse (b).}\label{fig:PAsim}
\end{figure}
Figure~\ref{fig:PAsim}, depicting the pressure-area relationship in the MPA, gives further insight into the behavior of the linear and nonlinear wall mechanics. As expected, results show that for both groups the nonlinear wall model yields area predictions that are concave down implying increased stiffening at higher pressures. Higher compliance in control animals lead to more significant deformation despite the observation that the unstressed radius is larger in the hypoxic animals  ($r_0 = 0.051 \pm 0.005$ vs. $0.047\pm 0.002$\,cm, from Table~\ref{Tab:Network}).  

\subsection{Parameter estimates}\label{sec:EstPar}
\begin{figure}[h]
\centering
\includegraphics[scale = 0.23]{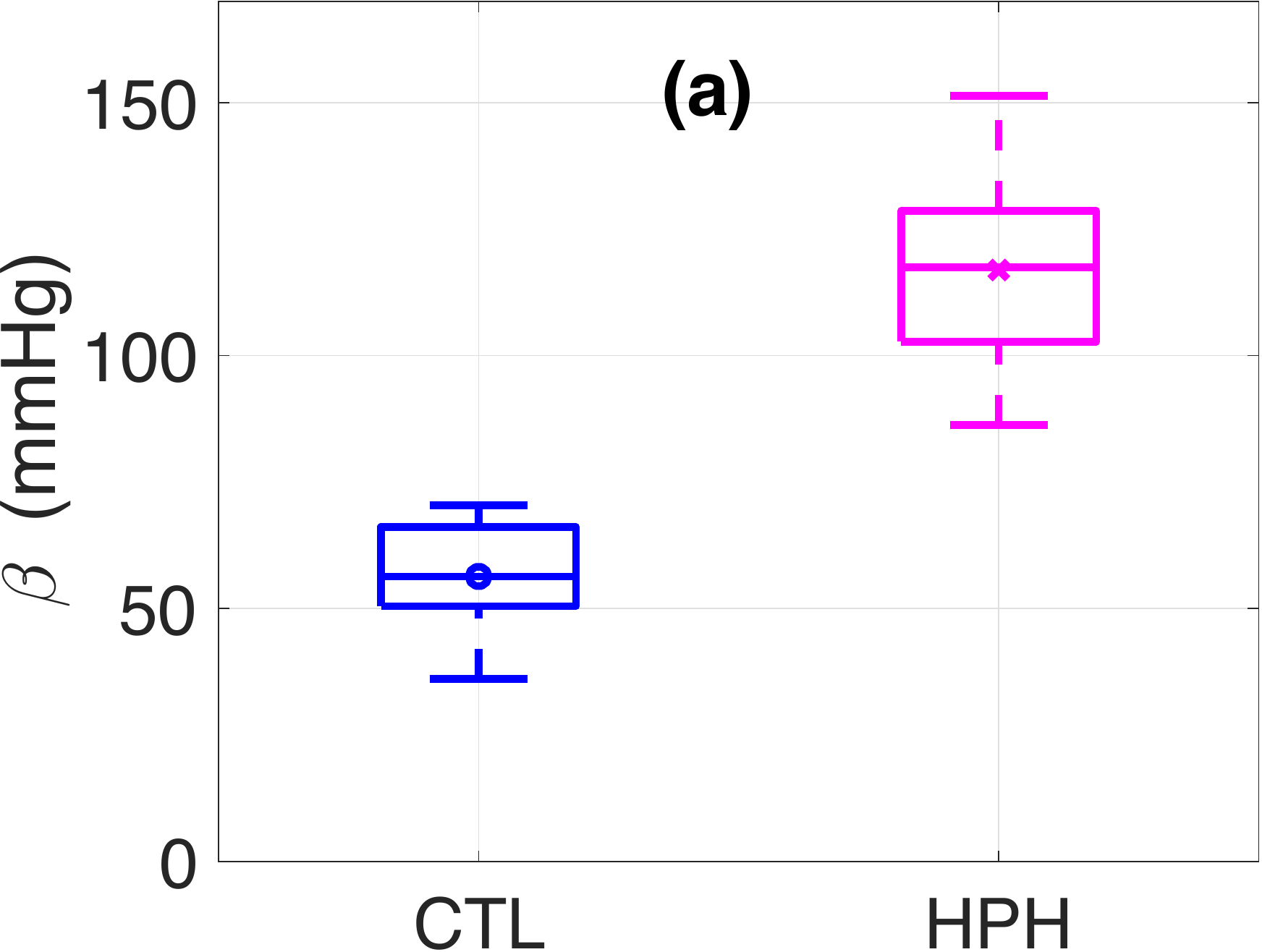}
\includegraphics[scale = 0.23]{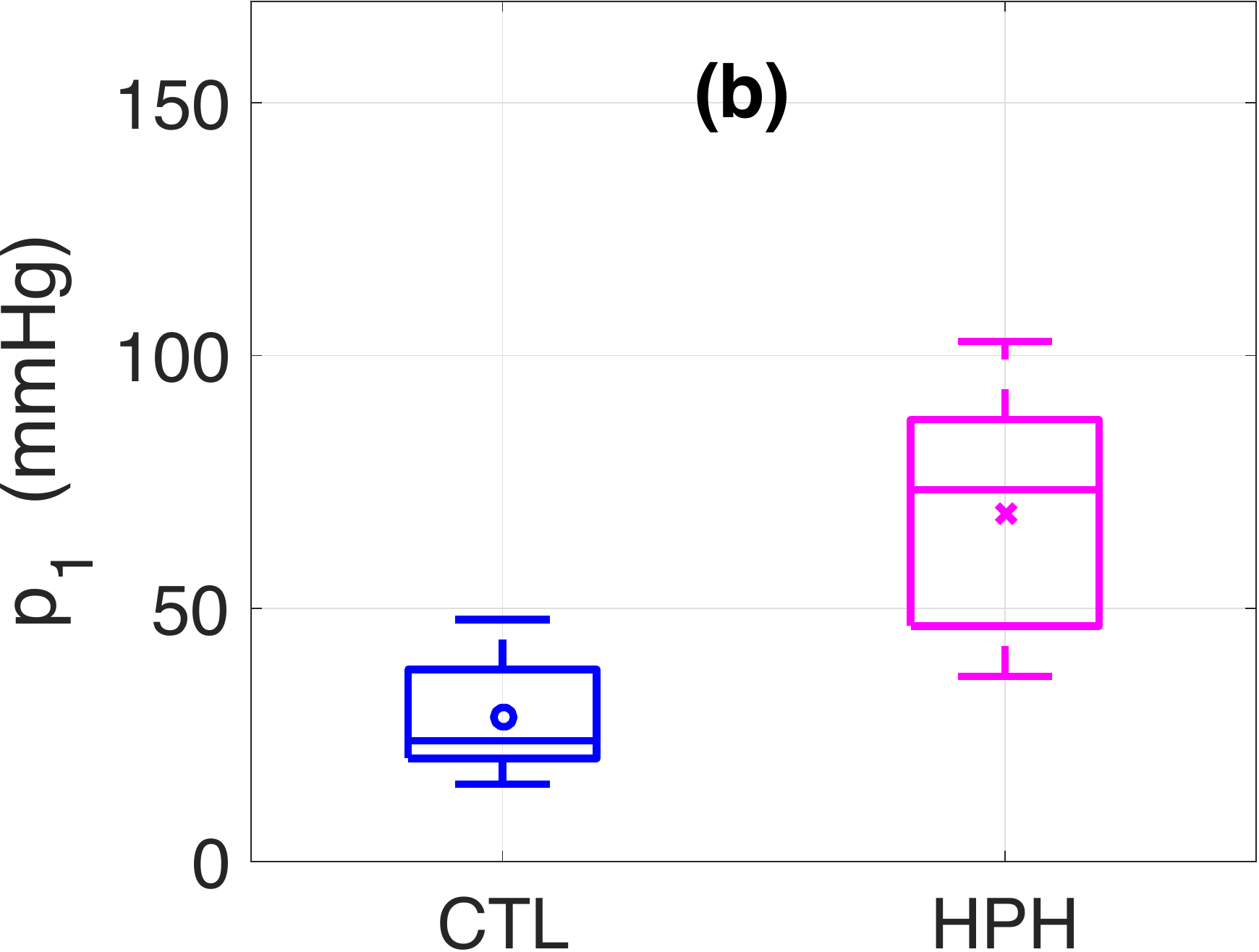} 
\includegraphics[scale = 0.23]{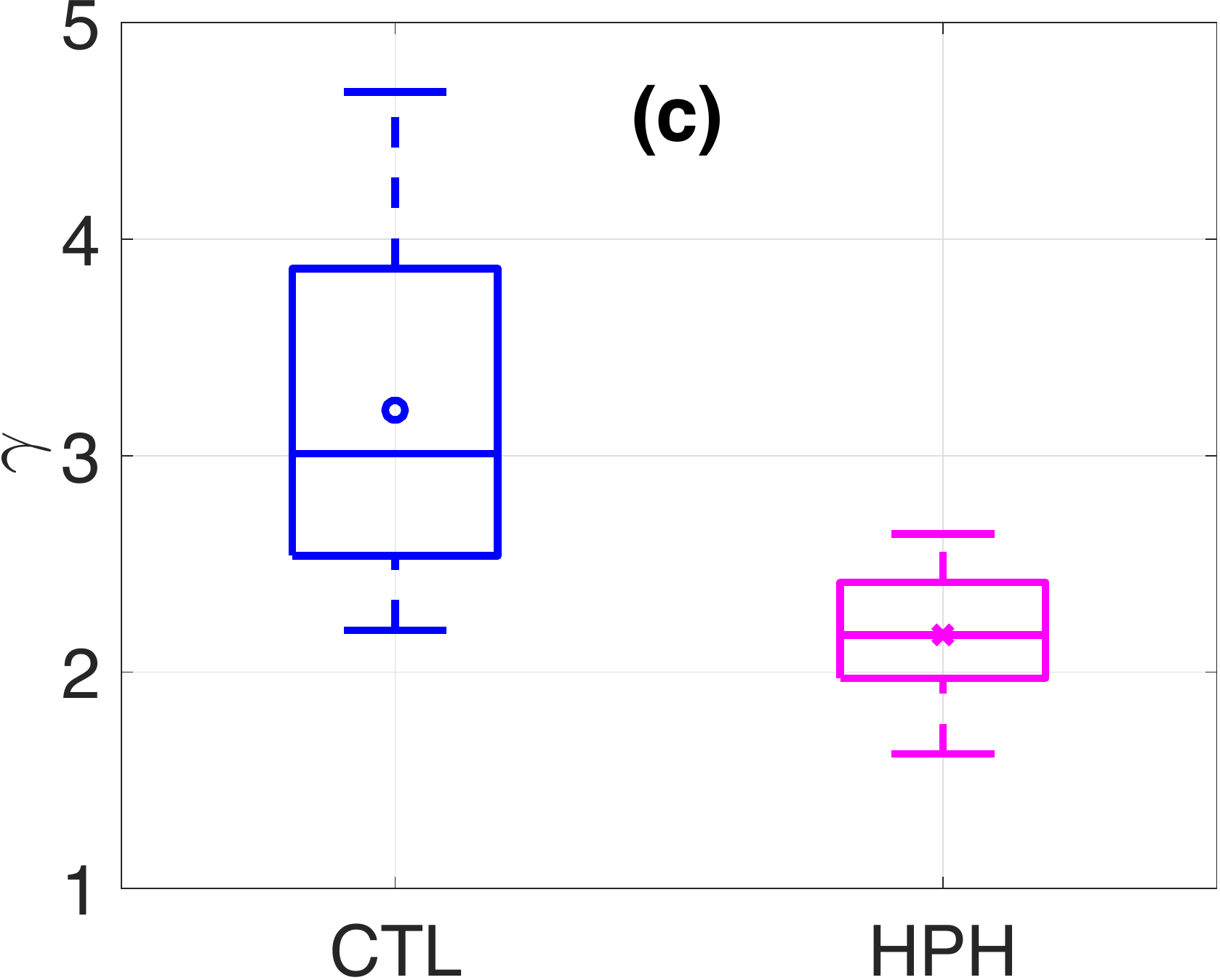}
\includegraphics[scale = 0.23]{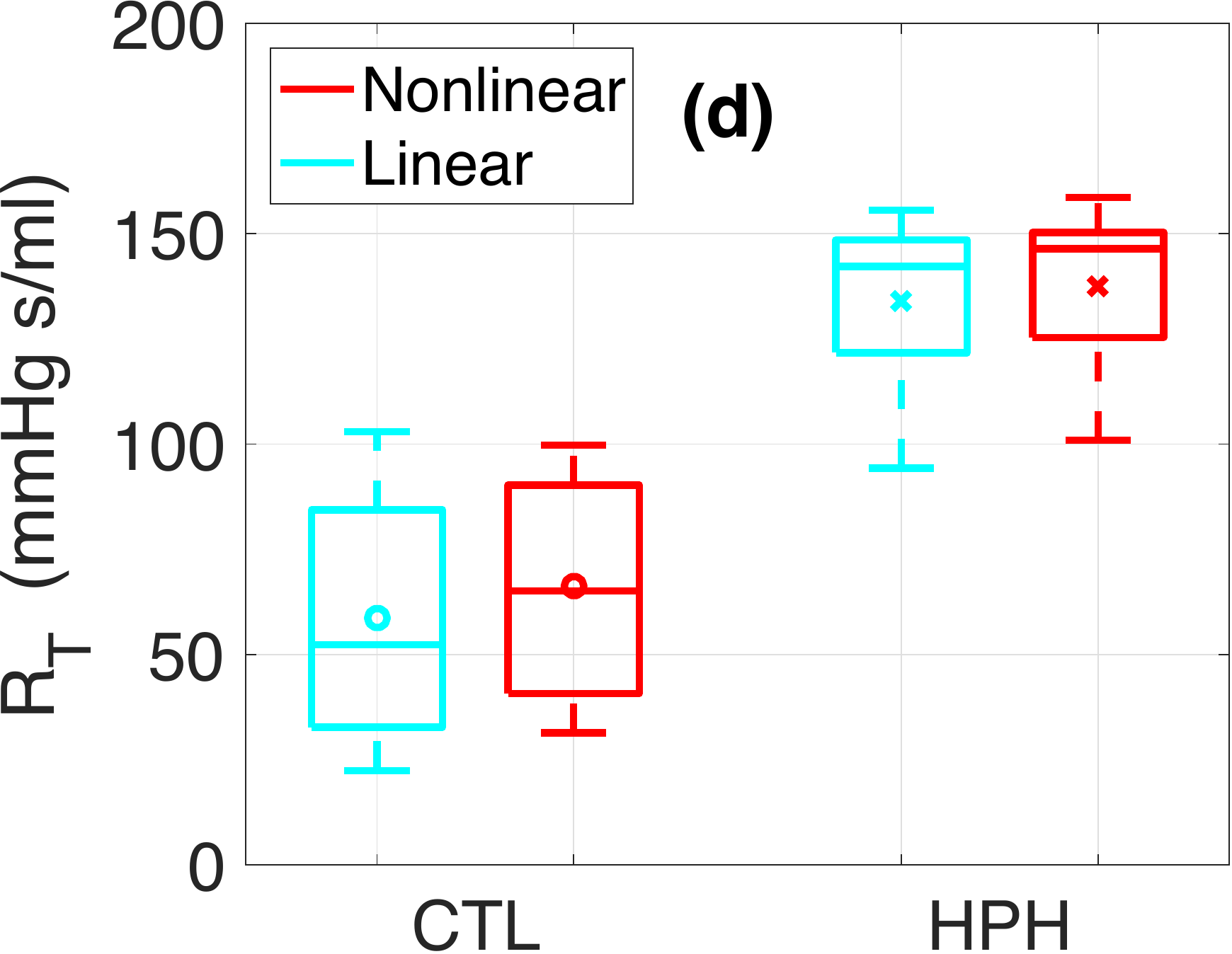}\\
\includegraphics[scale = 0.23]{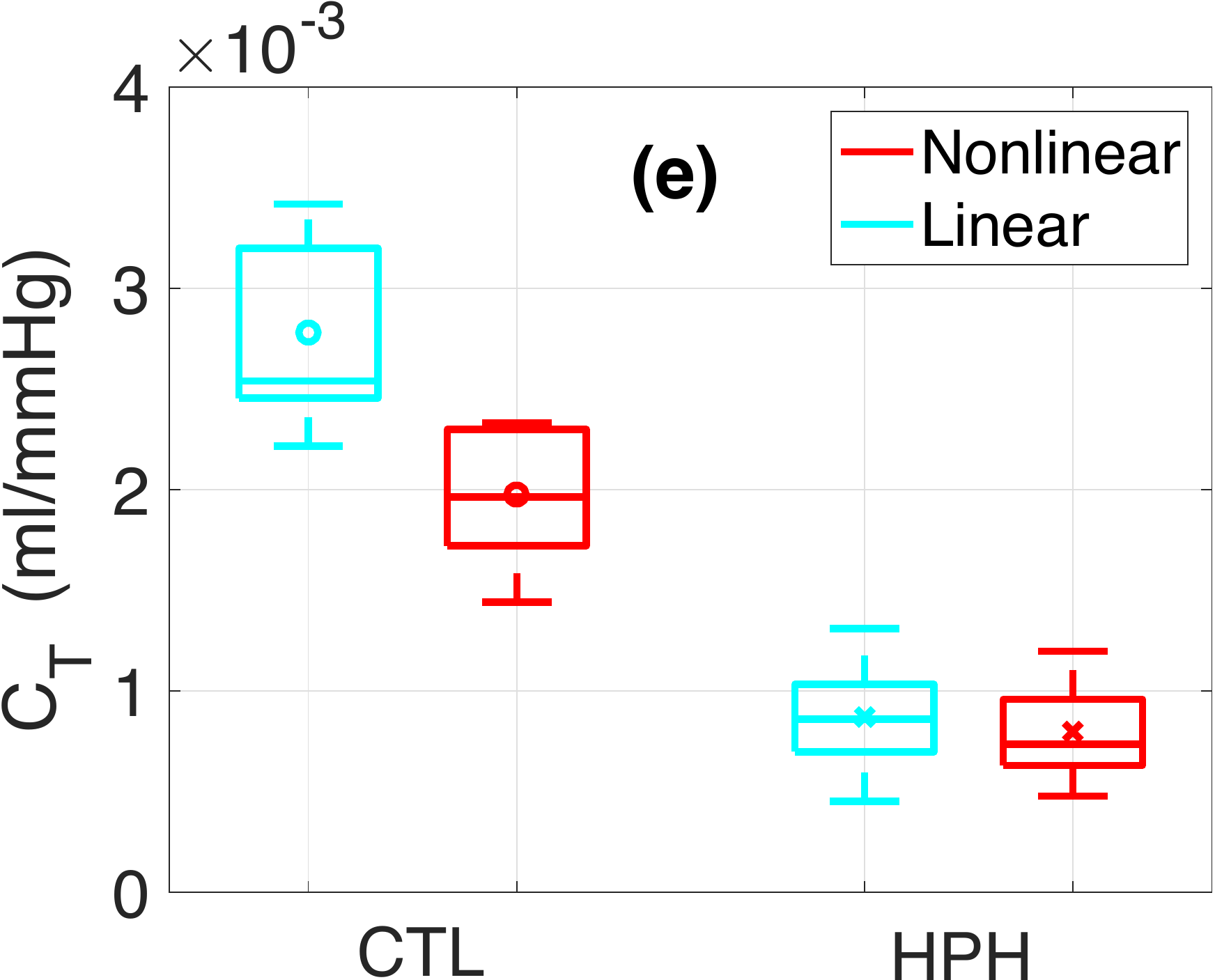}
\includegraphics[scale = 0.23]{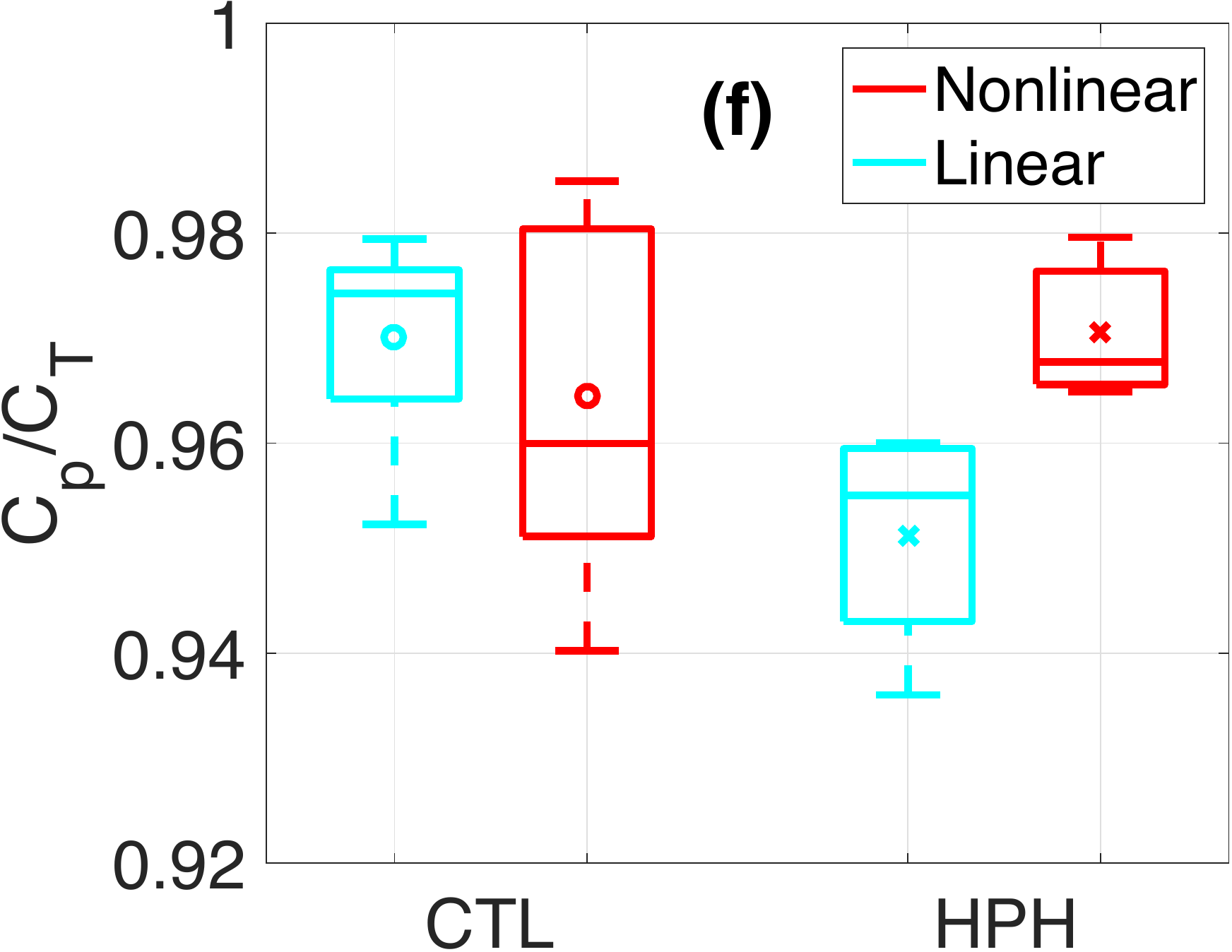}
\includegraphics[scale = 0.23]{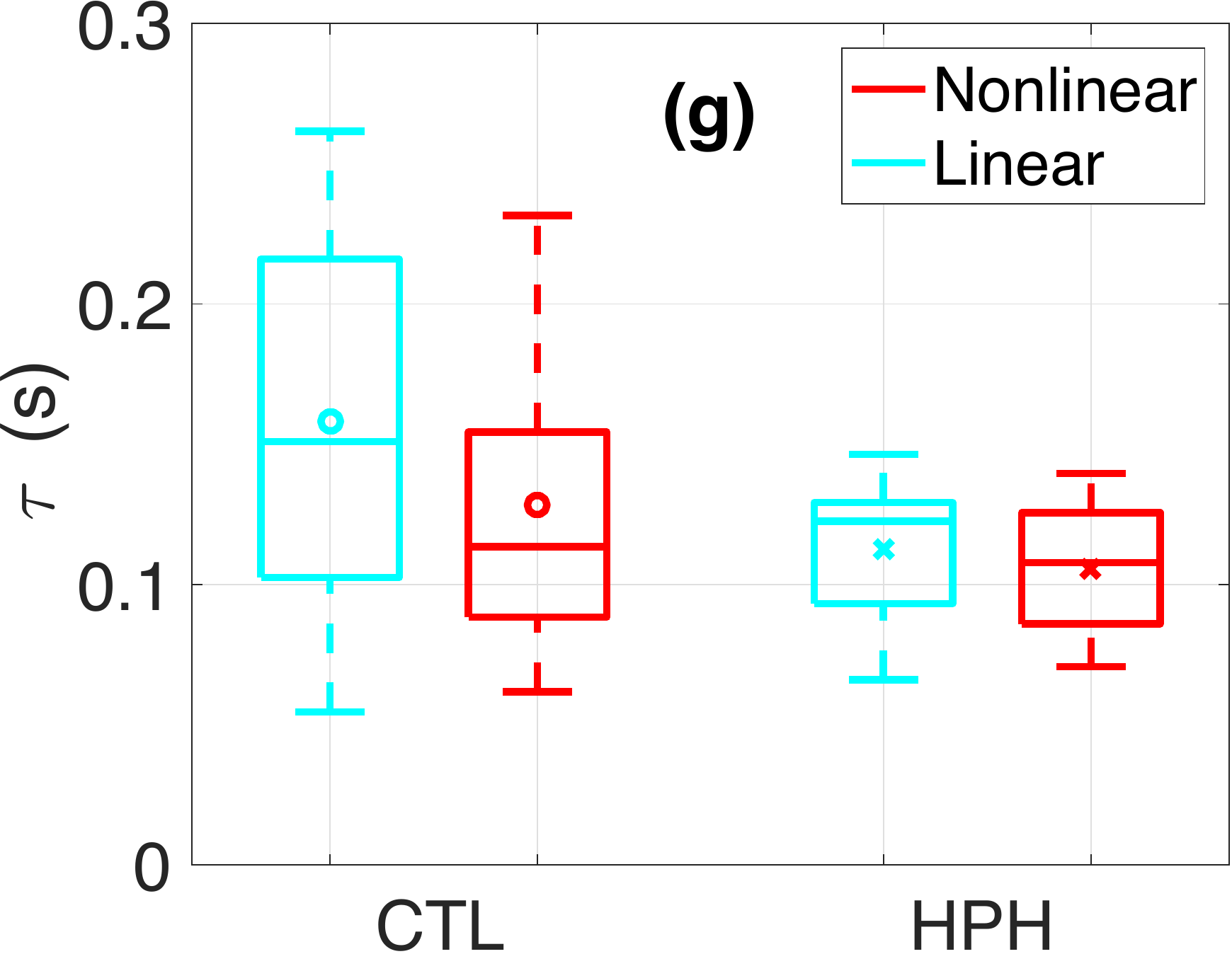}
\includegraphics[scale = 0.23]{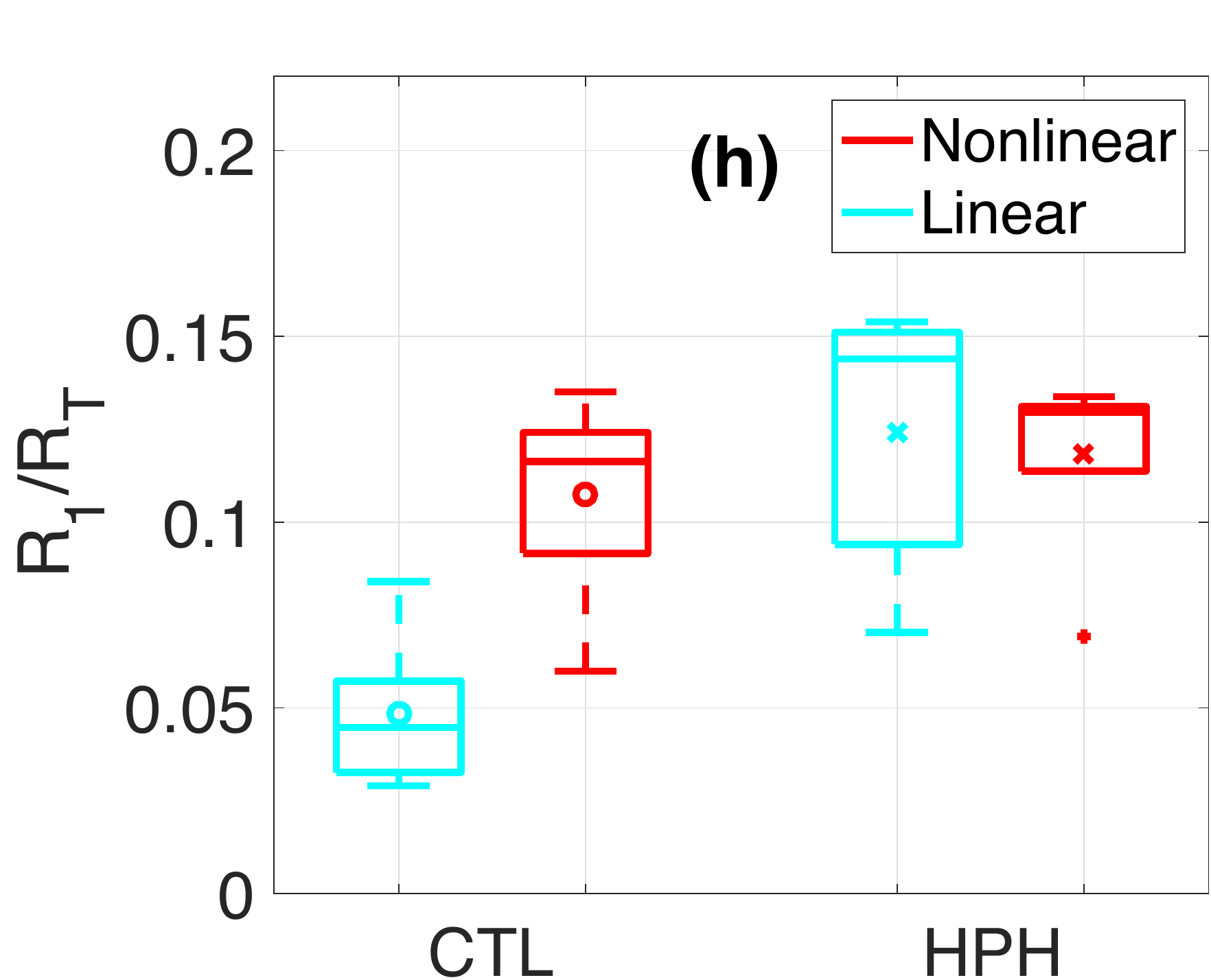}
\caption{\footnotesize Box and whisker plots summarizing the optimized parameter values for the CTL (n = 7) and HPH (n=5) groups. On each box, the horizontal bar represents the population median whereas the marker represents the population mean, the edges of the box are the 25th and 75th percentiles, the whiskers extend to the most extreme data points the algorithm, with the exception of outliers, which are plotted with red stars.   Top: (a) stiffness ($\beta$), (b) half width pressure ($p_1$), (c) maximal area ($\gamma$), (d) pulmonary vascular resistance ($R_T$). Bottom: (e) total vascular compliance ($C_T$), (f) peripheral to total compliance ratio ($C_p/C_T$), (g) characteristic time-scale $\tau$, and (h) resistance ratio $a=R_1/R_T$.}\label{fig:ParOpt}

\end{figure}

Figures~\ref{fig:ParOpt} summarizes the variation in estimated parameters for the two groups. Parameters, used to to generate Figures~\ref{fig:ParOpt}, are given in Table~\ref{Tab:Parameters} in the Appendix. 

\paragraph{The arterial wall stiffness} is larger in mice with HPH (a--c). For the linear model, this claim is supported by larger estimates of  the stiffness parameter $\beta$ (a). For the nonlinear wall model increased stiffness is predicted by a statistically significant increase in the parameter representing the half width pressure $p_1$ (b). This is not surprising since $\beta$  and $p_1$ have the same units and  are proportional  (see eq.~\ref{eq:nomnLin}). In addition, the constant $\gamma$ (estimated using the nonlinear wall model) is smaller leading to a smaller maximal area $A_\infty$ (see~ eq. (\ref{eq:Am})). 

\paragraph{Compliance distributions} between the network ($C_v$) and the vascular beds ($C_p$) are shown in Figure~\ref{fig:ParOpt}\,(e--f). Here the total compliance ($C_T$) for the entire vasculature is computed as
\[C_T =C_v+ \sum_j \widehat{C}_{p,j},\]
where $C_v$ is determined by recursively adding local compliance estimates $C_{v,i}$ (eqs.~(\ref{eq:volCompliance}) and (\ref{eq:Compliance}) in Appendix~\ref{app:compliance}), computed as  functions of  the diastolic blood pressure and area for each vessel $i$, while $\widehat{C}_{p,j}$ denote the weighted Windkessel capacitance (eq.~(\ref{eq:Zwk})) associated with each terminal vessel $j$. For simplicity, we omit the $\widehat{.}$ symbol in the rest of the manuscript. 

Results show that the total vascular compliance $C_T$ decreases with hypertension (Figure~\ref{fig:ParOpt}e). Given the reciprocal relation between compliance and stiffness (eqs.~(\ref{eq:volCompliance}) and~(\ref{eq:Compliance})), this behavior is anticipated, shown in Figure~\ref{fig:ParOpt}(a)--(c). The nonlinear model predicts a smaller compliance for both groups due to an increased stiffening with pressures.  Finally, Figure~\ref{fig:ParOpt}f shows the compliance distribution via the ratio $C_p/C_T$. Although the majority of the compliance is located in the vascular beds (i.e.~$C_p/C_T\approx 1$)(f), it decreases along the vasculature (from larger to smaller vessels). Finally, results with the linear wall model reveal that $C_p$ reduces to 95\% of $C_T$ under hypertensive conditions, indicating that large vessels have been remodeled by the disease.

\paragraph{The total vascular resistance} $R_T$ is computed as
\[
  \frac{1}{R_T} = \sum_j\frac{1}{R_{T,j}}\quad\text{where}\quad R_{T,j}=R_{1,j}+R_{2,j},
\]
where $j$ is the terminal vessel index.  Its distribution within the vascular beds is depicted in Figure~\ref{fig:ParOpt}\,(d,h). 
In this study, we estimated  the proximal and distal components $R_{1,j}$ and $R_{2,j}$  as described in Sec.~\ref{sec:Simulation}. Results show that  for both wall models $R_T$  increases under hypertensive conditions (d). This increase is dominated by an increase in $R_1$, evidenced by the increase in the resistance ratio $R_1/R_T$ (h).

\paragraph{The Characteristic time scale} $\tau=R_TC_T$, decreases under hypertensive conditions (Figure~\ref{fig:ParOpt}g). However, the nonlinear model predicts a smaller $\tau$ for all animals.

\subsection{Wave intensity analysis.} Results of the wave intensity analysis (WIA) allow us to investigate the behavior of the incident and reflected waves. Since WIA requires measurement of dynamic cross-sectional area and PWV, this analysis is done on the simulated waveforms.
\begin{figure}[h]
\centering
\includegraphics[scale = 0.23]{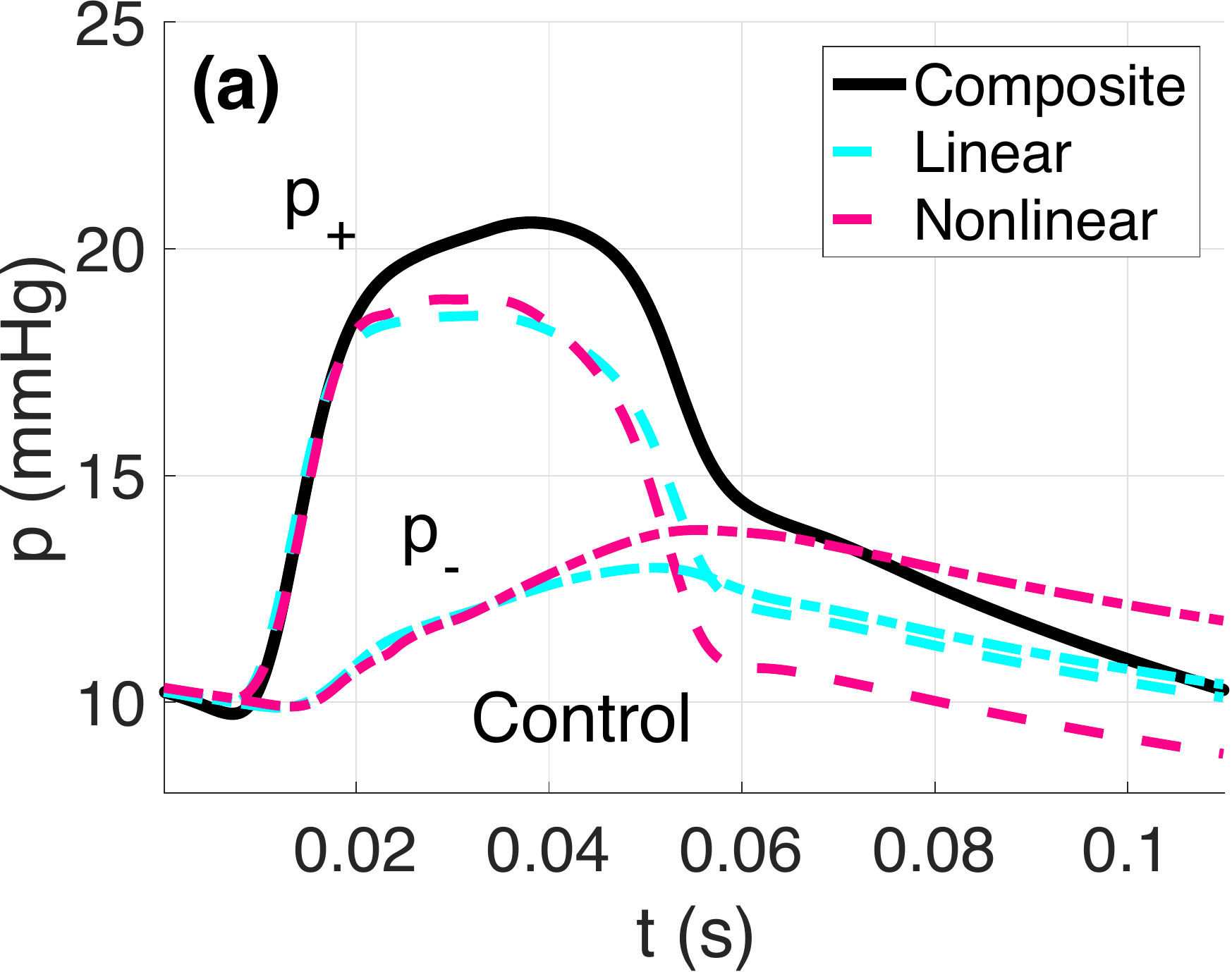}
\includegraphics[scale = 0.23]{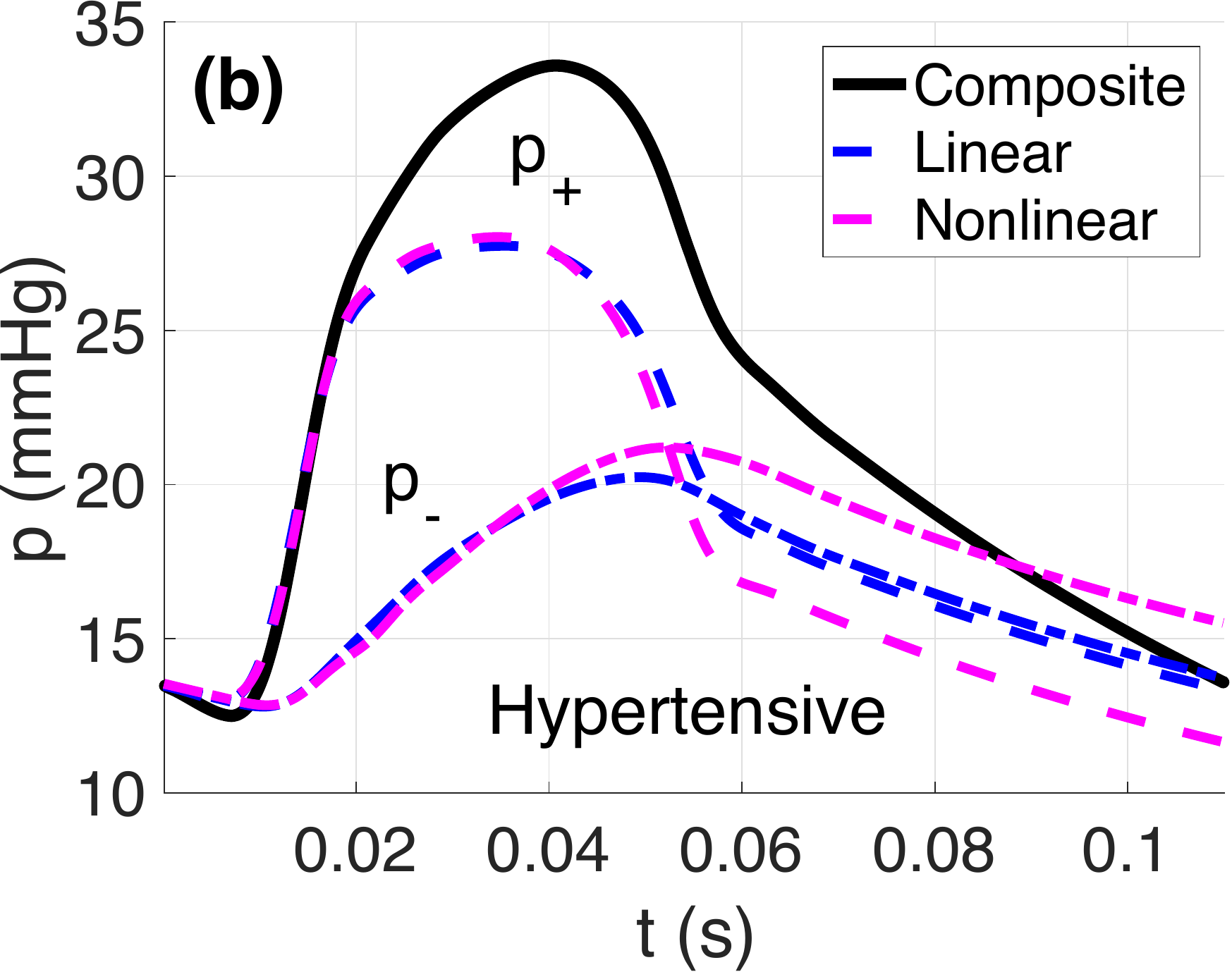}
\includegraphics[scale = 0.23]{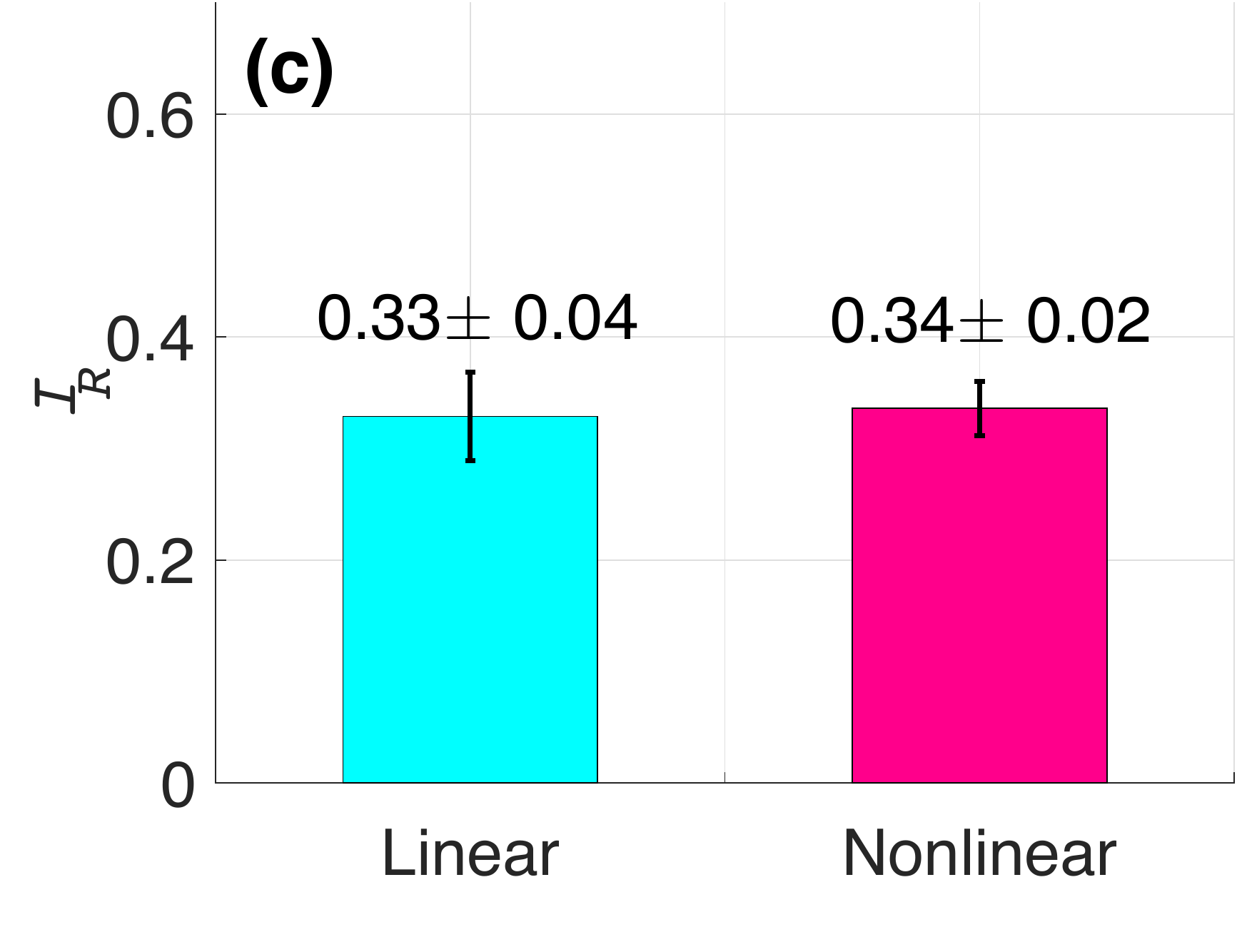}
\includegraphics[scale = 0.23]{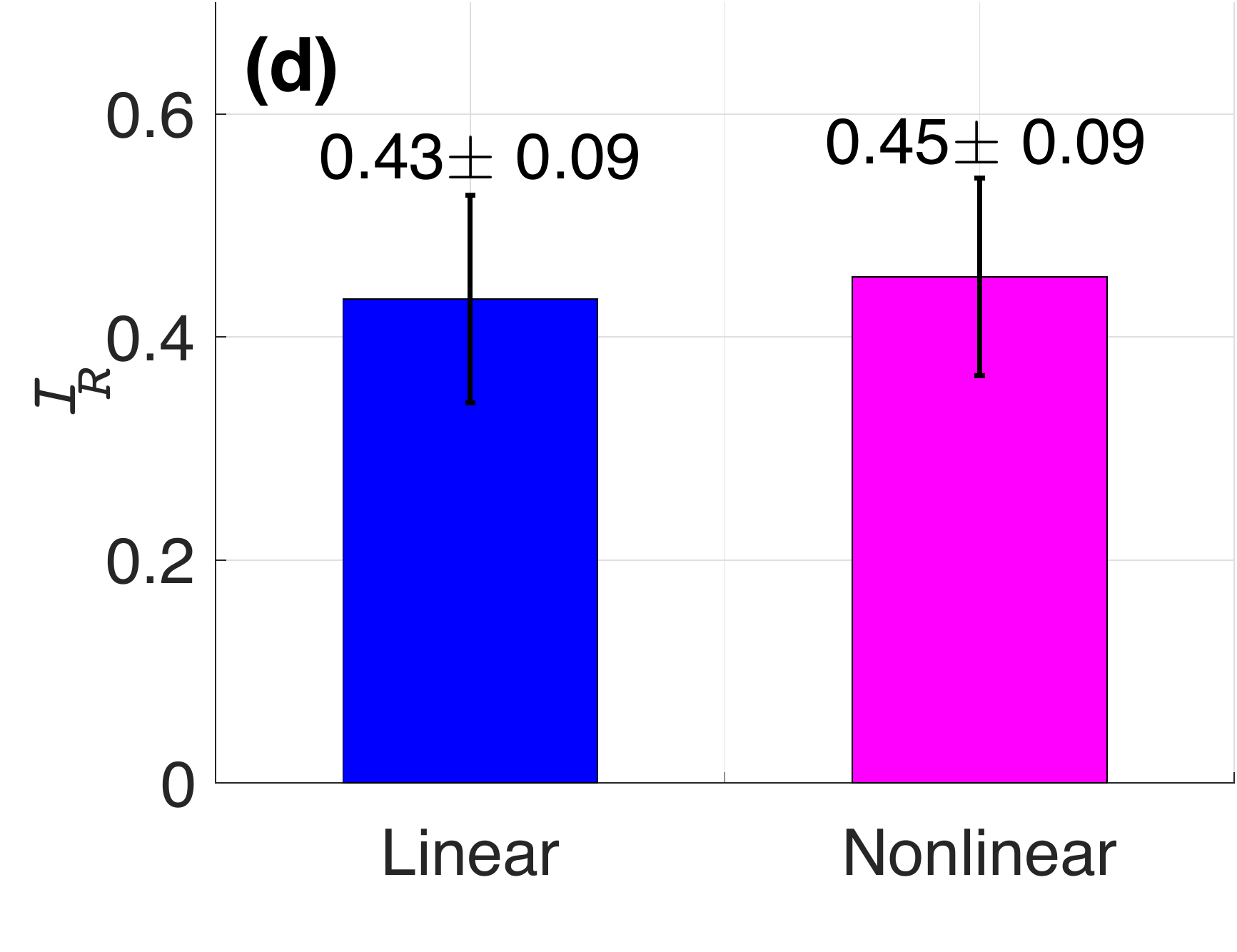} 
\includegraphics[scale = 0.23]{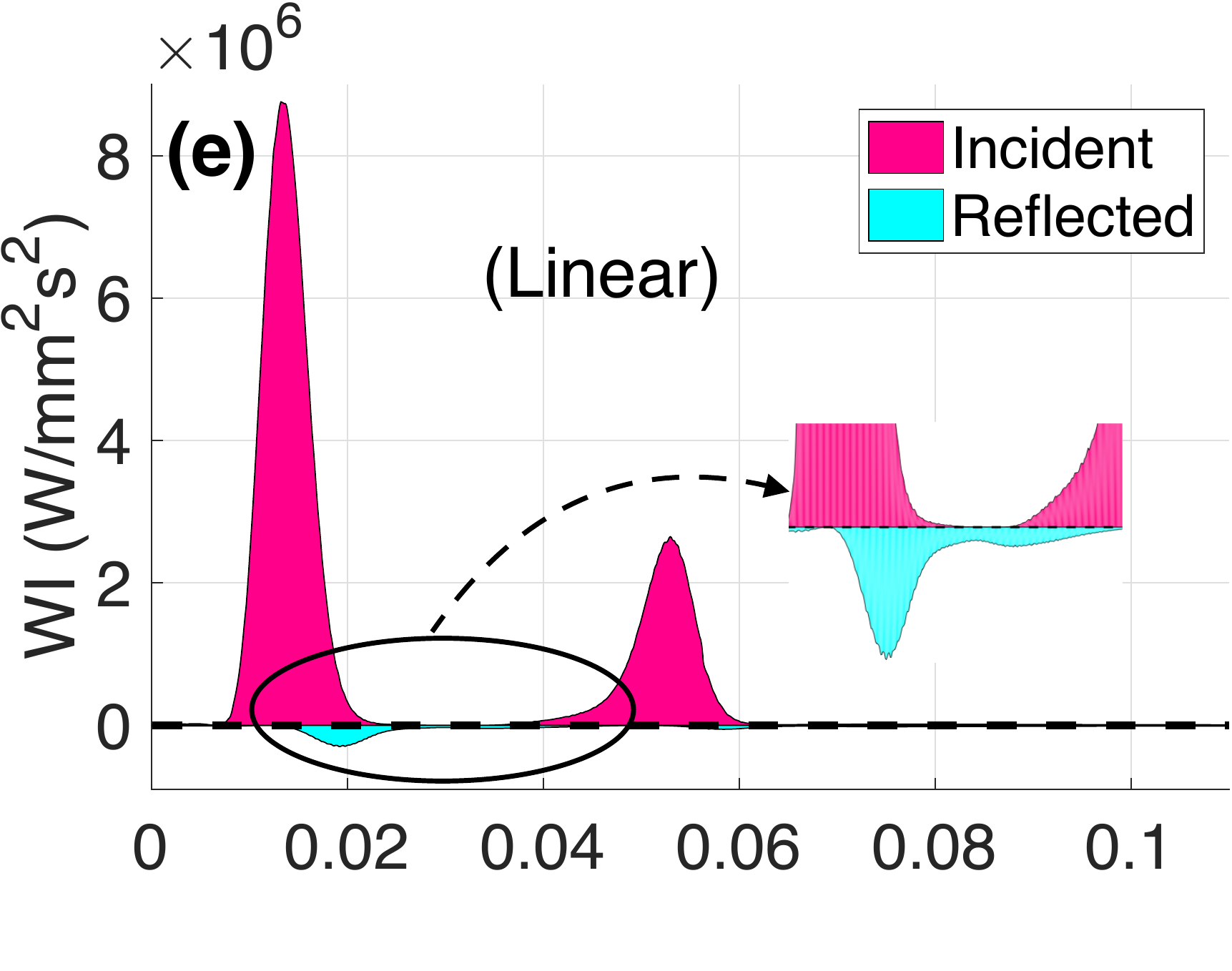}
\includegraphics[scale = 0.23]{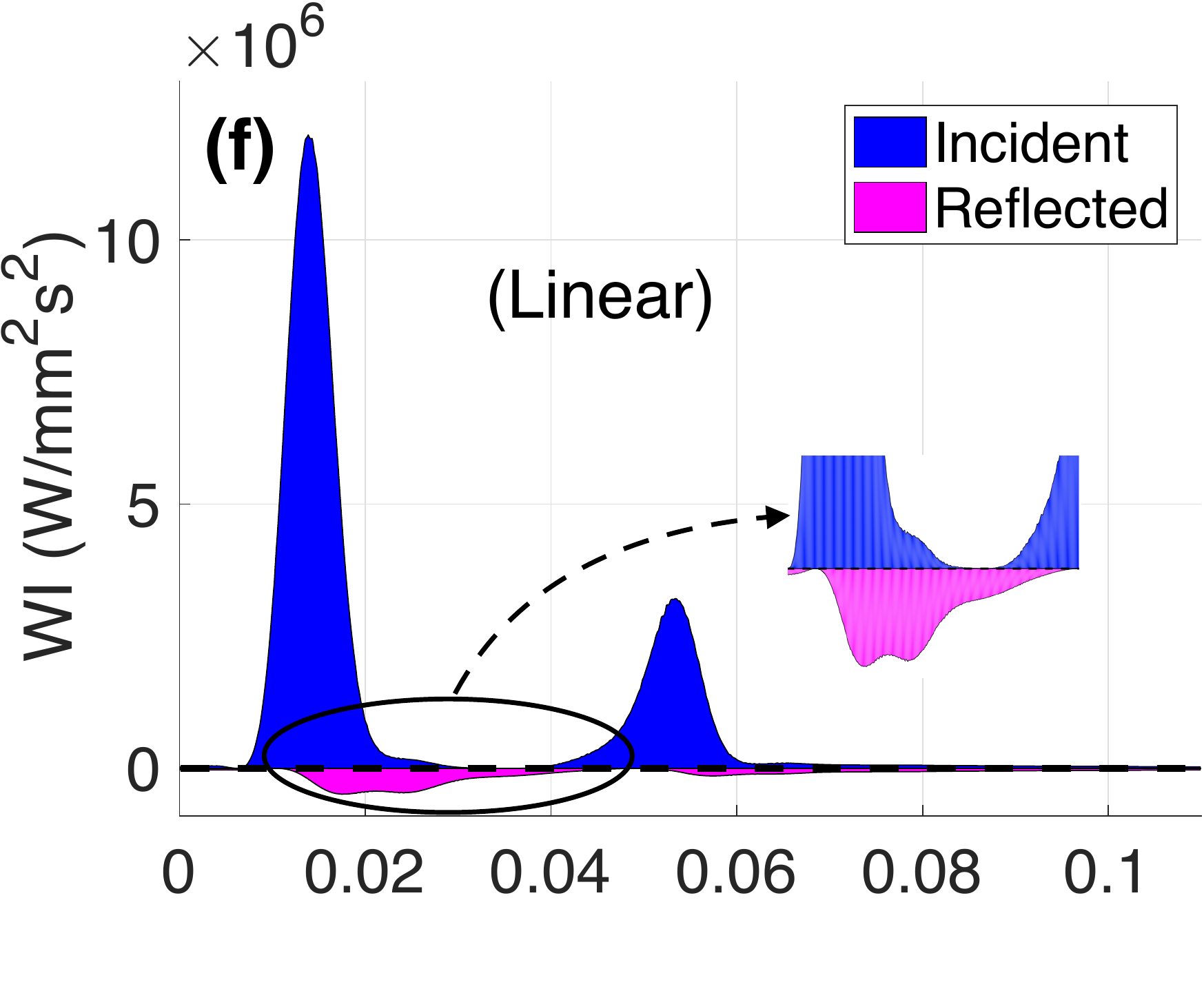}
\includegraphics[scale = 0.23]{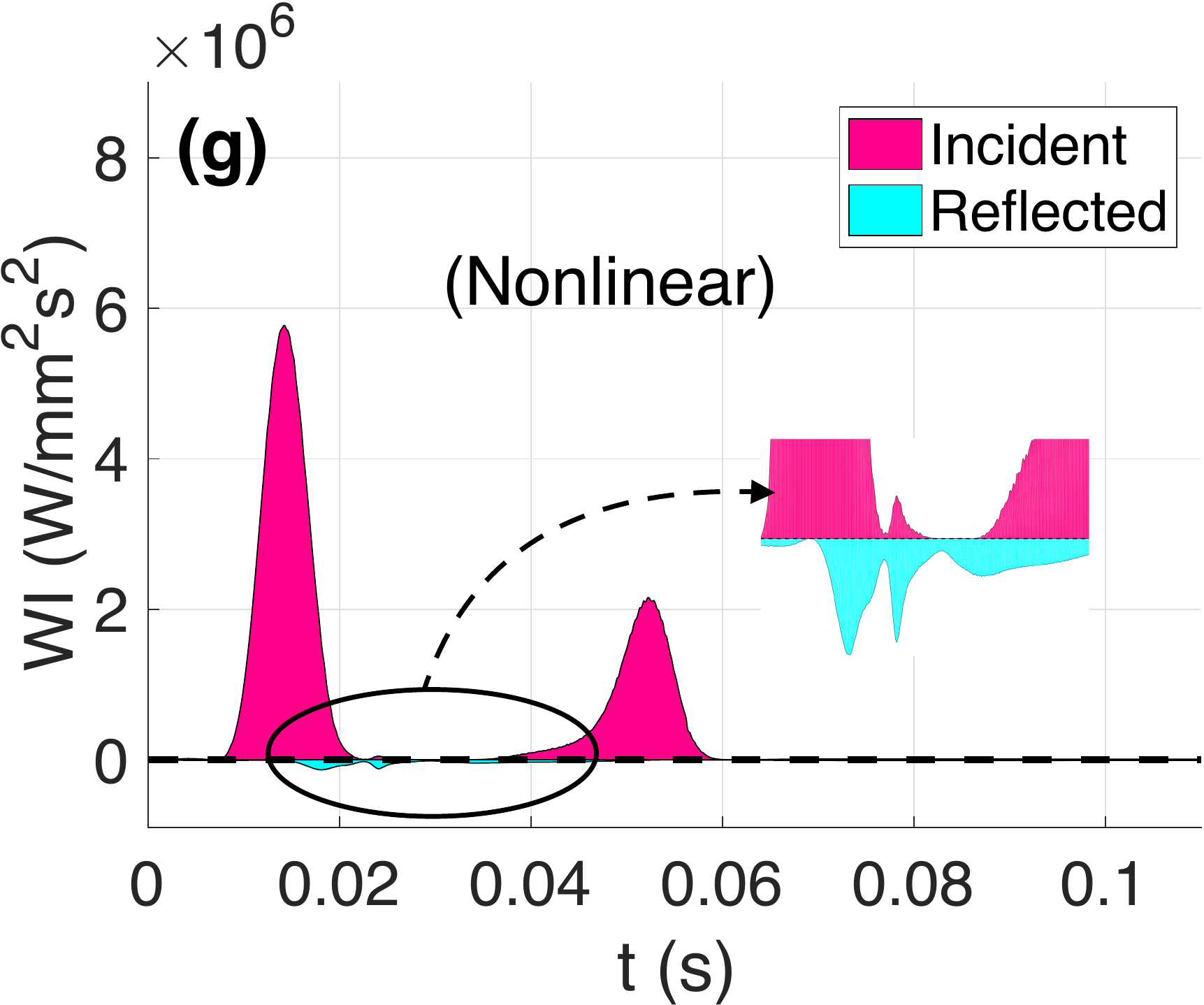}
\includegraphics[scale = 0.23]{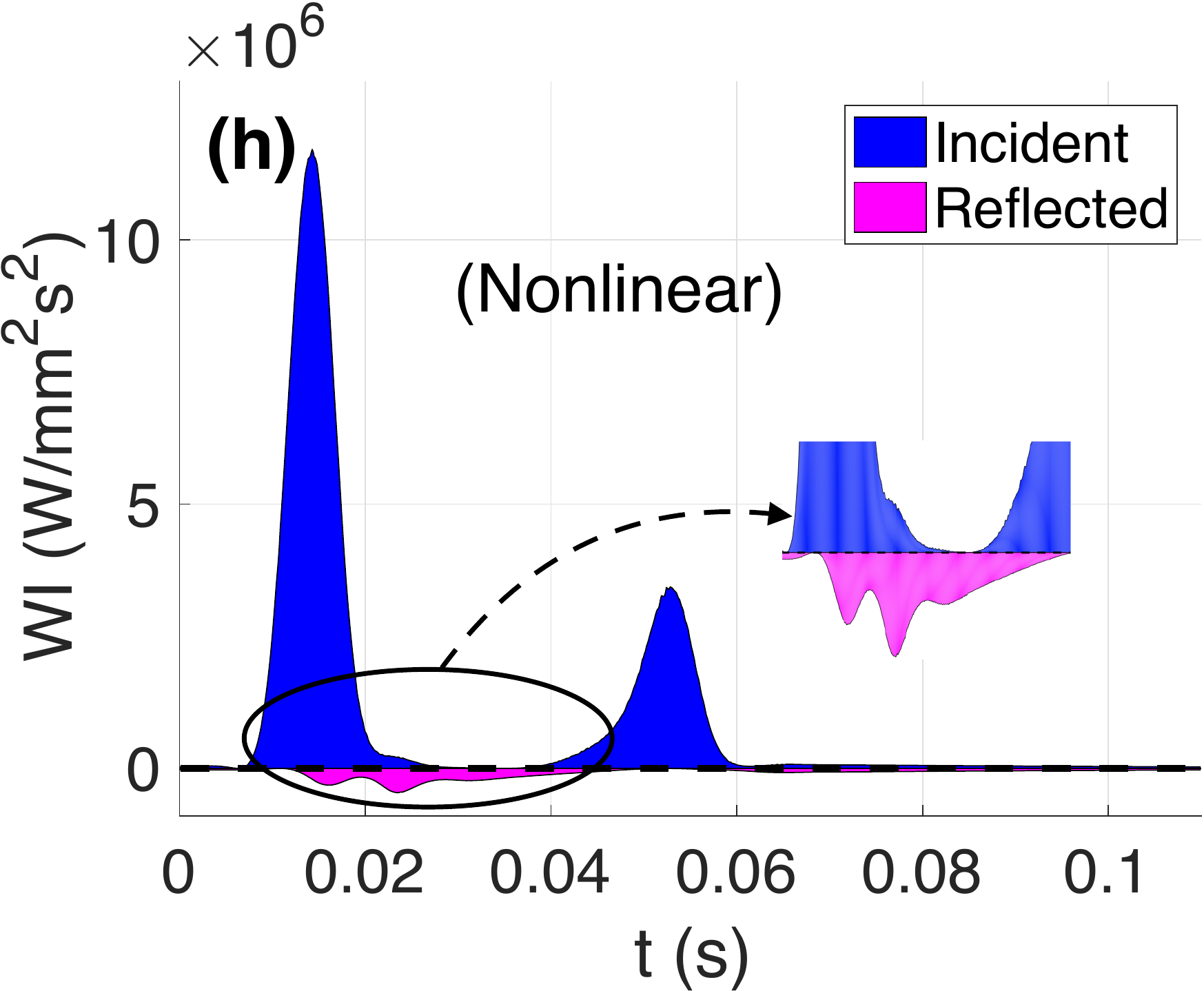}
\caption{\footnotesize Wave intensity analysis  comparing the linear and nonlinear wall models for the control (a,c,e,g) and the hypertesive (b,d,f,h) mice. (a,b) separation of the pressure wave into its incident and reflected components. The solid black curves represent the composite pressure waveforms, (c,d) the reflection coefficients averaged across the CTL and HPH groups. (e--h) the associated wave intensity profiles.}\label{fig:WIA}
\end{figure}
 
Figure~\ref{fig:WIA} shows separation of the pressure waveforms (a,b) into their incident and reflected components and the associated wave intensities (e-h). The group averaged reflection coefficients $I_R$, quantifying the magnitude of the reflections, are shown in panels (c) and (d). Results show that the amplitudes of the reflected waves are bigger in the hypertensive mice (b) leading to a bigger wave reflection coefficient, shown in panel (d). The effects of elastic nonlinearities on the amplitudes of the incident and reflected waves are minor (the increase in $I_R$ is negligible) (c,d).

Figure~\ref{fig:WIA}(e)-(h) present the wave intensity profiles associated with the separated pressure and velocity waves. The inner panels show zoomed intensities around  peak systole. Again, the amplitude of the incident wave is higher for hypertensive mice  (compare panels (e,g) with (f,h)). For both groups the nonlinear wall model gives rise to a smaller amplitude incident wave (compare panels (e,f) to (g,h)). Similar patterns are observed for the reflected waves. One notable difference, for both groups, is that the nonlinear model gives rise to more oscillations of the reflected wave. 

\subsection{Impedance analysis}
\begin{figure}[h]
\centering 
\includegraphics[scale = 0.23]{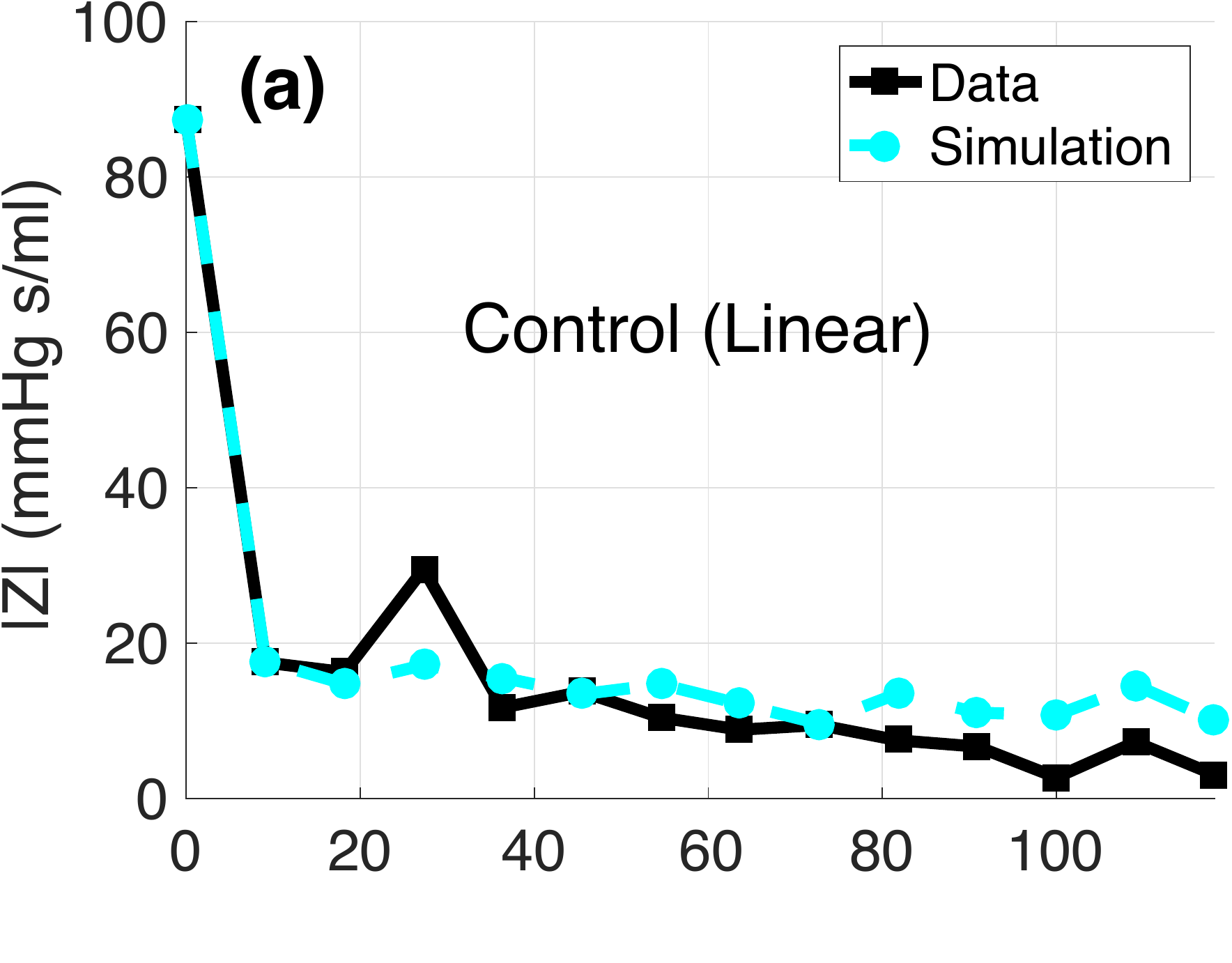}
\includegraphics[scale = 0.23]{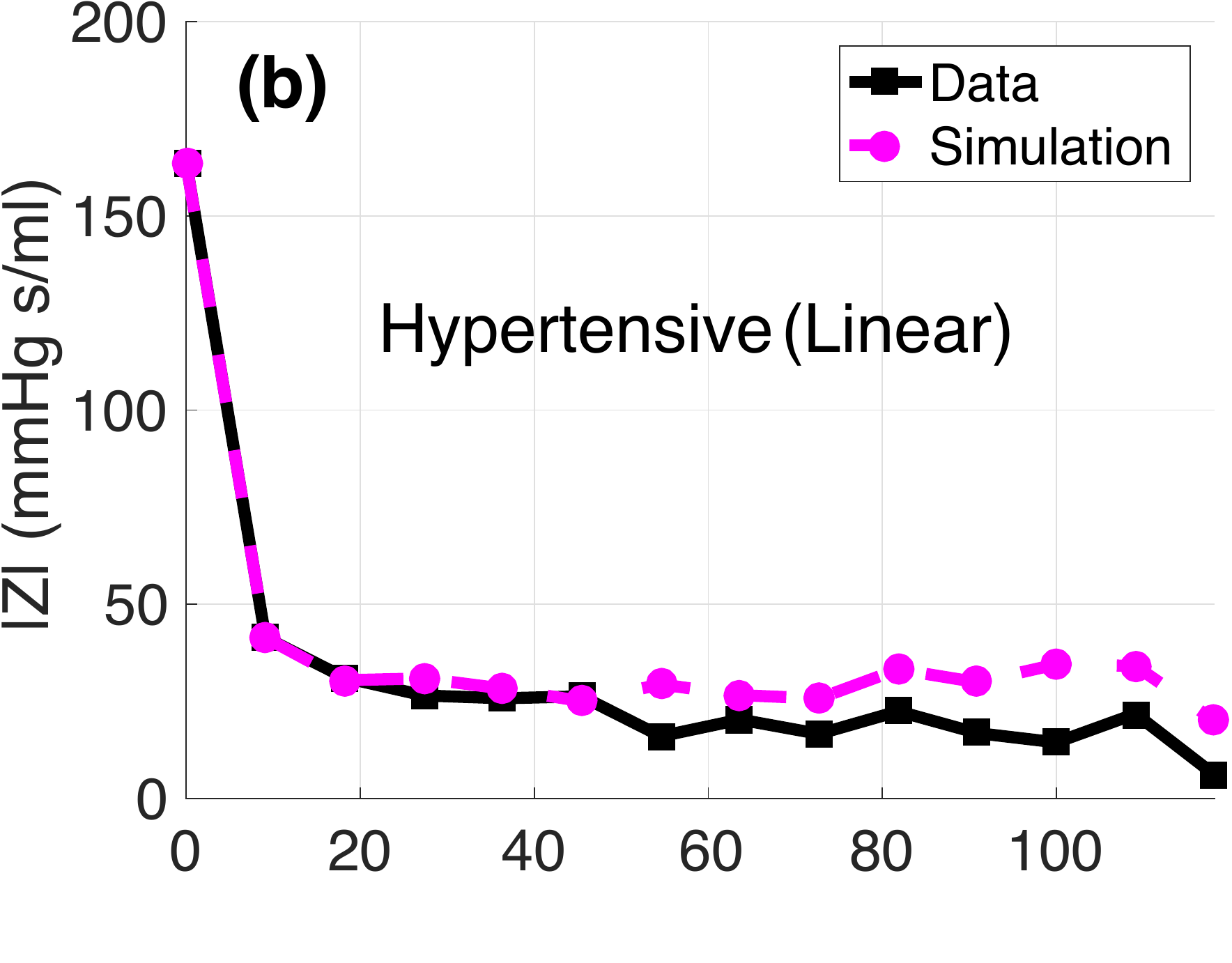}\vspace{-0.2cm}
\includegraphics[scale = 0.23]{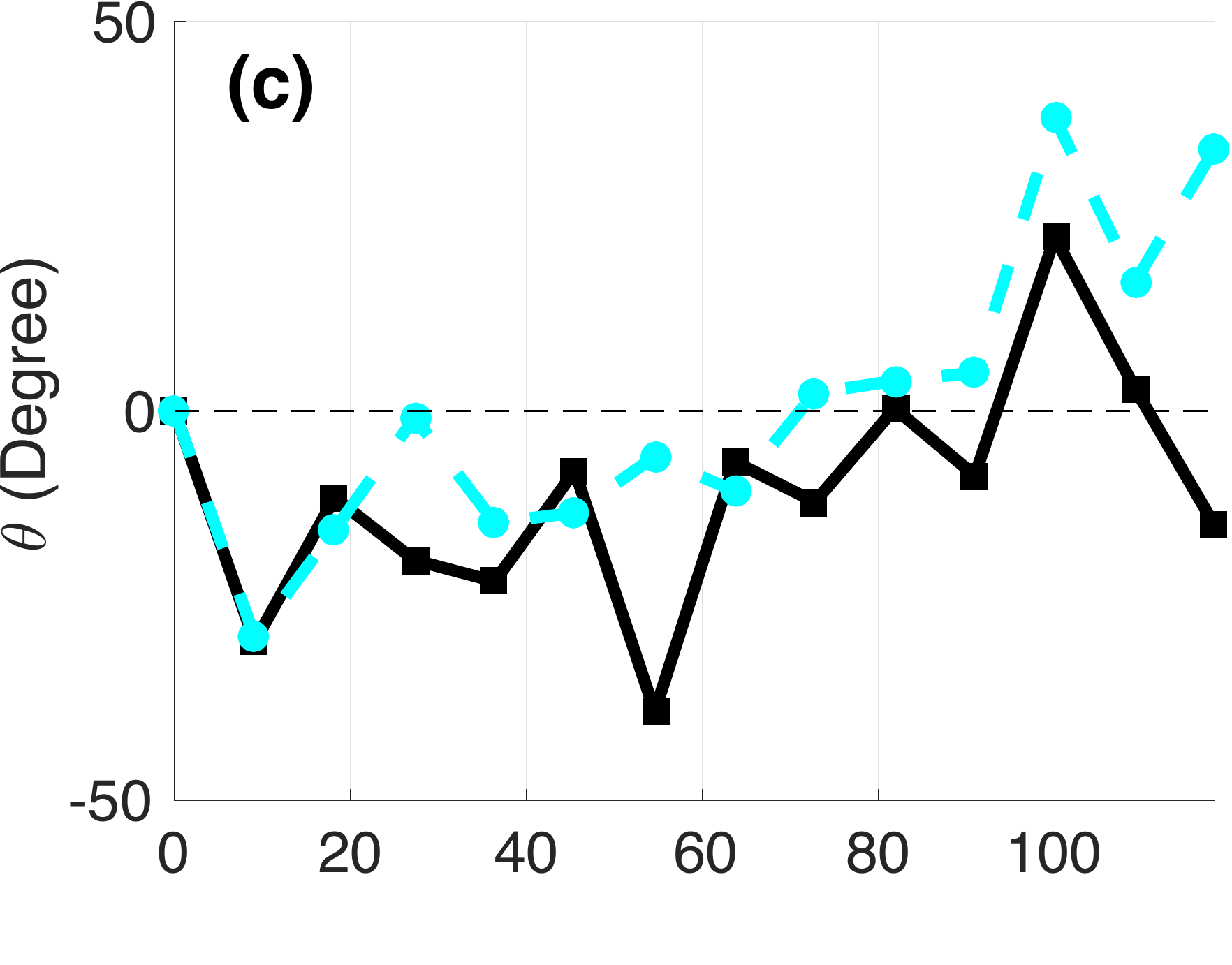}
\includegraphics[scale = 0.23]{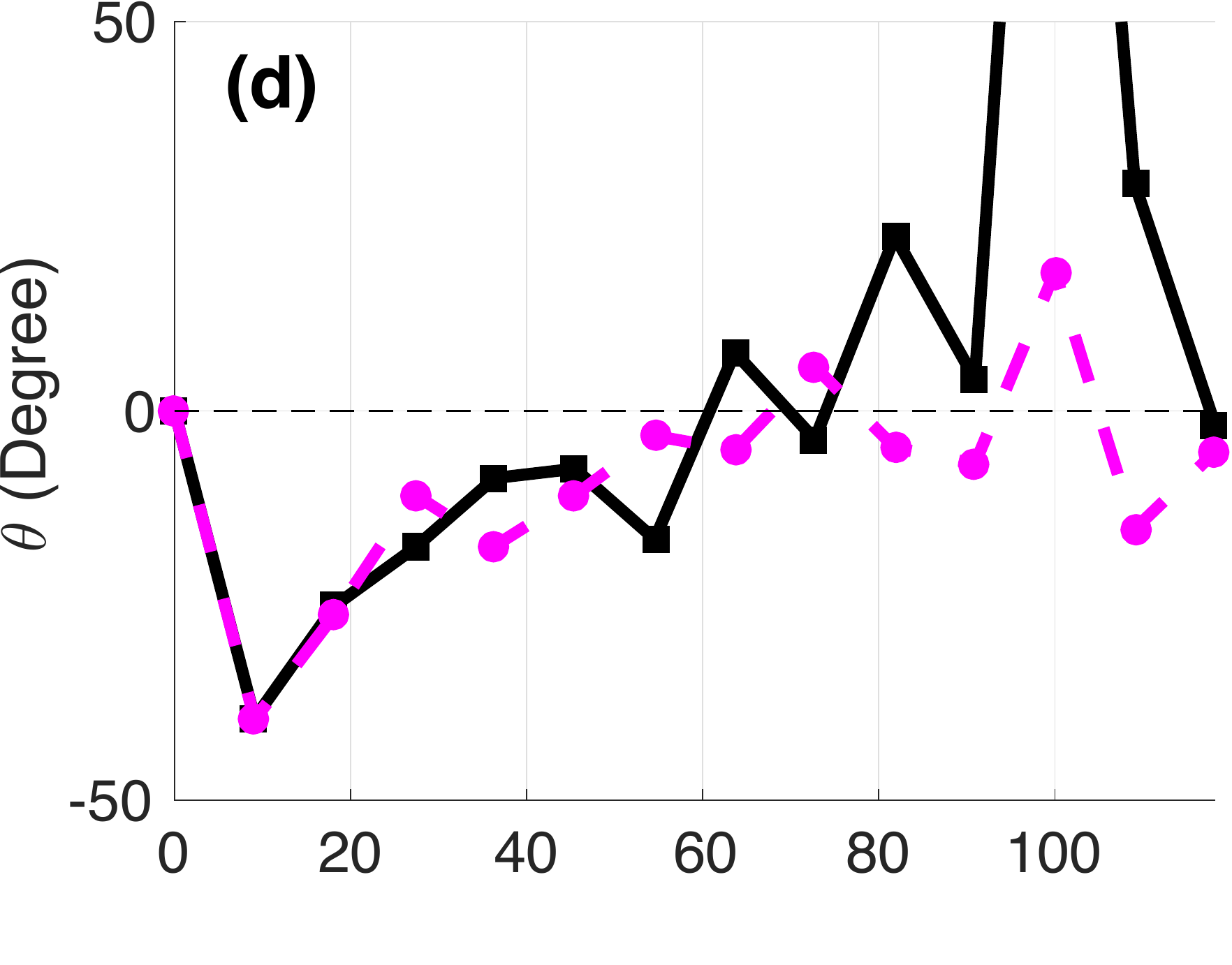}\vspace{-0.2cm}
\includegraphics[scale = 0.23]{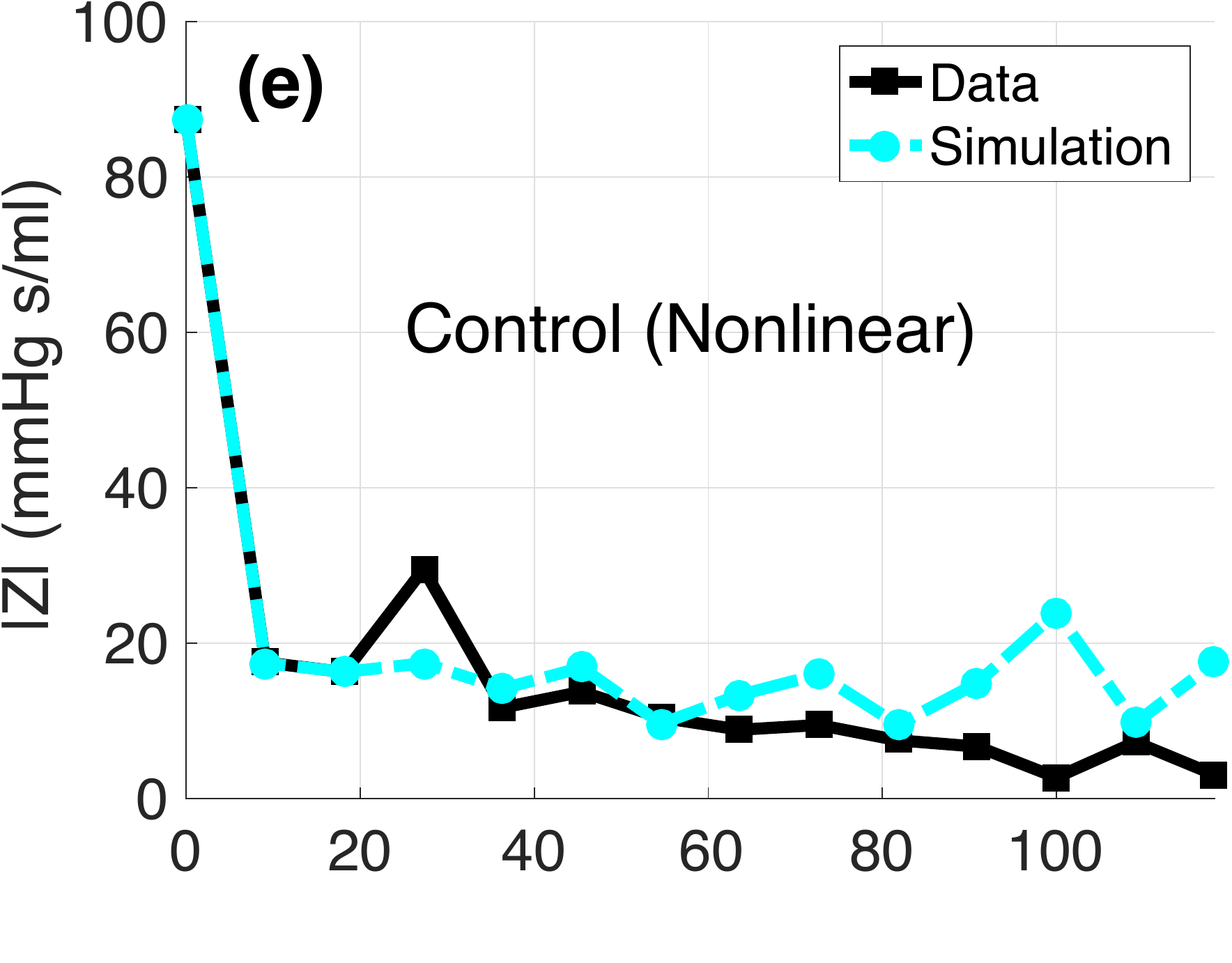}
\includegraphics[scale = 0.23]{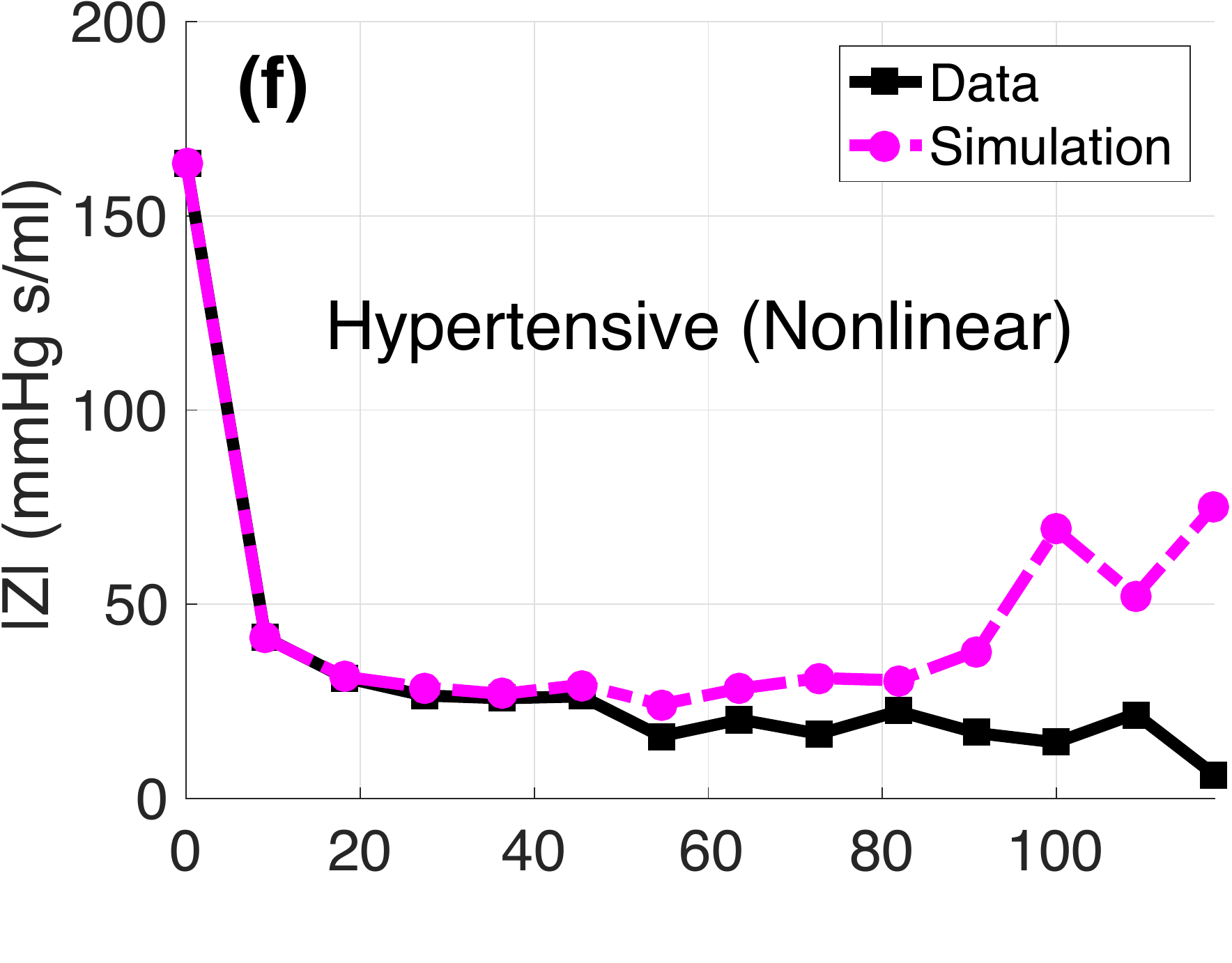}\vspace{-0.2cm}
\includegraphics[scale = 0.23]{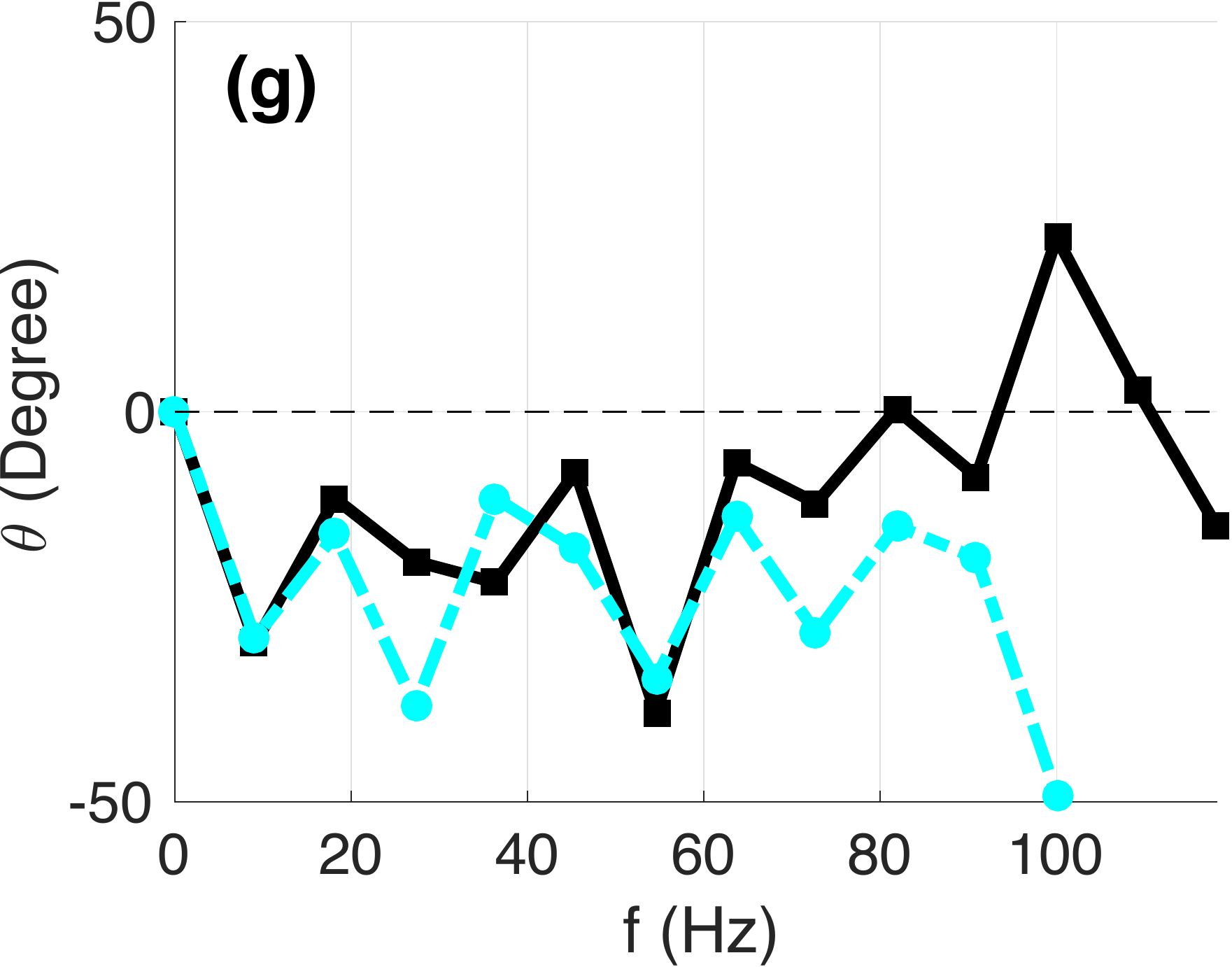}
\includegraphics[scale = 0.23]{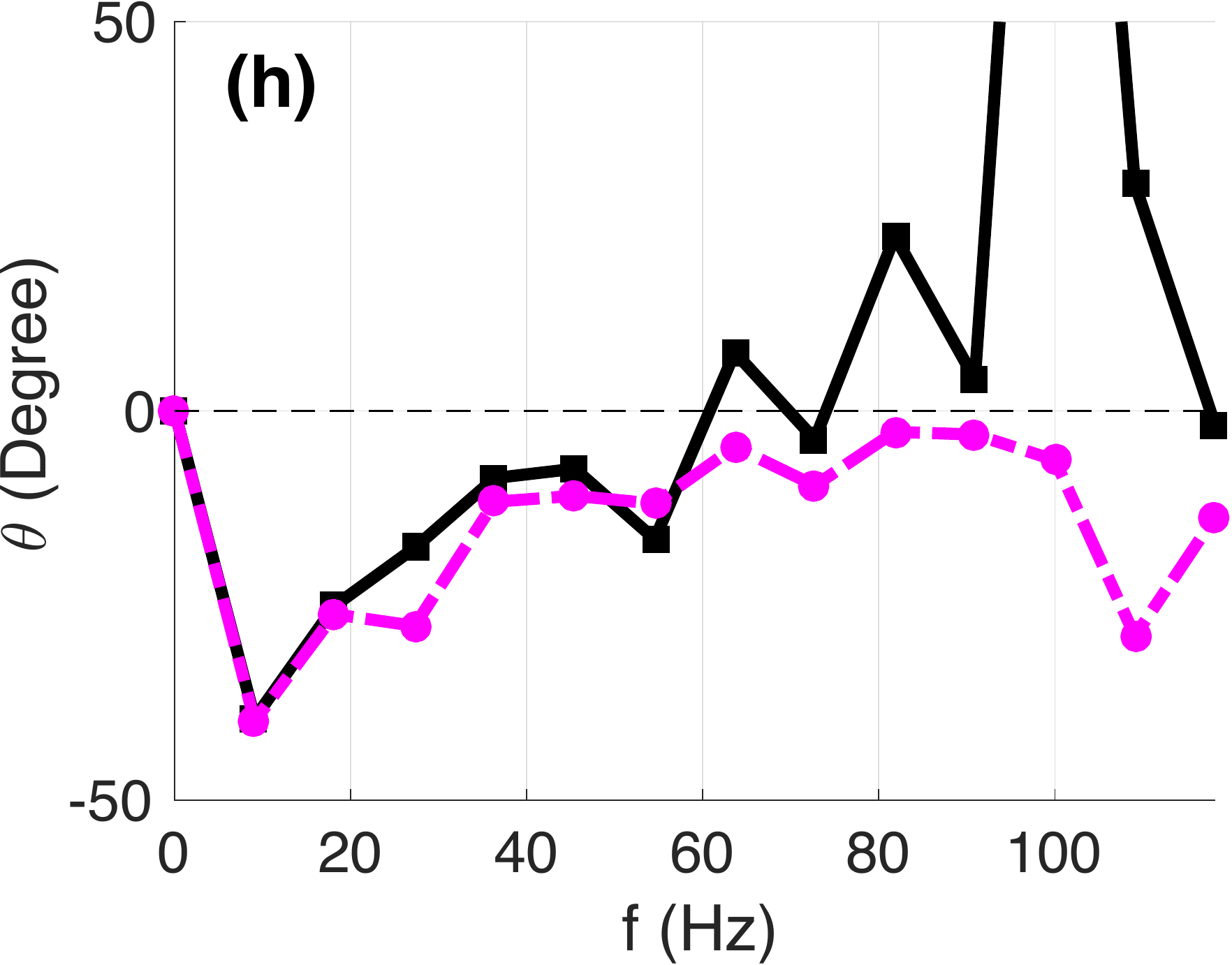}
\caption{\footnotesize Impedance spectra comparing the effects of linear (dashed line `--') and nonlinear wall models (dashed dotted line `-.') compared to data (solid black lines)  for a control (cyan) and a hypertensive (magenta) animal. Impedance moduli (a,b,e,f) and phase spectra (c,d,g,h) are plotted for the first 14 harmonics.}\label{fig:Impd}
\end{figure}

Figure~\ref{fig:Impd} depicts the impedance moduli $|Z|$ and phase spectra $\theta$ computed from the measured and simulated waveforms. Dashed lines show simulation results and solid black lines show data. Panels (a,c,e,g) show results from a control mouse and panels (b,d,f,h) show results from a hypertensive mouse. The impedance spectra were generated using~(\ref{eq:Zin}) and plotted for the first 14 harmonics including the mean component (zeroth harmonic). 

While time-varying simulations  fit the data well, Figure~\ref{fig:PQsim} shows characteristic differences in frequency domain signatures between the linear and nonlinear wall models. First,  the zeroth frequency components do not vary between models. Comparison of moduli spectra (a-d) show that the linear wall model better captures the frequency response of the original system. However, both models miss the spike in the impedance moduli at the 3rd harmonic (about 30\,Hz) observed for the control mouse (a,c). This should be contrasted with results from the hypertensive animal, where both models predict the low frequency behavior well (b,f). At higher frequencies, particularly after the 9th harmonic, the nonlinear wall model deviates from the measured impedance. In addition, the associated phase ($\theta$) dips below zero indicating persistence of pressure harmonics, which recede the flow harmonic. Again, the linear wall model deviates less at higher frequencies and its phase oscillates about zero. 

\subsection{Statistical analysis }

In this section we compare estimated  hemodynamics quantities pertinent to analysis of disease progression in HPH mice. To do so we calculate an importance index ($\eta$) computed from~(\ref{eq:sensetivity}) using quantities averaged across the CTL and HPH groups. 
\begin{figure}[h]
\centering
\includegraphics[scale = 0.23]{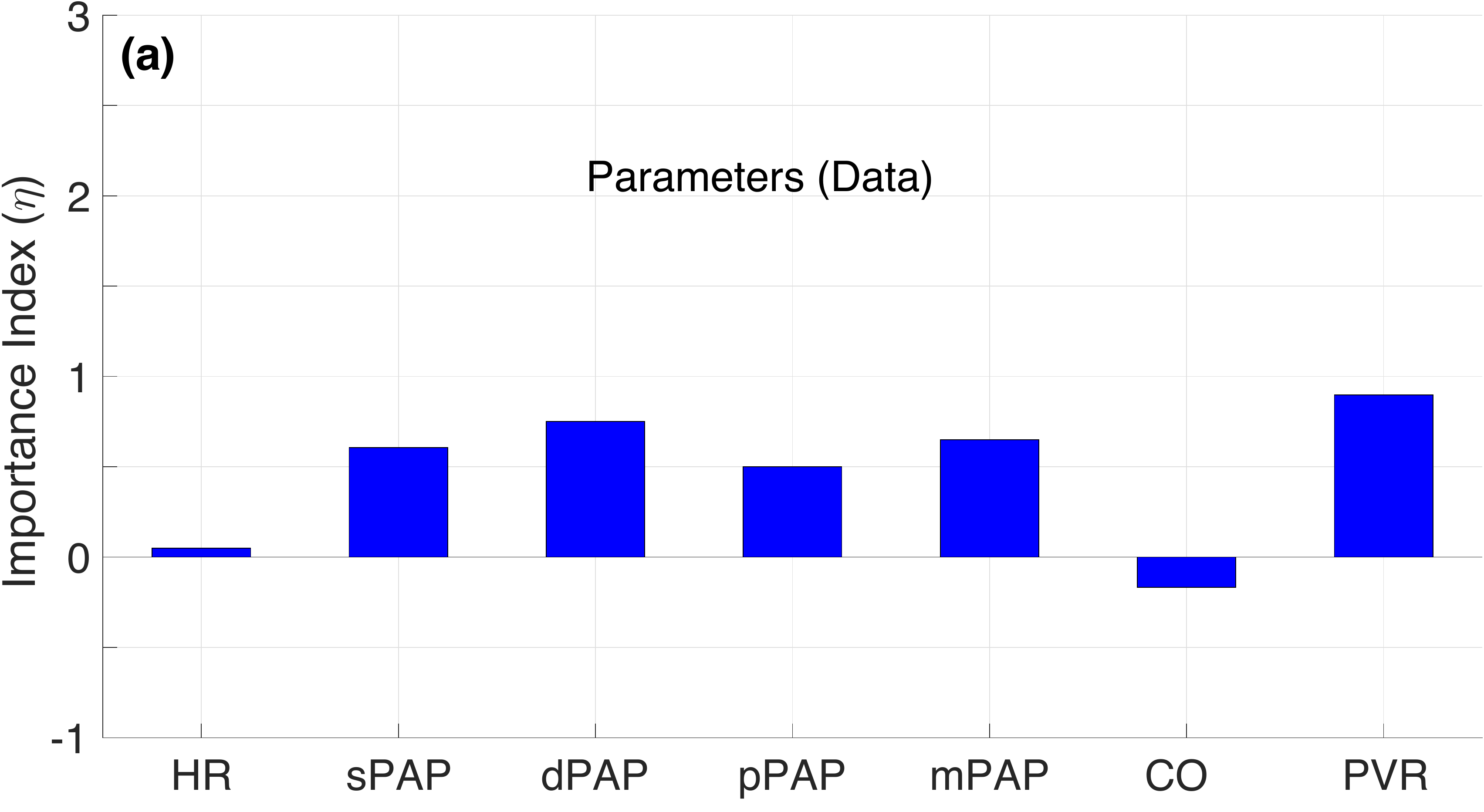}
\includegraphics[scale = 0.23]{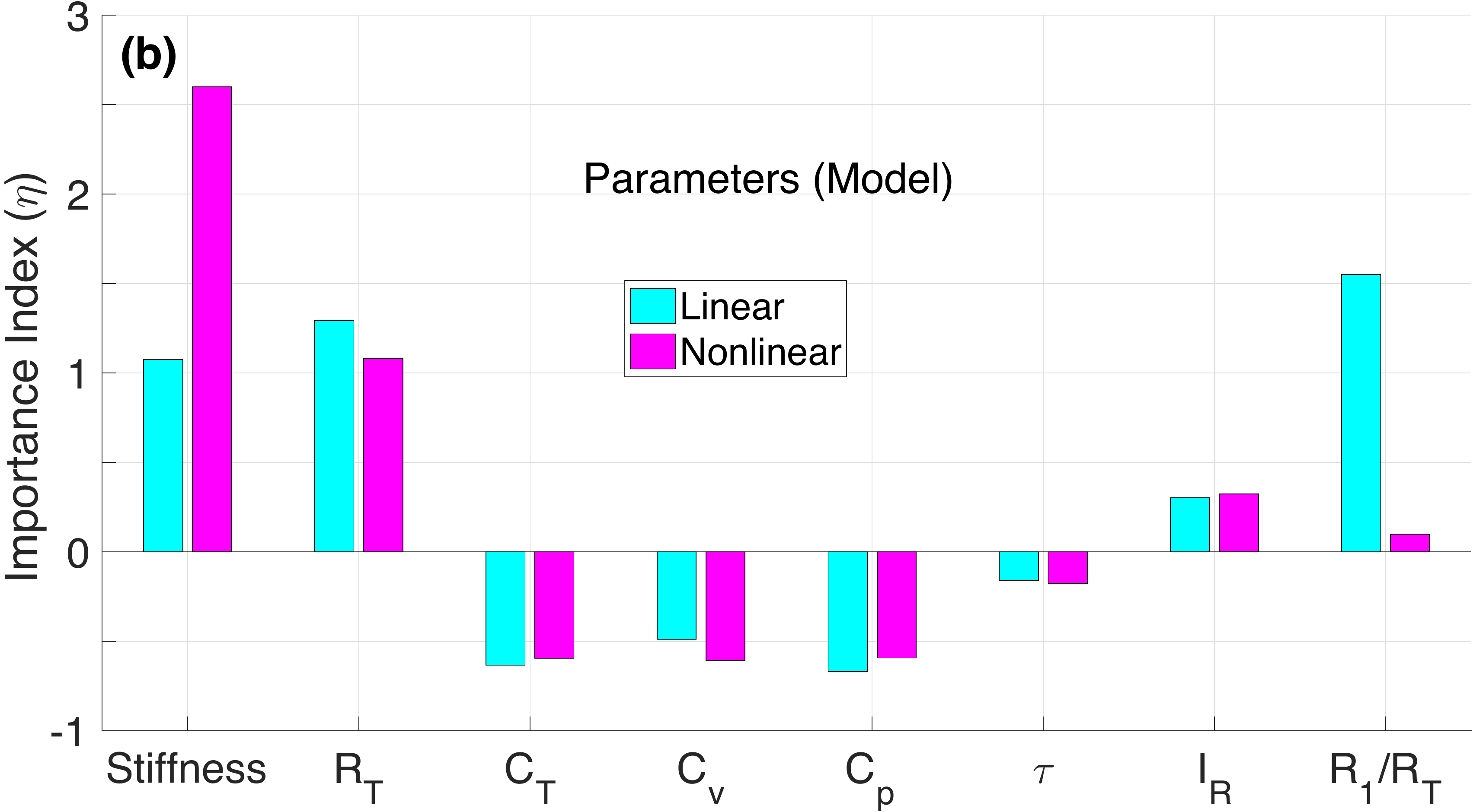}
\caption{\footnotesize (a) Importance  of essential hemodynamic parameters computed comparing model predictions with data. See Table~\ref{Tab:HemoStats} for abbreviations on the abscissa. The figure show predictions of arterial stiffness $\beta$ for the linear and $p_1/\gamma$ for the nonlinear model. In addition, we compare predictions of total vascular resistance ($R_T$), total vascular compliance ($C_T$), total network (vessel) compliance ($C_v$), total peripheral (vascular bed) compliance ($C_p$), characteristic time-constant ($\tau$), wave reflection coefficient ($I_R$), and resistance ratio ($R_1/R_T$).}\label{fig:ParAll_change}
\end{figure}

Figure~\ref{fig:ParAll_change}(a) shows predictions of $\eta$ for the essential cardiovascular quantities (summarized in Table~\ref{Tab:HemoStats}), whereas Figure~\ref{fig:ParAll_change}(b) does the same for the model parameters. Positive or negative values indicate an increase or a decrease in the quantity due to HPH. Moreover, a value of 1 (or -1) denote a 100\% change in the quantity.

Results show that peripheral vascular resistance ($R_T$) and compliance ($C_T,C_v,C_p$) significantly contribute to differentiating between control and disease for the hypertensive animals  ($\eta>1$), whereas the resistance ratio $R_1/R_T$ is important for predictions with the linear model ($\eta=1.5$), but not for predictions with the nonlinear model $\eta=0.1$. A similar observation was made for the nonlinear stiffness parameters,  
$p_1/\gamma$ that is more important than the linear stiffness parameter $\beta$. Though both of these contributed significantly to distinguishing control vs. hypertension. Finally, the wave reflection coefficient $I_R$ and the time constant $\tau$ also differ between control and disease but at a smaller scale ($\eta<0.5$). 

\begin{figure}[h]
\centering
\includegraphics[scale = 0.35]{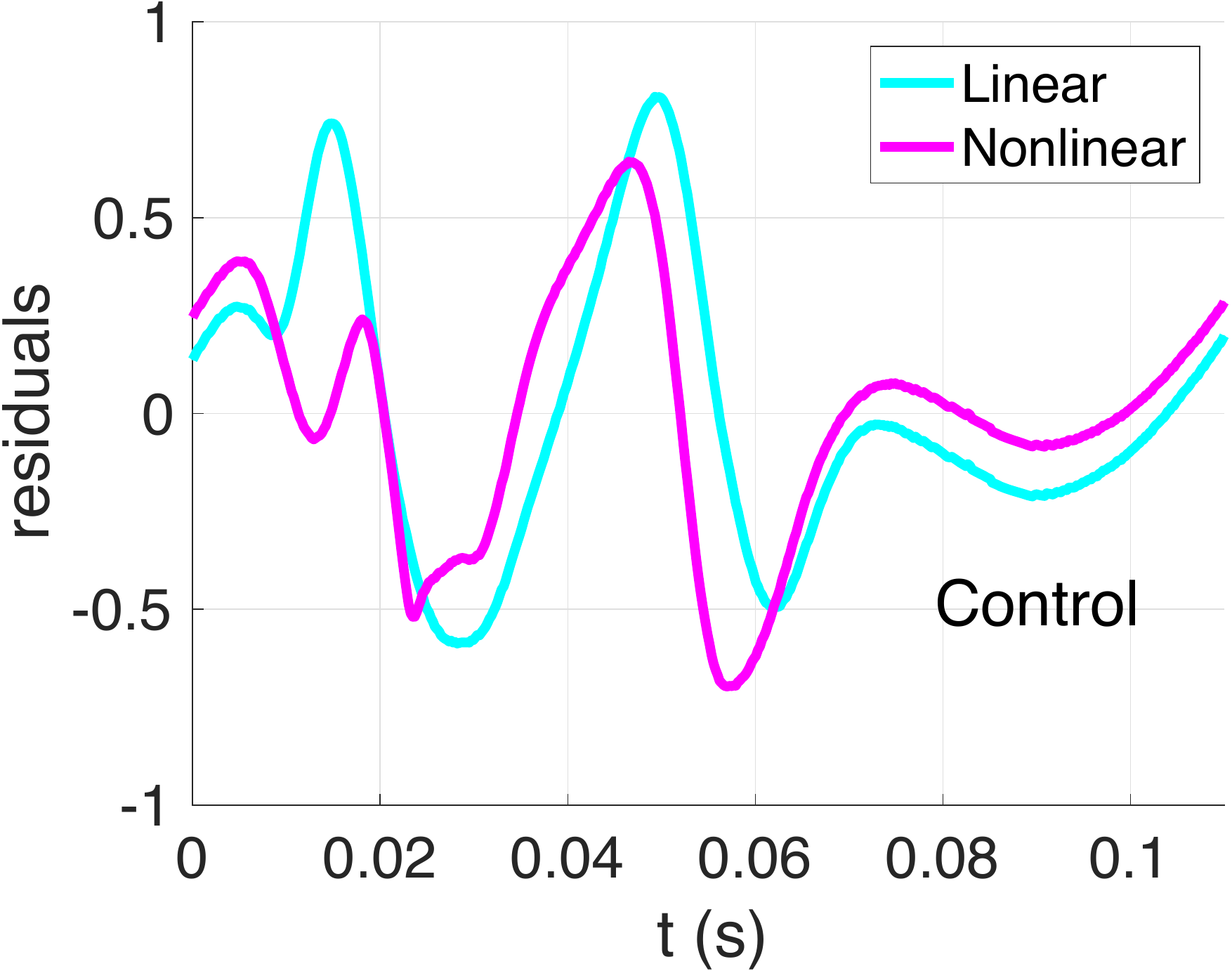}
\caption{Residual time series, as given by the difference between the measured and the simulated pressure signal corresponding to the linear and nonlinear wall model.}\label{fig:Residuals}
\end{figure}
Finally, we use statistical model selection criteria to determine which wall model is more consistent with the available hemodynamic data. Since the simple least squares error shows no difference between the two wall models for the hypertensive mouse, the model selection was done for the representative control mouse only. Specifically, we employ two model selection criteria (AICc and BIC), and model the residual correlation using the statistical ARMA model, and the GP models with different kernels (squared exponential, Mat\'ern 3/2, Mat\'ern 5/2, neural network and periodic kernel), as described in Sec.~\ref{sec:MethStat}. We estimated the hyperparameters of the GP covariance matrices from the time series of residuals plotted in Figure~\ref{fig:Residuals} by maximum likelihood. This was done by either using standard optimization algorithms involving maximizing the combined likelihood for the linear and the nonlinear models \citep{Rasmussen2006}, or by separately finding the hyperparameters for the linear and the nonlinear models and then averaging the covariance matrices. Since the correlation structure of the noise depends on the experimental protocol and is independent of the model assumptions, the later approach can be implemented under the constraint that the covariance matrices are the same for both wall models. We found that both approaches yield similar results to optimizing the GP hyperparameters for the residuals of both wall models. For simplicity, we applied the latter procedure to the ARMA model as well.

\begin{table}[h]
\centering
\caption{{AICc and BIC scores for the linear and the nonlinear wall model for the control mouse, with the covariance matrices obtained using three Gaussian Process (GP) models with Mat\'ern 3/2, Mat\'ern 5/2 and neural network covariance kernels. Lower AICc and BIC scores (in bold) indicate the better model.}}\label{Tab:AICcBIC}
{\footnotesize
 \setlength\tabcolsep{4.0pt}
\begin{tabular}{ccccc}
\hline\noalign{\smallskip}
\textbf{\makecell{Score\\ type}} & \textbf{\makecell{Wall \\ model}} &$\textbf{\thead{GP \\ \textrm{Mat\'ern 3/2}}}$ & $\textbf{\thead{GP \\ Mat\'ern 5/2}}$ &$\textbf{\thead{GP \\ neural network}}$ \\
\noalign{\smallskip}\hline\noalign{\smallskip}
AICc & Linear & \textbf{-621} & \textbf{-693} & \textbf{-652} \\ 
 AICc & Nonlinear & -607 & -681 & -641 \\ 
BIC & Linear & \textbf{-604} & \textbf{-676} & \textbf{-635} \\ 
BIC & Nonlinear& -586 & -660 & -620 \\ 
\noalign{\smallskip}\hline
\end{tabular}}
 \end{table}

We found that for three covariance matrices (GPs with squared exponential and periodic kernels and the ARMA model), the amount of regularization required for numerical stabilization was high ($\lambda>0.1$). Therefore, we discarded these results due to the high expected bias (although the results obtained with the ARMA model were consistent with our findings). We obtained the AICc and BIC scores for the covariance matrices that only required a low amount of numerical stabilization, $\lambda < 0.1$ (GP with Mat\'ern 5/2, Mat\'ern 3/2 and neural network kernel), and found that they were consistent with each other indicating lower AICc and BIC scores for the linear model compared to the nonlinear model, see Table~\ref{Tab:AICcBIC}.

\section{Discussion}\label{sec:discussion}
In this section we discuss our findings in physiological (CTL vs. HPH) and modeling (Linear vs. Nonlinear wall mechanics) contexts. 

\subsection{Inference of Disease Progression: CTL vs. HPH}\label{sec:diseaseinference}
In this study we used micro-CT images to set up anatomical networks and combined them with hemodynamic predictions from a 1D model. We calibrated the model for seven control and five hypertensive mice, estimating key parameters (large vessel stiffness, peripheral vascular resistance, peripheral vascular compliance, wave reflection coefficient) that characterize vascular remodeling due to  pulmonary hypertension.

\paragraph{Stiffness, compliance, and resistance} Comparison between the two groups (independent of the wall model) shows that hypertensive mice exhibit stiffer vessels both proximally (within the network), and in the microcirculation (represented by the Windkessel models) reflecting a rather advanced stage of HPH.  In this study, we attribute changes in  total resistance $R_T$ and compliance $C_T$ to both functional (vasoconstriction) and structural (rarefaction) microvascular remodeling, known to elevate the mPAP \citep{Qureshi2014,Wang2011}. Changes in the large vessel parameters, $p_1$ and $\gamma$ for the nonlinear model and $\beta$ for the linear model as well as the unstressed vessel radius $r_0$, contribute to the stiffness of the proximal vessels.

The results presented here show that the parameter with the largest impact is $R_T$. This is expected physiologically, as the disease progresses from the microvasculature to the large vessels. Almost as significant are changes in vessel stiffness ($p_1/\gamma$ and $\beta$) reflecting that the large vessels also stiffen, but with a relatively small change in $r_0$ indicating that the large vessels have not decreased in size. These results also reveal a significant decrease in compliance with hypertension. 

In summary, these findings suggest that the hypertensive animals analyzed here exhibit  remodeling of the entire vasculature, with distal vascular remodeling playing a dominant role in elevating the larger arterial blood pressure.  


\paragraph{Wave reflection.} Analysis of computed waveforms show that the wave reflection coefficient $I_R$ is significantly higher in hypertension (Figure~\ref{fig:WIA}\,(c,d) and~\ref{fig:ParAll_change}). This observation can be explained by hypothesized proximal and distal remodeling. As shown in Figure~\ref{fig:WIA}\,(a,c,e,g) the system has some degree of compressive wave reflections under control conditions. Our results show that the coefficient is amplified in hypertension (Figure~\ref{fig:WIA}\,d) as the remodeled (stiffened) proximal arteries require a higher ejection pressure from the heart to maintain the cardiac output, which gets reflected as an augmented pressure wave. 

The advantage in using $I_R$ is that it provides insight into systemic effects rather than focusing on either small or large vessels, and as such $I_R$ could be an important indicator of disease progression. However, its usefulness cannot be fully analyzed in the current model since it employs a fixed inflow, simulating the cardiac compensation scenario during which the cardiac output is maintained with hypertension \citep{Acosta2017}. A model including a heart component (e.g. \cite{Acosta2017,Mynard2015}) and 1D vascular beds (e.g \cite{Chen2016,Olufsen2012,Qureshi2014}) would be ideal to analyze the complex role of wave reflections in disease progression.

\paragraph{Impedance analysis} Overall, the characteristic features of the impedance spectra resemble those reported by \cite{Nichols2011} (Ch. 16). We note that the impedance moduli are higher in hypertension, whereas the phase show that it starts out negative and then crosses zero at higher frequencies. These results supplement observations from \cite{Vanderpool2011}, who studied the impedance spectra using {\it in vitro} pulsatile hemodynamics in isolated lungs of the same type of mice.

The results of low frequency components are more subtle. One observation is that the low frequency response is more dynamic in the control animals, reflecting that large vessels are more compliant. A characteristic feature that we were not able to reproduce was a pronounced third harmonic observed in control animals. To our knowledge, this feature has not been reported elsewhere, and could be a consequence of vessel tapering or wave propagation from the vascular beds (which are not included in this study), or it could be an artifact from cycle averaged data. Unfortunately we do not have access to the raw data to confirm or deny this characteristic. \\

In summary, pulmonary hypertension is characterized by a more resistive and less compliant vasculature with augmented wave reflections, leading to high blood pressure in the pulmonary arteries. This is in line with observations by \cite{Lankhaar2006} and \cite{Lungu2014} for  patients with and without pulmonary hypertension. \cite{Lankhaar2006} used a 0D Windkessel model neglecting the effects of wave propagation and arterial stiffness, whereas \cite{Lungu2014} used a coupled 0D-1D model representing the pulmonary vasculature by a single vessel. 

\subsection{Inference of Disease Progression: Linear vs. nonlinear wall mechanics}\label{sec:EstParDisc}
It is well known that arterial deformation acts nonlinearly imposing increased stiffening with increased pressure \citep{Valdez2011}. Although, most studies confirming this behavior are done in systemic arteries \citep{Eck2017,Langewouters1985,Valdez2011}, pulmonary arteries are composed of the same type of tissue and therefore should exhibit similar behavior under dynamic loading \citep{Pilhwa2016}. To understand the effects of nonlinearities on hemodynamic predictions and parameter estimation, we implemented a linear mechanistic and a nonlinear empirical wall model. A qualitatively reasonable outcome, i.e. increased stiffening with pressure, in the form of nonlinear pressure--area curve is evident from Figure~\ref{fig:PAsim}. However, in the absence of actual data for area deformation, the area predictions using the linear and nonlinear models cannot be validated.

\paragraph{The nonlinear wall model.}\label{sec:WallMechDisc}
Previous empirical nonlinear stress strain models are formulated using sigmoidal functions, relating pressure and area, saturating at both high and low pressures \citep{Langewouters1985,Valdez2011}. This type of model is characterized by an inflection point determined by a parameter representing the half-saturation (or maximum compliance) pressure. \cite{Valdez2011} validated these models against dynamic loading data, from ovine thoracic descending aorta (TDA) and carotid artery (CA) under {\it in vivo} and {\it ex vivo} conditions, showing that pressure-area dynamics lie on the upper (concave down) part of the sigmoidal curves. While these models were able to fit data well, their estimates of the half-saturation parameter $p_0$ lie outside the known physiological range. 

The advantage of the model used here, is that it provides similar estimates, but uses one less parameter, eliminating the need to estimate a parameter that likely is unidentifiable.  Moreover, our model provides a  basis for theoretical comparison with the linear model at a reference state $(A,p)=(A_0,0)$, since under the small strain assumption, the nonlinear wall model (\ref{eq:nlinWall}) approximates to
\begin{equation}\label{eq:LinnLin}
p \approx \frac{2\pi p_1}{\gamma}\left(\sqrt{\frac{A}{A_0}}-1\right)+O\left(\sqrt{\frac{A}{A_0}}-1\right)^2,
\end{equation}
which is the linear model with $\beta = 2\pi p_1/\gamma$ (see (\ref{eq:nomnLin})). This property justifies the interchangeable use of $c_{0.\text{lin}}$ and $c_{0.\text{nlin}}$ in (\ref{eq:Zc}), used to estimate nominal values for $p_1$ and $\gamma$. 

In summary,  simulations with the linear wall model are predominantly governed by the 0D boundary conditions, i.e. vascular beds, whereas the nonlinear model modulates both the 1D, i.e. large vessels and 0D domain to predict the hypertensive hemodynamics. This suggests that the linear wall model predicts greater remodeling of the vascular beds due to HPH. However, the general inferences about control and hypertensive hemodynamics, decreased compliance, increased stiffness, resistance and amplitudes of wave reflections, remain the same. 

The results shown here indicate that measurements in the MPA can be predicted with either wall model, but predictions in the small vessels differ. Without more data confirming behavior in small vessels it is not possible to conclude which model is better. Moreover, we showed that in hypertensive animals both models provide comparable predictions, likely a result of increased vessel stiffness. 

\subsection{Model selection} 
To our knowledge no previous 1D wave propagation studies have implemented statistical criterion for model selection. 
In this study we have carried out statistical model selection to discriminate between the linear and the nonlinear wall model given available data in the main pulmonary artery. Using AICc and BIC scores. While we have made a parametric assumption about the measurement noise, we have taken its correlation structure into consideration by fitting a set of standard time series models to the residuals. We have focused on the control mice, for which the difference between the linear and the nonlinear model was significant. Our results suggest that the linear model is preferred. One study by \cite{Valdez2010}, examining wall properties used the Akaike Information Criterion (AIC) for selecting the wall model. Their results showed that for control animals (they studied sheep) the nonlinear wall model performs better, whereas for stiffer vessels, the linear model performs better. Their former conclusion is contradictory to our model selection analysis of control mouse but the later is consistent with our findings, which show no difference in the predictions using the linear and the nonlinear wall models.  However, \cite{Valdez2010}only examined the stress-strain relation in single vessels in absence of fluid dynamics. It should be noted that results presented in this study were done using classical AICc and BIC selection criteria, which have an asymptotic justification. Better approximations that are less reliant on asymptotics are the Watanabe-Akaike information criterion (WAIC) \citep{Watanabe2010} or the Watanabe-Bayesian information criterion (WBIC) \citep{Watanabe2013}. However, these more accurate model selection methods, which we have explored in a recent proof-of-concept study \citep{Paun2018}, are based on Monte Carlo simulations and are thus computationally considerably more expensive.

\subsection{Limitations}\label{sec:Limitations}  In this study, we did not account for uncertainties in network dimensions and topology associated with image segmentation. We also ignored the uncertainty in the hemodynamic data resulting from  ensemble encoding of pressure and flow waveforms over multiple cardiac cycles.  Although uncertainty quantification is beyond the scope of this study, it is an important aspect that may have significant impact on the parameter inference. Moreover, it is not clear from image studies if the individual vessels taper, thus the effects of vessel tapering were not considered in this study. Yet, it is known that tapering introduces significant augmentation of pressure along each vessel and throughout the network, which makes it a sensitive model parameter, which could be investigated. Other modeling limitations of this work include the use of a fixed inflow and lumped 0D model for the vascular bed. Finally, the model selection outcome is only valid for the current hemodynamic data, which is available from one location in the main pulmonary artery. The AICc and BIC scores may vary significantly if the waveforms are available from multiple location within the network. 

\section{Conclusions}\label{sec:conclusion}
We found that the hypertensive mice display significant disease progression associated with remodeling within both large and small vessels. Microvascular remodeling includes reduced compliance and increased resistance, augmenting wave reflections, that combined with increased stiffening of the large vessels lead to significant increase in blood pressure. We also conclude that both linear and nonlinear models can be used to predict the control and hypertensive hemodynamics in the MPA with high accuracy, yet the prediction in the smaller vessels differ. Without more data, it is not possible to select which model better reflects wave propagation along the entire network. These differences were only displayed for control mice, with more compliant vessels. For the hypertensive mice, both large and small vessels are almost rigid and the two models predict the same behavior. Although, the model selection criteria pick the linear wall model for the control muse, these results should not be considered in the statistical context alone as the availability of more physiological data for optimization may alter the present outcome. Finally, analysis of network hemodynamics, wave intensities and impedance moduli indicates an increased presence of wave reflections using the nonlinear model. For this reason, parameter inference and characterization of normal and remodeled vasculature should be regarded as qualitative while using our nonlinear model. 

\paragraph{Acknowledgements\\}

{\small{\bf Funding} This study was supported by the National Science Foundation (NSF) awards NSF-DMS \# 1615820, NSF-DMS \# 1246991 and Engineering and Physical Sciences Research Council (EPSRC) of the UK, grant reference number EP/N014642/1.\\

\noindent\textbf{Conflict of interest} The authors declare that they have no conflict of interest.}

\appendix
\section*{Appendix}
\section{Total Vascular Compliance}\label{app:compliance}

The volumetric compliance, defined as $C= dV/dp$ (ml/mmHg) for a cylindrical vessel with volume $V$, is computed from the linear ($C_\text{lin}$) and nonlinear ($C_\text{nlin}$) models. For a longitudinally tethered vessel $i$ in the network
\begin{equation}\label{eq:volCompliance}
C = \frac{dV}{dp} \equiv L\frac{dA}{dp},
\end{equation}
where $L$ is the fixed length of the vessel and $dA/dp$ is computed from~(\ref{eq:linWall}) and (\ref{eq:nlinWall}), giving 
\begin{equation}\label{eq:Compliance}
C_\text{lin}= C_{0.\text{lin}}\left(\frac{p}{\beta}+1\right),\qquad \text{and}\qquad
C_\text{nlin}= C_{0.\text{nlin}}\left(\frac{p_1^2}{p^2 + p_1^2}\right), 
\end{equation}
where $C_0$ denotes the reference compliance at $p=0$, given by
\begin{equation}\label{eq:RefCompliance}
 C_{0.\text{lin}} = \frac{2A_0L }{\beta},\qquad\text{and}\qquad C_{0.\text{nlin}}=\frac{\gamma A_0 L}{\pi p_1}.
\end{equation}

Computation of the total compliance $C_v$ in a bifurcating network requires compliances to be added in series and in parallel. The total compliance of two vessels $A$ and  $B$ connected in parallel ${A||B}$ and in series ${A\leftrightarrow B}$ are given by
\begin{equation}\label{eq:Cparser}
C_{A||B} = C_A + C_B,\quad \text{and}\quad C_{A\leftrightarrow B} = \frac{C_AC_B}{C_A+C_B}.
\end{equation}
As an example, the total compliance in a single bifurcation network shown in Figure \ref{fig:parEst}c, is given by 
\begin{equation}\label{eq:Cnet}
C_v = C_{{p}\leftrightarrow({d_1||d_2})}.
\end{equation}

The total vascular compliance $C_T$ in a vascular region consisting of single bifurcation network and two vascular beds is given by
\begin{equation}
C_T = C_v + \sum_{j = 1} ^2\widehat{C}_{p,d_j},
\end{equation}
where $C_v$ is computed for the linear or nonlinear wall models using (\ref{eq:volCompliance})--(\ref{eq:Cnet}), while $\widehat{C}_{p,j}$ is the weighted compliance, as suggested by \cite{Alastruey2016}, i.e. $\widehat{C}_{p,d_j}=(R_{2,d_j}/R_{T,d_j})C_{p,d_j}$ where $C_{p,d_j}$ is estimated from the Windkessel models attached to the terminal daughter vessels $d_j$ in the example network. 

\section{Pulse wave velocity}\label{app:PWV}

The pulse wave velocity (PWV), $c$ (cm/s), is computed from the eigenvalues of the hyperbolic system of equations (\ref{eq:mom}), from. $\lambda_{1,2} = q/A \pm c$ where
\begin{equation}\label{eq:PWV}
 c=\sqrt{\frac{A}{\rho}\frac{d p}{dA}} = \sqrt{\frac{AL}{\rho\, C_v}}.
\end{equation}
Setting $C_v = C_\text{lin}$ and $C_\text{nlin}$ in (\ref{eq:PWV}) gives the squared PWV computed for the linear and nonlinear wall models, respectively 
\begin{eqnarray}
c^2_\textrm{lin}&=&c^2_{0.\text{lin}}+\frac{p}{2\rho},\nonumber\\
c^2_\textrm{nlin} &=&c^2_{0.\text{nlin}}\left( \frac{\gamma}{\pi}\tan^{-1}(p/p_1)+1\right)\left(1+(p/p_1)^2\right)\label{eq:PWVnlin},
\end{eqnarray}
where $c_0^2$ is the square of the reference PWV at $p=0$, given by
\begin{equation}\label{eq:c0}
c^2_{0.\text{lin}}=\frac{\beta}{2\rho},\qquad \textrm{and}\qquad
c^2_{0.\textrm{nlin}}=\frac{\pi p_1}{\gamma\rho}.
\end{equation}
We use PWV in wave intensity analysis, described in Sec.~\ref{sec:WIA&IA}, for separating  the incident and reflected waves.

\section{Nominal parameter values}\label{app:ParametersNom}

\begin{table}[h]
\centering
\caption{Nominal values for wall parameters and compliance for individual mice in each group.}
\label{Tab:ParNom}    
\begin{tabular}{llll}
\hline\noalign{\smallskip}
{\bf Control} & $\beta_0$ (mmHg) & $R_{T_0}$ (mmHg s/ml)&$\tau_0$ (s)  \\
\noalign{\smallskip}\hline\noalign{\smallskip}
1 & 37.5 &108 & 0.15   \\
2 & 44.5 &56   & 0.06  \\
3 & 37.7 &47  &  0.06 \\
4 & 40.7 &70  &  0.10\\
5 & 15.6 &69  &  0.13 \\
6 & 26.0 &87  &  0.14 \\
7 & 31.7 &101&  0.13\\
\noalign{\smallskip}\hline\noalign{\smallskip}
{\bf Hypertensive} &  $\beta_0$ (mmHg) & $R_{T_)}$ (mmHg s/ml)&$\tau_0$ (s)  \\
\noalign{\smallskip}\hline\noalign{\smallskip}
1 & 150.6 &164   &  0.09  \\
2 & 56.8 &107   &  0.11 \\
3 & 123.8 &163   &  0.15 \\
4 & 100.7 &154   &  0.13 \\
5 & 78.5 &143   &  0.11 \\
\noalign{\smallskip}\hline
\end{tabular}
\end{table}

As described in Sec.~\ref{sec:ParInit}, nominal values for nonlinear model are set as $p_1=\beta/\pi$ and $\gamma = 2$ for all cases. Moreover, nominal values for $R_{T,j}$ are computed from $R_T$ reported in the Table above using methods described in Sec.~\ref{sec:ParInit} and the network dimensions stated in Table~\ref{Tab:Network}. For all cases, the resistance ratio $a\equiv R_1/R_T = 0.2$.

\section{Optimized parameter values}\label{app:Parameters}
For all cases, we optimized $\beta, \gamma,$ and $ p_1$ for the wall models, and the global scaling parameters $r_1, r_2, c_1$ for the Windkessel model, such that
\[{R}_{1,j} = r_1R_{10,j}, \quad {R}_{2,j} = r_2R_{20,j},\quad {C}_{p,j} = c_1C_{p0,j}\]
where $0$ indicate the nominal quantity. Upper and lower bounds for the optimization intervals are given in Table~\ref{Tab:Bounds}.

\begin{table}[h]
\centering
\caption{Bounds for optimization.}\label{Tab:Bounds}
      
\begin{tabular}{lllll}
\hline\noalign{\smallskip}
Parameter & $\beta$ & $p_1 $  &$\gamma$ & ($r_1,r_2,c_1$)\\
\noalign{\smallskip}\hline\noalign{\smallskip}
Lower bound & $\beta_0$ &$\beta_0/2\pi$ &$1$&0.05\\
Upper bound & $2.5\beta_0$ & $2\beta_0$&$2\pi$&2.5 \\
\noalign{\smallskip}\hline
\end{tabular}
\end{table}
 \begin{landscape}
 \begin{table}[h]
\caption{Optimized values for individual mice in each group. Parameters were computed back from optimized values of $\beta$, $p_1$, $\gamma$, $r_1$, $r_2$ and $c_1$.}\label{Tab:Parameters} 
 {\footnotesize 
 \begin{tabular}{l lllllll | llllllllll}
\hline\noalign{\smallskip}
  \multicolumn{11}{c}{\bf Linear} & \multicolumn{2}{c}{\bf Nonlinear}\\
  \noalign{\smallskip}\hline\noalign{\smallskip}
{\bf Control} & $\beta$ & $R_T$&$C_T\times10^{-3}$ &$I_R\times10^{-2}$ & $C_p/C_T$ & $a$ (\%) &$S$& $p_1$ & $\gamma$  &$R_T$ & $C_T\times10^{-3}$ &$I_R\times10^{-2}$ &$C_p/C_T$& $a$ (\%)&$S$\\
\noalign{\smallskip}\hline\noalign{\smallskip}
 1& 67.5&90.7  &2.52& 29.6 & 97.1 &2.9&199&    28.6&2.6&94.1&1.67& 31.7   &94.0  &8.6&147\\[2pt]
 2& 61.9&28.0  &3.15& 30.4 & 97.9 &4.5&640&    41.0&2.2&34.7&2.33& 31.8   &98.5  &12.4&654\\[2pt]
 3& 36.1&22.4  &2.43& 40.3 &95.2 &8.4&587&    15.3&2.5&34.4&1.96 & 33.2  &95.9  &6.0&376\\[2pt]
 4& 56.3&46.9  &3.22& 29.8 &97.7 &3.0&369&     23.9&3.0&58.9&1.88& 33.8   &97.7  &11.0&348\\[2pt]
 5& 49.9&52.3  &3.42& 31.7 &97.4 &4.1&606&    19.8&4.7&65.1&2.24 & 33.4   &96.0 &13.5&702\\[2pt]
 6& 52.1&65.6  &2.22& 36.0 &96.2 &5.4&125&     22.1&4.0&78.7&1.44& 38.8   &94.8  &12.3&99\\[2pt]
 7& 70.4&102.9&2.54& 32.3 &97.6 &5.8&737&    47.8&3.5&99.8&2.32 & 32.5   &98.2  &11.6&698\\
\noalign{\smallskip}\hline\noalign{\smallskip}
{\bf HPH} & $\beta$ & $R_T$ &$C_T\times10^{-4}$&$I_R\times10^{-2}$ & $C_p/C_T$ & $a$ (\%) &$S$& $p_1$ & $\gamma$  &$R_T$ & $C_T\times10^{-4}$ &$I_R\times10^{-2}$ &$C_p/C_T$& $a$ (\%) &$S$\\
\noalign{\smallskip}\hline\noalign{\smallskip}
 1& 151.3&146.2&4.52& 56.5  &93.6  &15.4&49  &      102.7&1.6&147.5&4.79& 58.4   &97.5 &12.9&32\\[2pt]
 2& 86.3 & 94.3  &0.13& 36.3  &95.9 &10.2&593&       36.6 &2.6&100.9&0.12 & 39.3   &96.6  &13.0&450\\[2pt]
 3&121.0&155.6 &9.41& 33.3  &96.0 &7.0&407  &       82.2  &2.1&158.7&8.79& 35.7   &98.0  &6.9&467\\[2pt]
 4&117.4&142.2 &8.61& 42.9  &95.5 &14.4&63  &       49.8  &2.2&146.4&7.36& 44.3  &96.5  &13.4&41\\[2pt]
 5&108.2&130.9 &7.81& 48.0  &94.5 &15.0&113&       73.5  &2.3&133.5&6.82& 49.2  &95.8  &13.0&42\\
 \noalign{\smallskip}\hline
  \end{tabular}}\vspace{0.2cm}\\
   {\footnotesize  $\beta$ (mmHg), $R_T$ (mmHg\,s/ml), $C_T$ (ml/mmHg), $I_R, \ C_p/C_T \ ,a$ (dimensionless), $p_1$ (mmHg), $\gamma$ (dimensionless), $S$: least square error. \\}
\end{table}
\end{landscape}

\end{document}